\documentclass{article}

\usepackage[utf8]{inputenc} 
\usepackage[T1]{fontenc}    
\usepackage{hyperref}       
\usepackage{url}            
\usepackage{booktabs}       
\usepackage{amsfonts}       
\usepackage{nicefrac}       
\usepackage{microtype}      
\usepackage{xcolor}         
\usepackage{tabularx}
\usepackage{graphicx}
\usepackage{amssymb}
\usepackage{amsmath}
\usepackage{enumitem}
\usepackage[margin = 1 in]{geometry}
\usepackage{natbib}

\usepackage{comment}
\usepackage{multirow}
\usepackage{lmodern}

\newtheorem{definition}{Definition}
\newtheorem{proposition}{Proposition}

\title{Topo-Geometric Analysis of Variability in Point Clouds using Persistence Landscapes}

\author{
  James Matuk \\
  Epidemiology Data Center\\
  University of Pittsburgh\\
  Pittsburgh, PA, USA \\
  \texttt{JAM925@pitt.edu} \\
   \and
   Sebastian Kurtek \\
  Department of Statistics\\
  The Ohio State University\\
  Columbus, OH, USA \\
   \texttt{kurtek.1@stat.osu.edu} \\
   \and
   Karthik Bharath \\
   School of Mathematical Sciences \\
   University of Nottingham\\
   Nottingham, UK \\
   \texttt{Karthik.Bharath@nottingham.ac.uk} 
}
\date{}

\begin{document}

\maketitle
\begin{abstract}
Topological data analysis provides a set of tools to uncover low-dimensional structure in noisy point clouds. Prominent amongst the tools is persistence homology, which summarizes birth-death times of homological features using data objects known as persistence diagrams. To better aid statistical analysis, a functional representation of the diagrams, known as persistence landscapes, enable use of functional data analysis and machine learning tools. Topological and geometric variabilities inherent in point clouds are confounded in both persistence diagrams and landscapes, and it is important to distinguish topological signal from noise to draw reliable conclusions on the structure of the point clouds when using persistence homology. We develop a framework for decomposing variability in persistence diagrams into topological signal and topological noise through alignment of persistence landscapes using an elastic Riemannian metric. Aligned landscapes (amplitude) isolate the topological signal. Reparameterizations used for landscape alignment (phase) are linked to a resolution parameter used to generate persistence diagrams, and capture topological noise in the form of geometric, global scaling and sampling variabilities. We illustrate the importance of decoupling topological signal and topological noise in persistence diagrams (landscapes) using several simulated examples. We also demonstrate that our approach provides novel insights in two real data studies.
\end{abstract}


\section{Introduction}\label{sec:intro}

\begin{figure}[!t]
\begin{center}
\begin{tabular}{cc|cc}
      \includegraphics[width = 1.25 in]{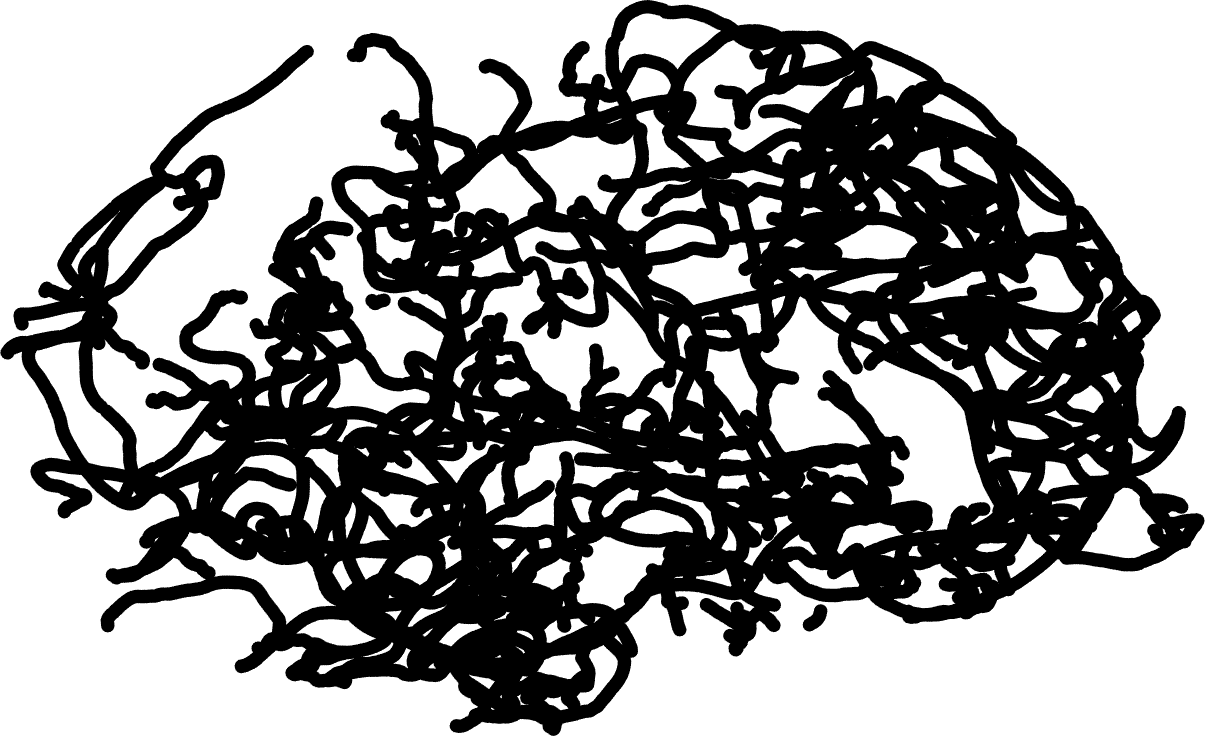} & \includegraphics[width = 1.25 in]{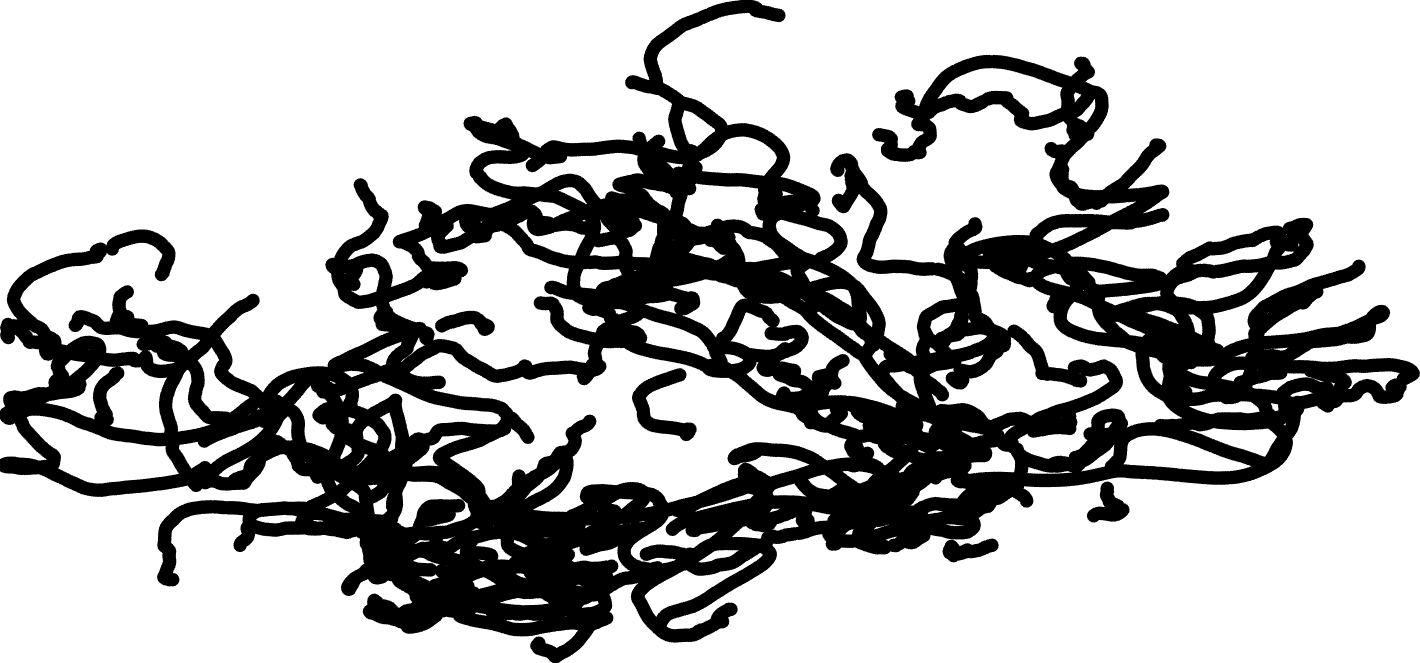} & \includegraphics[width = 1.1 in]{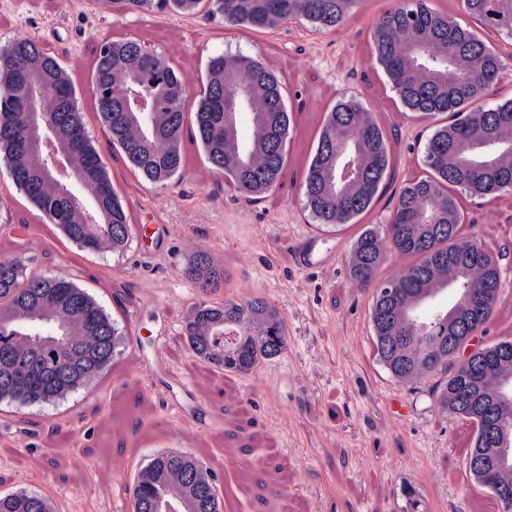}  & \includegraphics[width = 1.1 in]{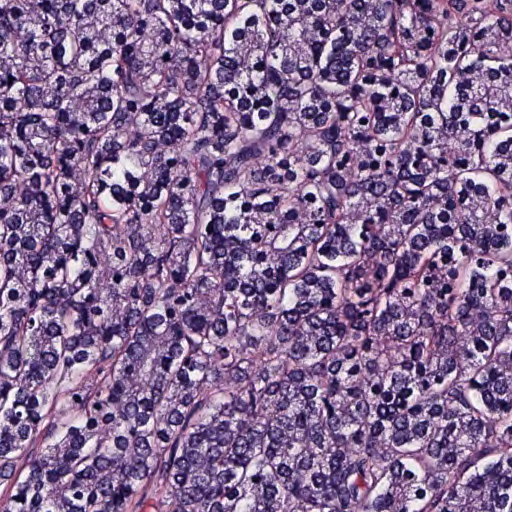}\\
      (a) & (b) & (c) & (d)
\end{tabular}
\end{center}
\caption{Left: Brain artery trees for (a) 20 year old and (b) 79 year old subjects from \cite{bendich_2016}. Right: Images of a prostate biopsy with (c) benign and (d) malignant carcinoma from \cite{berry_2020}.}\label{example_data}
\end{figure}

It is difficult to draw statistical insights from datasets where each observation corresponds to an object with rich structure. Consider the data shown in Figure \ref{example_data}, where panels (a) and (b) correspond to three-dimensional brain artery trees of different subjects, previously studied by \cite{bendich_2016} to understand how demographic factors are associated with brain structure; panels (c) and (d), on the other hand, display two example prostate gland biopsy images, which were studied by \cite{berry_2020} to aid prostate cancer prognosis. The visual differences between the observations within each of the two studies are striking. However, formally quantifying the differences between these objects, to enable statistical analysis, is a challenging and important problem. Topological Data Analysis (TDA) focuses on applying tools from (algebraic) topology to summarize and quantify the structure in complex data objects. In essence, TDA can be viewed as a general toolbox that enables discovery of topological and geometric features in complex data that can be used for subsequent analysis using existing statistical and machine learning methods.

Persistence homology is a prominent tool within TDA that provides a multi-resolution view of the topological and geometric features of data represented as point clouds in $\mathbb{R}^d$, e.g., samples of points on the brain artery trees or outlines of prostate glands shown in Figure \ref{example_data}. As a resolution parameter changes, so do the features of the data, and these changes are recorded in persistence diagrams. Statistical analysis of samples of persistence diagrams are based on distances, such as the Wasserstein distance or bottleneck distance, which enable computation of descriptive statistics \citep{mileyko_2011,turner_2014,wasserman_2018} and confidence regions \citep{fasy_2014}. Carrying out statistical analysis directly on the space of persistence diagrams is difficult since they are multisets of planar points. This motivates using functional representations (summaries) of diagrams that are more amenable for statistical analysis \citep{berry_2020} using tools from functional data analysis \citep{ramsay_silverman}. In this paper, we analyze persistence landscapes \citep{bubenik_2015}, although it will become clear that the proposed methods can be used on other functional summaries, e.g., silhouettes \citep{chazal_2014}, density estimates \citep{anirudh_2016}, rank functions \citep{robins_2016}, persistence entropy functions \citep{atienza_2020}, persistence intensity functions and images \citep{chen_2015,adams_2017}, with suitable modifications. Importantly, features extracted from persistence diagrams as well as the aforementioned functional summaries of diagrams have utility in modern machine learning approaches \cite{Hensel21}.

For point cloud data in $\mathbb R^d$ generated from a distribution with support on a lower-dimensional manifold $M$, persistence diagrams are typically computed by constructing geometric simplicial complexes based on an open cover of metric balls centered at data points with respect to the Euclidean distance on $\mathbb R^d$. The construction engenders noise in persistence diagrams that is complementary to topological signal in the point cloud, since distances between data points in $\mathbb R^d$ are sensitive to, mainly, three choices: (i) arbitrary (global) scaling of the point cloud; (ii) geometric configuration of the point cloud in $\mathbb R^d$ with respect to the manifold $M$; (iii) sampling variability, or density, of the points. The choices are linked to an implicit geometry of $M$: they imply an embedding $M \hookrightarrow \mathbb R^d$, under which a diffeomorphism of $M$, which preserves its topology, affects distances between points in $M$ when measured using the Euclidean distance in the image of the embedding in $\mathbb R^d$. This results in different persistence diagrams for point clouds sampled from topologically identical manifolds $M$. Since the map that takes a persistence diagram to a persistence landscape is invertible \cite{bubenik_2015}, it is natural to query how such `topological noise' manifests in a persistence landscape, and whether it is possible to exploit structure of the space of landscapes to mitigate noise and amplify topological signal. The resulting amplification of topological signal has the potential to enhance statistical or machine learning analyses.

We refer to topological noise as any variation in the data that is complementary to topological information, although we emphasize that the terminology does not imply that geometric features resulting from the above-mentioned choices are of no use in downstream statistical tasks. The situation is similar in spirit to Kendall's definition of landmark shape as all geometric information that remains after accounting for translation, scale and rotation variabilities \citep{Kendall84shapemanifolds}; in this setting, position, global scale and orientation of a set of landmark points are viewed as variation complementary to geometric shape. However, in many applications, these features of an object may be valuable descriptors \citep{kurtek2012statistical}. Thus, the main focus in Kendall's shape analysis lies in separating geometric shape information from the other sources of variability, and using them as complementary features of landmark configurations in downstream analyses. From this perspective, in the present setting, the only true source of nuisance variation is measurement error, which in general, is confounded with geometric and topological information.

A persistence diagram for a point cloud in $\mathbb R^d$ is a multiset of points on the plane that offers a multi-resolution summary of the homology of the data, constructed using geometric filtered complexes on $\mathbb R^d$ based on balls of radius $t>0$ around each datum; the radius $t$ acts as the resolution parameter in the sense that as it is increased, births and deaths of topological features of the point cloud are encoded in the corresponding persistence diagram. A persistence landscape is a collection of triangular functions $t \mapsto \lambda(t) \geq 0$, and is a  bijective multivariate functional summary of a persistence diagram. As such, how the value of $t$ is increased is a data analytic artefact, and should not affect the topological signal in the point cloud. In practice, however, increasing $t$ at different rates will result in different persistence diagrams, and hence, persistence landscapes. In particular, we focus on establishing a relationship between topological signal/noise and the two main sources of variation in a functional dataset consisting of persistence landscapes: amplitude or shape\footnote{The mathematical definitions of amplitude and shape are different in functional data analysis and shape analysis literatures. However, the two notions are very closely related for persistence landscapes, and we hence use them interchangeably.}, which captures $y$-axis variation, and phase, which tracks variation in the relative timing of shape features, e.g., extrema. We further show that phase variation in persistence landscapes is tied to the rate of increase of the resolution parameter $t$. In functional data analysis, the perils of not accounting for both sources of variation when computing summaries such as the mean or exploring dominant directions of variation via (functional) principal component analysis (PCA) are well-documented \citep[see e.g.,][]{marron_2015, srivastava_2016}. Evidently, such perils plague analysis of persistence landscapes, when viewed as points in a Banach space equipped with the $L^p$ norm: the pointwise mean of a sample of persistence landscapes can fail to be one, and this affects interpretability of the corresponding persistence diagram. Instead, computing a mean landscape using only the amplitude components of a sample of landscapes by registering, or aligning, them will better preserve shape, and mitigate effects of topological noise. In such a setting, our main contributions are as follows.
\begin{itemize}[leftmargin=3mm]
    \item We establish an explicit link between the rate of increase of the resolution parameter $t$ in a simplicial filtration that generates a persistence diagram and magnitude of phase variation present in the component functions $\lambda_k$, as captured through a reparameterization $s \mapsto \gamma(s)$ of the landscape $\Lambda(s) = (\lambda_1(s),\ldots,\lambda_K(s))$. Specifically, we show how $\gamma$ is related to variation in persistence diagrams induced by (i) (global) scaling of the data (Figure \ref{points2landscapes}), and (ii) sampling variability of data (Figure \ref{meanEstimationEg}).
    \item We show that alignment of landscapes $\{\Lambda_{i}\}_{i = 1}^n$ by determining optimal reparameterizations $\{\gamma_i\}_{i = 1}^n$ leads to an average landscape that better preserves the structure of the sample of landscapes. This induces a separation of variability in persistence landscapes into amplitude or shape, which captures the topological information in a dataset, and phase, which captures leftover variation due to geometric, global scaling and sampling variabilities. A key consequence is the `denoising' of points in the corresponding persistence diagrams by transforming them using $\{\gamma_i\}$; therefore, computing persistence diagrams for datasets $\{X_i\}$ using simplicial filtrations with balls of transformed radii $\{t \to \gamma_i(t)\}$ enhances topological information in persistence diagrams (Figure \ref{pcaEg}).
    \item We demonstrate that the proposed approach for statistical analysis of landscapes offers new insight, and adds substantially, to the analyses of the brain artery tree data in \cite{bendich_2016} (Section \ref{sec:brainEg}) and prostate cancer data in \cite{berry_2020} (Section \ref{sec:gleason}).
\end{itemize}

To the best of our knowledge, this is the first work in the literature to establish a concrete link between misalignment of persistence landscapes, and topological noise in persistence diagrams. However, in a certain sense, our approach in \emph{determining} an optimal rate of increase of $t$, given by $\gamma(t)$, falls between the standard approach of fixing a $t$ for each $x_i$ and having $t$ change with $x_i$, as with a multiscale approach that allows each ball $B_{x_i}(t_i)$ to have a possibly different radius $t_i$ \citep{yoon2020persistence}. 

The remainder of the paper is organized as follows. In Section \ref{sec:methods}, we further motivate alignment of persistence landscapes and provide technical details of our approach. In Section \ref{sec:simulations}, we use simulation studies to illustrate the importance of amplitude-phase separation in persistence landscapes for topological denoising. In Sections \ref{sec:brainEg} and \ref{sec:gleason}, we analyze the brain artery tree \cite{bendich_2016} and prostate cancer \cite{berry_2020} datasets, respectively. In both cases, we focus on the scientific questions that motivated the two studies, and illustrate the benefits and novel insights gained from separate statistical analysis of the amplitude and phase components of persistence landscapes. In Section \ref{sec:Discussion}, we provide a short discussion. Appendices A-C in the supplement contain additional simulated examples and real data analysis results.

\section{Elastic Functional Data Analysis of Persistence Landscapes}\label{sec:methods}

\begin{figure}[!t]
\begin{center}
\includegraphics[width = 5 in]{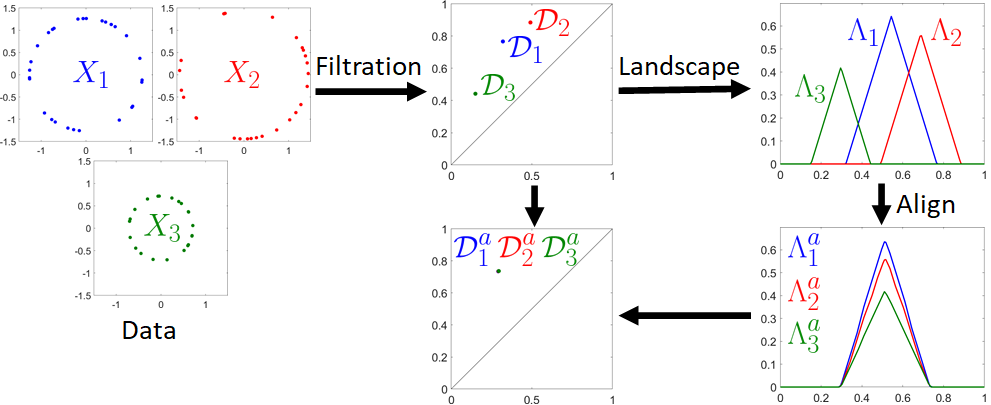}
    \caption{\emph{Example of topological noise}: Point clouds $X_1,X_2,X_3$ (with different sampling) from topologically identical spaces (differing only in scale) lead to different persistence diagrams $\mathcal D_1, \mathcal D_2, \mathcal D_3$ and hence landscapes $\Lambda_1,\Lambda_2,\Lambda_3$. \emph{Our approach}: Construct aligned landscapes $\Lambda^a_1,\Lambda^a_2,\Lambda^a_3$ and use alignment information to get transformed/denoised diagrams $\mathcal{D}^a_1,\mathcal{D}^a_2,\mathcal{D}^a_3$; use aligned landscapes for statistical analysis.} 
    \label{pipeline}
    \end{center}
\end{figure}

Before providing the technical details of our approach, we provide a summary of the proposed analysis pipeline in Figure \ref{pipeline} using three point clouds $X_1,X_2, X_3$ with degree-1 (loops), one-dimensional persistence landscapes $\Lambda_1,\Lambda_2,\Lambda_3$. Topological noise is induced purely through scale (radii of circles) and sampling variability. Notice how transforming the diagrams $\{\mathcal D_i\}$ using $\{\gamma_i\}$ from alignment of $\{\Lambda_i\}$ collapses the three points to a single one (denoising), as it should be since spaces from which $\{X_i\}$ are sampled are topologically identical. In this setting, it is evident that the global scale and sampling variability are captured purely in the phase component of the persistence landscapes. The aligned landscapes, in turn, capture the topological information about the underlying spaces from which the data was sampled: they all contain a single maximum corresponding to a single cycle, a topological feature that arises at the same exact time across the three point clouds. Further, the aligned landscapes are identical to each other up to a uniform scaling of the function values, i.e., they have the same shape. In the remainder of this section, we review the basics of persistence diagrams and landscapes, discuss distinct sources of variability in persistence landscapes, and specify a statistical framework to analyze these sources of variability using tools from elastic functional data analysis.

\subsection{Persistence diagram and landscape}
For a point cloud $X = \{x_1,\ldots,x_N\}$ in $\mathbb R^d$, equipped with the standard Euclidean norm, generated from a distribution with support on a lower-dimensional manifold $M$, persistence homology is a tool that tracks homological features, such as connected components (degree-0), loops (degree-1), voids (degree-2), etc., of the point cloud at different resolutions \citep{edelsbrunner_2002}. Homology is computed using geometric filtered complexes constructed from the union of balls $\cup_{i=1}^N B_{x_i}(t)$ around each point, each with the same radius $t>0$. The \v{C}ech complex, $\text{\v{C}ech}(X,t)$, consists of $k$-simplices whose nodes have $k+1$ many balls with a non-empty intersection. In contrast, the Vietoris-Rips complex, or just Rips complex, $\text{Rips}(X,t)$, is easier to compute and consists of $k$-simplices whose nodes have $k+1$ many balls with a non-empty pairwise intersection. At each fixed radius $t$, the homology of the simplicial complex is a snapshot of the features of the point cloud. Considering all radii, $t>0$, provides a multi-resolution view of the features of the point cloud where features are born and die at different values of $t$. Persistence homology tracks the features with a persistence diagram, a function from a countable set to $\{(x,y) \in \mathbb R^2| x<y \}$, consisting of birth-death pairs, $(b_j,d_j)$, the times at which the $j$\textsuperscript{th} feature was born and its corresponding death time. A persistence diagram is thus a multiset consisting of these points and represents a multi-resolution summary of the homology of the point cloud; see \cite{edelsbrunner_2002} for more general treatments of persistence homology and persistence diagrams.

A persistence landscape is an invertible functional representation of persistence homology computed from a persistence diagram \citep{bubenik_2015}. For $X = \{x_1,\ldots,x_N\}$, let $\mathcal{D}^p(X)$ denote its degree-$p$ persistence diagram consisting of $m$ birth-death pairs $\{(b_j,d_j)\}_{j = 1}^m$. The basic units of persistence landscapes are triangular functions computed using coordinates of points in a persistence diagram, $\ell_j^p(t) =  (t - b_j)\mathbb I_{\{b_j \leq t \leq \frac{1}{2}(b_j+d_j)\}}
  +(d_j-t) \mathbb I_{\{\frac{1}{2}(b_j+d_j) \leq t \leq d_j\}}$. For $k \in\mathbb{N}$, the $k$\textsuperscript{th} landscape function is defined as $\lambda_k^p(t) = \underset{j = 1,\ldots,m}{k^\text{th} \max}\ \ell_j^p(t)$, which is the $k$\textsuperscript{th} maximum of the triangular functions with $ \lambda_k(t) = 0$ for all $k>m$ by definition. Each function $\lambda_k^p$ thus begins and ends at zero. In practice, we truncate the number of landscape functions used for data analysis to the $K$ many that have some positive values along their domain. The degree-$p$ persistence landscape for $X$ is defined as the collection of landscape functions $ \Lambda^p_X(t) = \{\lambda^p_k(t)\}_{k = 1}^K$. In this work, we consider degree-$p$ persistence landscapes of samples of point clouds $X_1,\ldots,X_n$, simply denoted by $\Lambda_1(t),\ldots,\Lambda_n(t)$; we explicitly specify the degree $p$ under consideration in all simulated and real data examples.

\subsection{Effects of global scaling, sampling and geometric variabilities}
The construction of geometric simplicial complexes on $X$ engenders noise in a persistence diagram that is complementary to the topological signal in the point cloud, since distances between points in $\mathbb R^d$ are sensitive to (i) (global) scaling, (ii) geometric configuration, and (iii) sampling variability of the point cloud. Our interest lies in studying how changes in (i)-(iii) result in different persistence diagrams constructed using Rips or \v{C}ech simplicial filtrations.

\begin{figure}[!t]
\begin{center}
\begin{tabular}{cccc}
      \includegraphics[width = 0.7 in]{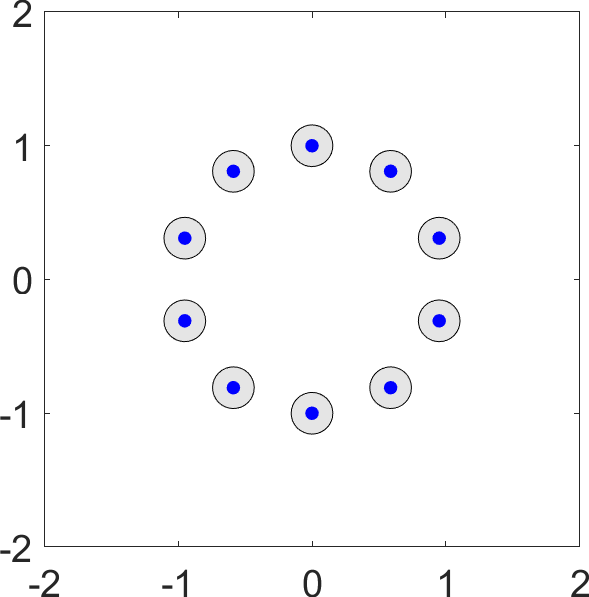} & \includegraphics[width = 0.7 in]{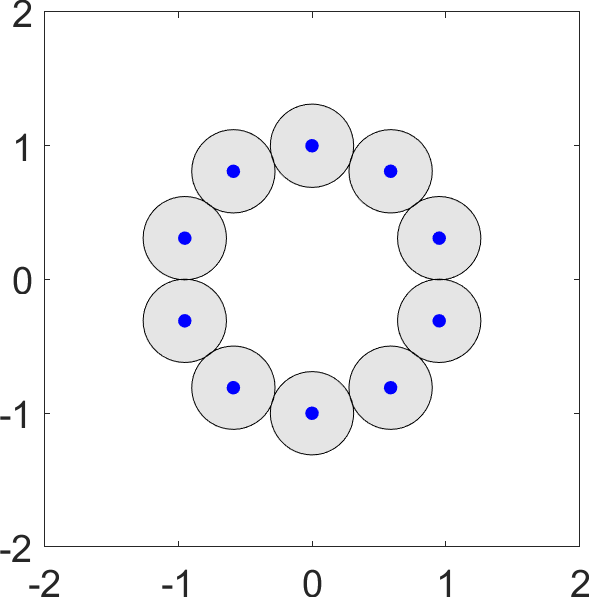} & \includegraphics[width = 0.7 in]{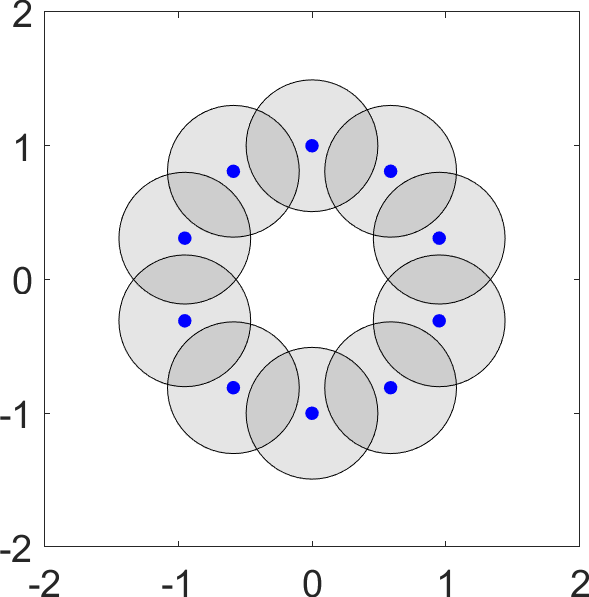}  & \includegraphics[width = 0.7 in]{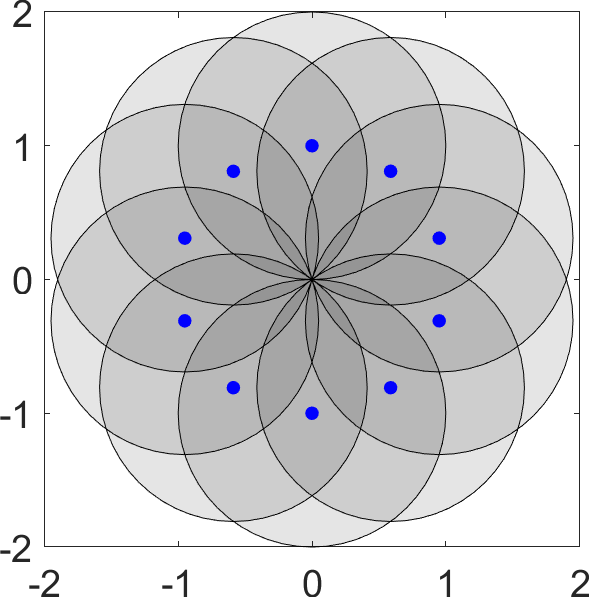}\\
      \includegraphics[width = 0.7 in]{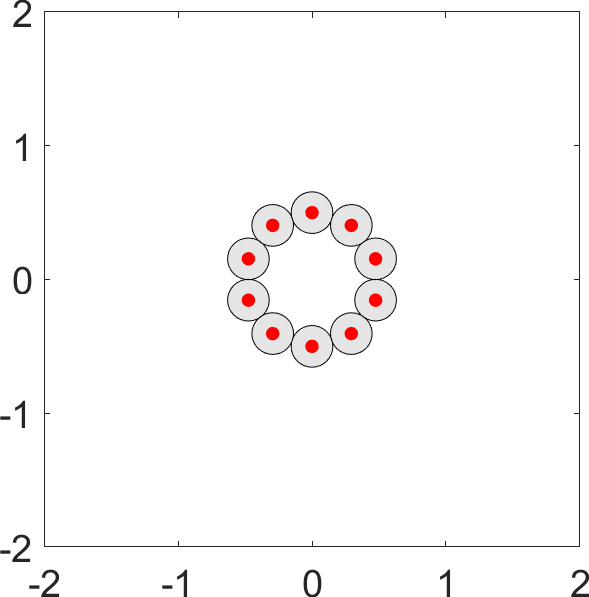} & \includegraphics[width = 0.7 in]{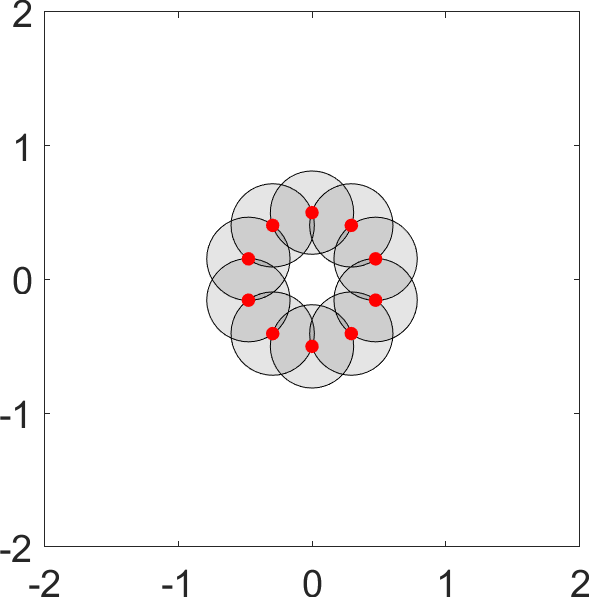} & \includegraphics[width = 0.7 in]{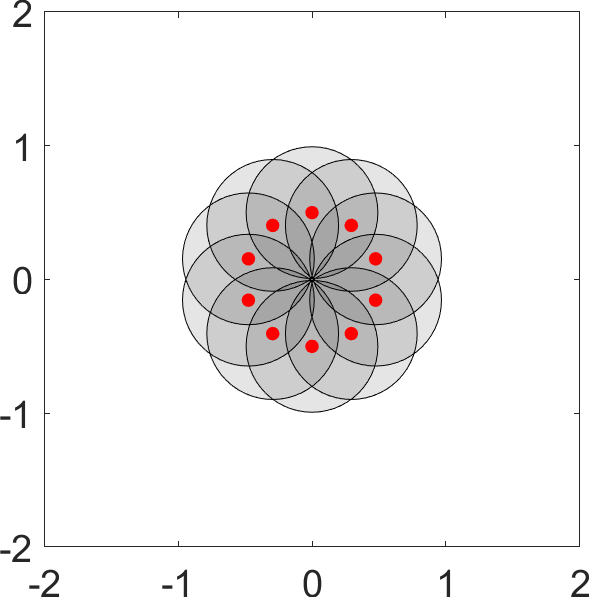}  & \includegraphics[width = 0.7 in]{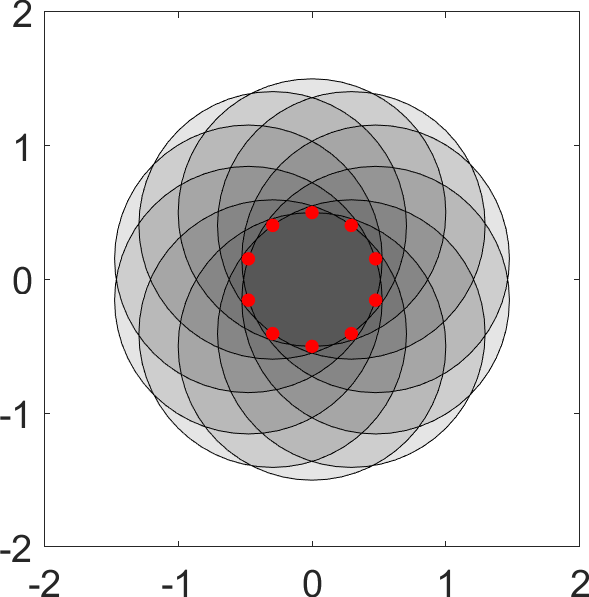}\\
      (a) & (b) & (c) & (d)
\end{tabular}
\begin{tabular}{cc}
      \includegraphics[width = 1 in]{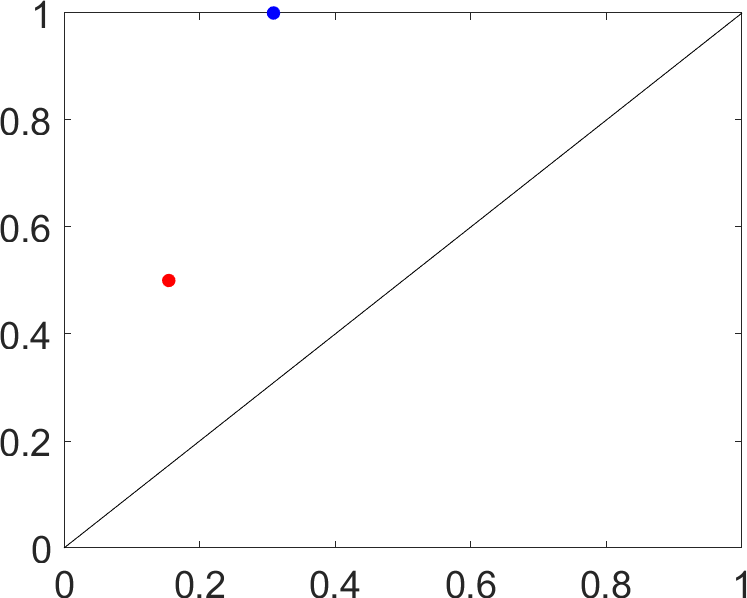} & \includegraphics[width = 1 in]{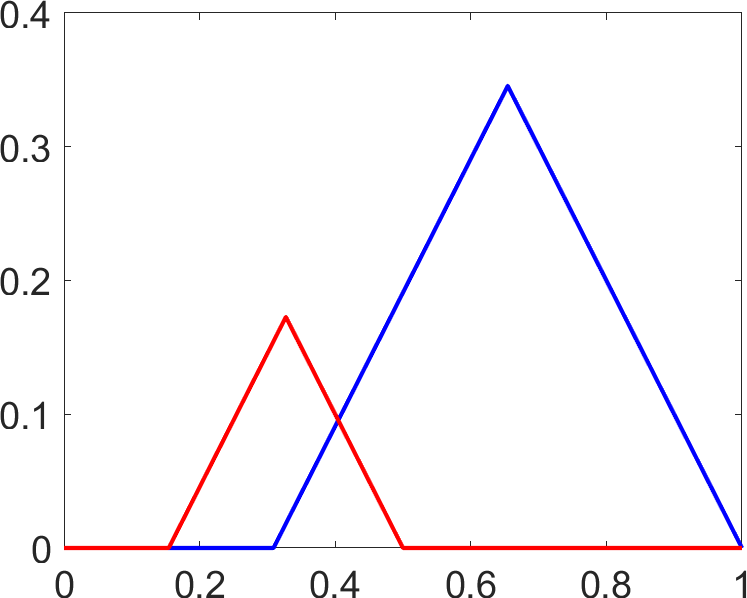}\\
      (e) & (f)
\end{tabular}

    \caption{\emph{Same topology with scale variability only}: Construction of Rips filtration for two point clouds on circles with radii 0.5 (red) and 1 (blue) at resolutions (a) $t = .1545$, (b) $t = .3090$, (c) $t = .5$, and (d) $t = 1$.  (e) Corresponding persistence diagrams and (f) landscapes.}
    \label{points2landscapes}
    \end{center}
\end{figure}

Consider the simple setting where the manifold $M$ is a circle of radius $t>0$, embedded into $\mathbb R^2$ as $ \theta \mapsto (t \cos \theta, t \sin \theta)$. Note that all circles with radius $t\geq 0$ are topologically identical. However, changing the radius $t$ changes the metric $\text{d}t^2+t^2\text{d}\theta^2$ on $\mathbb R^2$, and hence its restriction to the circle; this amounts to changing the embedding that provides coordinates for the observed points (geometric configuration), which ultimately changes the distance between points, as measured in $\mathbb R^2$, used to construct the filtration. Figure \ref{points2landscapes} illustrates this by considering two point clouds consisting of ten equidistant points along circles with radii 1 (blue) and 0.5 (red), respectively. We consider degree $p=1$ persistence homology (loops) with $K=1$-dimensional landscapes when topological noise is entirely due to scale effects. In Figure \ref{points2landscapes}(a)-(d), balls of different radii $t$ are drawn around the points in the point clouds. In (a), when $t = .1545$, a loop forms for the red point cloud, while there is no loop present for the blue point cloud. In (b), when $t = .3090$, a loop forms for the blue point cloud, and the loop persists for the red point cloud. In (c), when $t = .5$, the loop closes for the red point cloud, and persists for the blue point cloud. Finally, in (d), when $t = 1$, the loop closes for the blue point cloud. The loops can be summarized by the birth-death pairs $(.1545,.5)$ and $(.3090,1)$ for the red and blue point clouds, respectively. Panel (e) shows them in the persistence diagram, while panel (f) displays the corresponding misaligned persistence landscapes.

In the persistence diagram, $(b,d)$ coordinates of the red point are half of those for the blue, and this matches the ratio of the radii of the two circles; this implies that the persistence landscape for the red point cloud is shorter and shifted to the left by a commensurate amount as compared to the landscape for the blue point cloud. This scale-induced topological noise thus arises by a common scaling of the triangular function $\ell_1^1$, given by 
\begin{equation}
\label{eq:scale_s}
  \ell_1^1(t) =  (t - \alpha b)\mathbb I_{\{\alpha b \leq t \leq \frac{\alpha}{2}(b+d)\}}
  +(\alpha d-t) \mathbb I_{\{\frac{\alpha}{2}(b+d) \leq t \leq \alpha d\}},
\end{equation}
where $\alpha=2$ (when blue point cloud is derived from red). If $\alpha<1$, the triangular function will be shifted to the left along the domain and will be shorter relative to $\alpha = 1$; if $\alpha>1$, the function will be shifted to the right and taller. 

In this example based on the circle, we are able to explicitly link topological noise to a single parameter, the radius $t$ of the circle, which governs the magnitude of both scale and geometric configuration of points. In essence, \emph{when topological noise is due to scaling, alignment of peaks of the persistence landscapes will move points in a persistence diagram toward each other, and thus amplify the topological signal}. In other words, such topological noise manifests entirely through phase variation in the landscapes. In higher dimensions, it is not possible in general to carry out this program since the restriction of the metric on $\mathbb R^d$ induced by the embedding of the manifold $M$ is more complicated. Nevertheless, we demonstrate through numerical examples in Section \ref{sec:simulations} and Appendices A and B in the supplement that alignment of persistence landscapes acts as a denoising mechanism for the corresponding diagrams, even when topological noise is not only due to scaling. 

From the discussion above, it is clear that changing the embedding (e.g., $x \mapsto (x,\sqrt{t-x^2})$) would have generated similar topological noise in the persistence diagram. This of course amounts to a change in the metric which ultimately results in a change in the simplicial filtration. When sampling variability is present (e.g., non-equispaced points on the circle), the situation can be viewed as one involving \emph{local} scaling of the point cloud, and alignment of peaks of the landscapes will again induce points on the persistence diagrams to move toward each other (Section \ref{sec:sim2}, Figures \ref{meanEstimationEg} and \ref{pcaEg}); similar comments apply to the situation involving measurement error (points do not lie exactly on $M$; Appendix A in the supplement, Figures 1-3).  

\subsection{Reparameterizing a landscape and denoising a diagram}

The discussion in the previous subsection suggests alignment of persistence landscapes, by lining up peaks and valleys of the component functions, as a viable denoising mechanism for persistence diagrams. We propose to do this through shape analysis of landscapes, viewed as parameterized curves. Our approach is based on the elastic metric to compare shapes of curves \citep{srivastava_2016}, based on a convenient transform of the landscape curves. Specifically, the transformation maps landscapes into a Hilbert space, where geometric computations become simplified; this stands in contrast to the Banach space setting typically used for persistence landscapes. An important consequence of this is that we are able to compute a mean landscape based on its amplitude which, in contrast to the pointwise landscape currently computed in a Banach space setting, better preserves the shape of a landscape; moreover, the Hilbert space structure provides an inner product to carry out PCA on the amplitude or shape component of landscapes, which enables one to study dominant modes of variation in samples of point clouds.

By virtue of its definition, a landscape is parameterized by the resolution (filtration) parameter $t$, used to construct the Rips or the \v{C}ech simplicial filtrations, which in principle can be any positive real. In order to choose a closed interval of $\mathbb R$ as a parameter domain, note that, given $n$ persistence landscapes, there always exists an $0<s<\infty$ such that $\Lambda_i (t) = 0,\ \forall\ t > s,\ i = 1,\ldots,n$. Given this, one can assume, without loss of generality, the parameter domain to be the unit interval obtained through rescaling by $1/s$. Then, for each $i=1,\ldots,n$, landscape $\Lambda_i$ is a $K$-dimensional piecewise linear parameterized closed curve $[0,1] \ni t \mapsto \Lambda_i(t) \in \mathbb R^K$ with $\Lambda_i(0)=\Lambda_i(1)=0$. This in turn results in scaled persistence diagrams $\{(b_{i,j}/s,d_{i,j}/s)\}_{i=1,j=1}^{n,m_i}$, so that the birth-death pairs are in $[0,1]^2$. 
\begin{definition}
\label{def: scaling}
A persistence landscape (diagram) $\Lambda$ ($\{(b_j,d_j)\}$) obtained by rescaling in the above manner is referred to as a scaled persistence landscape (diagram). 
\end{definition}
\noindent We will simply henceforth refer to a scaled persistence landscape (diagram) as a persistence landscape (diagram), unless explicitly mentioned to the contrary. 

Since a reparameterization of a landscape $\Lambda$ preserves its image, its shape, modulo scale, is preserved. As the set of reparameterizations, consider
$$\Gamma=\left \{\gamma:[0,1] \to [0,1]: \dot\gamma>0,\ \gamma(0)=0,\ \gamma(1)=1\right\},$$ 
the set of orientation-preserving diffeomorphisms of $[0,1]$, which forms a group under composition ($\dot\gamma$ is the derivative of $\gamma$). The group $\Gamma$ acts on the set of landscapes from the right as function composition: $(\Lambda, \gamma) \rightarrow \Lambda (\gamma)$. 
Alignment of landscapes $\{\Lambda_i\}$ thus amounts to establishing correspondence between ($K$-dimensional) points in the images $t\mapsto \Lambda_i(t)$, achieved by determining optimal $\gamma_i \in \Gamma$ such that the collection $\{\Lambda_i(\gamma_i)\}$ is `optimally' aligned, where optimality is defined with respect to a metric-based matching functional. For each $i$, $\gamma_i$ represents common phase variation in the component functions $(\lambda_{1_i},\ldots,\lambda_{K_i})$ of $\Lambda_i$. 

Since the resolution/filtration parameter $t$ is the parameter for a landscape when viewed as a closed curve, we observe the following.
\begin{proposition}
\label{prop:equivariance}
The map $\Lambda \mapsto \Phi(\Lambda):=\{(b_j,d_j)\}$ that takes a scaled persistence landscape $\Lambda$ to a unique scaled persistence diagram $\{(b_j,d_j)\}$ is equivariant with respect to the action of $\Gamma$ on the set of scaled persistence landscapes, i.e., $\Phi(\Lambda (\gamma))=\gamma^{-1}\{(b_j,d_j)\}=\{(\gamma^{-1}(b_{j}),\gamma^{-1}(d_{j}))\}$. 
\end{proposition}
\noindent A useful way to think of the induced transform $\{(\gamma^{-1}(b_j),\gamma^{-1}(d_j))\}$ on a persistence diagram is as a \emph{nonlinear local scale change} of the multiset of points that generalizes the global scale change described in \eqref{eq:scale_s}. This attempts to reverse the effects of (i) working with rescaled persistence landscapes, and (ii) potential topological noise induced through the geometric construction of simplicial filtrations. Indeed, this is tantamount to considering a geometric \v{C}ech or Rips filtration with parameter $\gamma(t)$: for a point cloud $X$, the corresponding Rips simplicial complex is defined as
\begin{align*}
    \sigma=[x_1,\ldots,x_k] \in \text{Rips}(X,\gamma(t)) & \iff |x_i-x_j| \leq \gamma(t)\\ 
    & \hspace*{-1cm} \iff \gamma^{-1}( |x_i-x_j|) \leq t, \quad \forall i,j
\end{align*}
since $\gamma$ is strictly increasing. Equivalently, the equivariant action on $\Phi$ implies that the metric induced by the Euclidean norm $|\cdot|$ on $\mathbb R^d$ is deformed by a differmorphism $\gamma^{-1}$ to track local scale changes needed to preserve the integrity of topological signal in the presence of topological noise. 

However, some care is needed with this interpretation since $\gamma^{-1}(|\cdot|)$ fails to be a metric on $\mathbb R^d$ if the function $\gamma^{-1}$ is not concave, since the triangle inequality will otherwise not be satisfied. Moreover, the chain of inclusions 
\[
\text{Rips}(X,t')\subset \text{\v{C}ech}(X,t) \subset \text{Rips}(X,t), \text{ when } \frac{t}{t'} \geq \sqrt{\frac{2d}{d+1}}
\]
that characterize Rips and \v{C}ech complexes with parameter $t$ \citep{SG} need not be preserved under $\gamma(t)$ for all $\gamma \in \Gamma$; since preservation would depend on the magnitude of the derivative of $\gamma$, it is difficult to provide a lower bound for $\gamma(t)/\gamma(t')$ that holds for \emph{all} $\gamma \in \Gamma$. In principle, it is possible to restrict attention to a subset of diffeomorphsims $\gamma$ with a concave inverse that preserve the chain of inclusions, but this may restrict how well one is able to denoise persistence diagrams by aligning persistence landscapes. 

We can summarize the practical consequence of the above discussion in the following manner: on point clouds $\{X_i\}$, the equivariant action of $\Gamma$ ensures that, in the optimally aligned landscapes $\{\Lambda_i(\gamma_i)\}$, the extrema (mainly peaks) of the component functions $(\lambda_{1_i}(\gamma_i),\ldots,\lambda_{K_i}(\gamma_i))$ line up, and the transformed points $\{(\gamma_i^{-1}(b_{ij}),\gamma_i^{-1}(d_{ij}))\}$, consequently, will tend to cluster, the number of which will depend on the topology of the underlying manifold. As a consequence, if a \v{C}ech or Rips filtration for $X_i$ is constructed with balls of radius $\gamma_i(t)$, the corresponding persistence diagram will be `denoised'. 

\subsection{Optimal reparameterizations, mean amplitude landscape and PCA}
The program described above rests on determining the optimal reparameterizations $\{\gamma_i\}$ from observed landscapes $\{\Lambda_i\}$. 
In principle, any registration or alignment procedure for curves in $\mathbb R^K$ can be used. Our choice is based on the highly successful Elastic Functional Data Analysis (EFDA) framework, a Riemannian-geometric approach that utilizes the elastic metric for curves in $\mathbb R^K$. This framework is characterized by two important theoretical considerations for the curve alignment problem: (i) isometry, i.e., invariance to simultaneous reparameterization of curves, and (ii) invariance of optimal reparameterizations to rescaling of curves. These are addressed through the use of the elastic Riemannian metric for comparing absolutely continuous curves, which is difficult to compute in practice. For ease of exposition, we refrain from providing the definition of the metric and its properties; see Chapter 10 in \cite{srivastava_2016} for details. 

Let $\mathcal{F}$ denote the space of absolutely continuous curves in $\mathbb{R}^K$ equipped with the elastic metric. Practical use of the metric is greatly simplified through use of the square-root velocity function (SRVF) representation. For any curve $\beta\in\mathcal{F}$, its SRVF is defined as
\[
\beta \mapsto Q(\beta)=q:=
\dot\beta(|\dot\beta|)^{-1/2},
\]
where $\dot \beta$ is the componentwise derivative and $|\cdot|$ is the Euclidean norm on $\mathbb R^K$. The map $Q: \mathcal F \to \mathbb L^2([0,1],\mathbb R^K)$ is a homeomorphism \citep{bruveris} with inverse $\beta(t) = \int_0^t q(u)|q(u)|\text{d}u$, and effectively `flattens' the complicated elastic metric: the distance $d(\beta_1,\beta_2)$ between two absolutely continuous curves  with respect to the elastic metric equals $\|Q(\beta_1)-Q(\beta_2)\|_2=\|q_1-q_2\|_2 = [\int_0^1|q_1(t)-q_2(t)|^2\text{d}t]^{1/2}$, and the standard $\mathbb L^2$ metric on SRVFs of curves possesses desiderata (i) and (ii) mentioned above.

Absolute continuity of a curve in $\mathbb R^K$ is defined via absolute continuity of its one-dimensional component functions. Absolutely continuous functions in one dimension have constant speed parameterization \citep{SS}. In the present setting, persistence landscapes $\Lambda:[0,1] \to \mathbb R^K$ are piecewise linear curves. It is known that the set of piecewise linear curves in $\mathbb R^K$ is dense in $\mathcal F$ \citep{LKR}, and continuity of the map $Q$ ensures that its image under $Q$, consisting of piecewise constant SRVFs, is dense in $\mathbb L^2([0,1],\mathbb R^K)$. 

Our definition of the amplitude (shape) of a curve and subsequent statistical analysis approach are analogous to the definitions presented in \cite{srivastava_2011} for univariate functions. The group $\Gamma$ acts on $\mathcal F$ through composition, and since the map $Q: \mathcal F \to \mathbb L^2([0,1],\mathbb R^K)$ is bijective, the action $(q,\gamma) = Q(\Lambda(\gamma)) = (q(\gamma))\sqrt{\dot \gamma}$ of $\Gamma$ can be defined on $\mathbb L^2([0,1],\mathbb R^K)$, under which, the amplitude of a landscape $\Lambda$ is its orbit $[q] := \{(q, \gamma) \,|\, \gamma \in \Gamma \}$. Since $\|(q,\gamma)\|_2=\|q\|_2$ for every $q \in \mathbb L^2([0,1],\mathbb R^K)$ and $\gamma \in \Gamma$, we note that $\Gamma$ acts by isometries on $\mathbb L^2([0,1],\mathbb R^K)$. Under this definition, two curves, $\Lambda_1,\Lambda_2$, have the same amplitude if their corresponding SRVFs are in the same orbit, i.e., there exists a $\gamma \in \Gamma$ such that $q_1 = (q_2,\gamma)$.  The set of all orbits forms a partition of $\mathcal{Q}$ and is the quotient space $\mathcal{Q}/\Gamma$. Hence, $\mathcal{Q}/\Gamma$ defines the amplitude space of persistence landscapes. 

The amplitude distance between two landscapes $\Lambda_1,\Lambda_2\in\mathcal{F}$ is defined as the distance between their corresponding SRVF orbits $[q_1],[q_2]\in\mathcal{Q}/\Gamma$:
\begin{equation}\label{eqdist}
    d_\text{a}(\Lambda_1,\Lambda_2)=d([q_1],[q_2]) = \underset{\gamma \in \Gamma}{\text{min}}\ \|q_1-(q_2,\gamma)\|_2.
\end{equation}
Key to the definition of the above distance is the invariance of the $\mathbb{L}^2$ metric, under the SRVF representation, to simultaneous reparameterization of curves. The function 
\[
\gamma^*=\underset{\gamma \in \Gamma}{\text{argmin}}\ \|q_1-(q_2,\gamma)\|_2
\]
is then the optimal reparameterization of $\Lambda_2$ to register or align it to $\Lambda_1$. Furthermore, $\gamma^* =\underset{\gamma \in \Gamma}{\text{argmin}}\ \|q_1-(q_2,\gamma)\|_2=\underset{\gamma \in \Gamma}{\text{argmin}}\ \|c_1q_1-(c_2q_2,\gamma)\|_2,\ c_1,c_2\in\mathbb{R}_+$, i.e., rescaling of the curves does not alter the optimal reparameterization.

Denoising persistence diagrams from point clouds $X_1,\ldots,X_n$ requires determining optimal reparameterizations $\gamma_1,\ldots,\gamma_n$ that \emph{jointly} align persistence landscapes $\Lambda_1,\ldots,\Lambda_n$. This requires a template landscape to align the individual ones to. We use a data-driven template given by the mean amplitude landscape. Denote by $q_1,\ldots,q_n$ the SRVFs of $\Lambda_1,\ldots,\Lambda_n$. This mean is defined as the quantity that minimizes the sum of squared amplitude distances: 
\begin{equation}\label{eq:elasticmean}
    [\hat \mu_q] = \underset{[q] \in \mathcal{Q}/\Gamma}{\text{argmin}}\ \sum_{i = 1}^n \underset{\gamma \in \Gamma}{\text{min}}\ \|q-(q_i,\gamma)\|^2_2.
\end{equation}
An orbit representative $[\hat \mu_q]$ is found by iteratively aligning $q_1,\ldots,q_n$ to the current estimate of the mean and averaging the aligned SRVFs to produce a new mean estimate; this is repeated until convergence. For identifiability, we use the center of the orbit of $[\hat\mu_q]$ as the representative element of the elastic mean; henceforth, we simply refer to this element of the mean orbit as $\hat\mu_q$. 
For additional algorithmic details and the orbit centering step, we refer to \cite{srivastava_2011}.  The corresponding mean amplitude landscape $\hat\mu\in\mathcal{F}$ is defined as $Q^{-1}(\hat \mu_q)$. 

The joint alignment of $\Lambda_1,\ldots,\Lambda_n$ can then be achieved via pairwise alignment of each $\Lambda_i,\ i=1,\dots,n$ to the mean $\hat\mu$ using \eqref{eqdist} by determining the optimal reparameterizations 
\[
\gamma_i = \underset{\gamma\in\Gamma}{\text{argmin}}\|\hat\mu_q - (q_i,\gamma)\|_2,\quad i = 1,\ldots,n,
\]
which can be used to study phase variability. Details of methods for statistical analysis of reparameterization functions, including computation of a distance, averaging and PCA are available in \cite{tucker_2013}, and are omitted here for brevity.

Since the aligned landscapes $\Lambda_i(\gamma_i),\ i = 1,\ldots,n$, or equivalently their SRVFs, $(q_i,\gamma_i)$, describe amplitude variability in the sample, a sample amplitude covariance function can be defined as
\begin{equation}
   \widehat{C_q(t,u)} := \frac{1}{n-1}\sum_{i = 1}^n((q_i,\gamma_i)(t) - \hat\mu_q(t))((q_i,\gamma_i)(u) - \hat\mu_q(u))^\top.
\end{equation}
Amplitude-based PCA is carried out via eigendecomposition of $\widehat{C_q(t,u)}$,
\begin{equation}
   \widehat{C_q(t,u)} = \sum_{b=1}^\infty \hat{\tau}_b\hat\phi_b(t)\hat\phi_b(u)^\top,
\end{equation}
where $\hat\phi_b,\ b\in \mathbb{N}$ are the primary directions of amplitude variability (amplitude PCs) and $\hat\tau_b,\ b\in \mathbb{N}$ are variances in the corresponding directions. Typically, one selects a finite number, $B$, of principal components that describe a large portion of amplitude variability. The aligned SRVFs can then be projected onto the $B$ directions of amplitude variability with largest variance, $\beta_{i,b} := \int_0^1 \langle (q_i,\gamma)(t)-\hat\mu_q(t),\hat\phi_b(t)\rangle dt,\  b = 1,\ldots,B,\ i = 1,\ldots,n$,
where $\langle\cdot,\cdot\rangle$ is the Euclidean inner product in $\mathbb{R}^K$. The PC scores, $\boldsymbol{\beta_i} = (\beta_{i,1},\ldots \beta_{i,B})^\top,\ i = 1,\ldots,n$ serve as a low dimensional Euclidean representation of the amplitude of landscapes. To visualize the primary directions of amplitude variability, we compute $\mathcal{F}\ni Q^{-1}(\hat\mu_q + \nu\sqrt{\hat\tau_b}\hat\phi_b)$, i.e., a landscape that is $\nu$ standard deviations from $\hat\mu_q$ in the direction of $\hat\phi_b$.  

\section{Simulation Studies}\label{sec:simulations}
In this section, we present simulation examples which demonstrate (i) denoising of persistence diagrams, obtained under scale and sampling variabilities in point clouds, through alignment of landscapes, and (ii) benefits of computing the mean landscape and PC directions on the set of aligned landscapes as opposed to computing a pointwise mean with unaligned ones, as currently done in practice. 
The supplement contains additional examples that (i) consider data sampled with additive noise (Appendix A), and (ii) illustrate mean estimation (Appendix B). 


For these examples, we use the \verb|ripsDiag| function to compute persistence diagrams using the Vietoris-Rips simplicial complex for point clouds, and the \verb|landscape| function to compute landscapes from persistence diagrams; both functions are part of the \verb|TDA| \verb|R| package \citep{fasy_2014R}. In the EFDA framework, registration, mean estimation and PCA for a sample of landscapes are implemented in \verb|MATLAB|. Code and data to reproduce simulated examples are available here: \url{https://github.com/jamesmatuk/EFDA-of-Persistence-Landscapes}. 

\subsection{Examples 1 and 2: Mean from aligned landscapes}
\label{sec:sim1}

\begin{figure}[!t]
\begin{center}
\setlength{\tabcolsep}{1pt}
      \begin{tabular}{ccc}
    \includegraphics[width = 1.1 in]{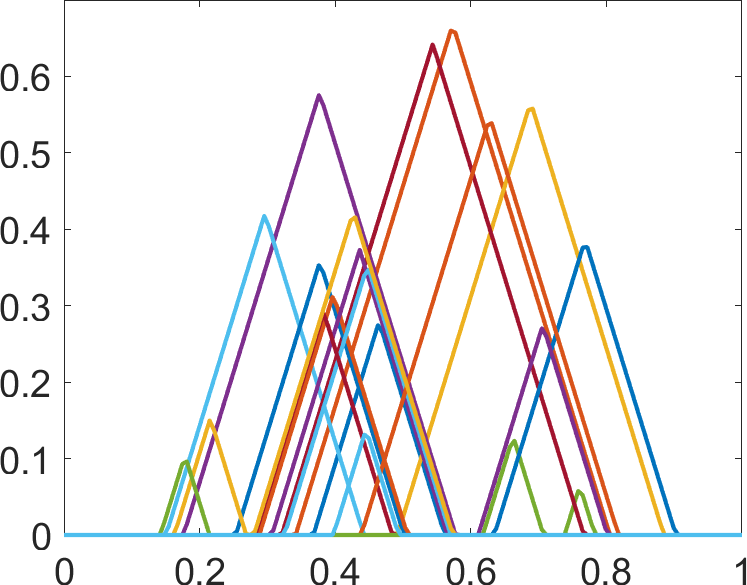} & \includegraphics[width = 1.1 in]{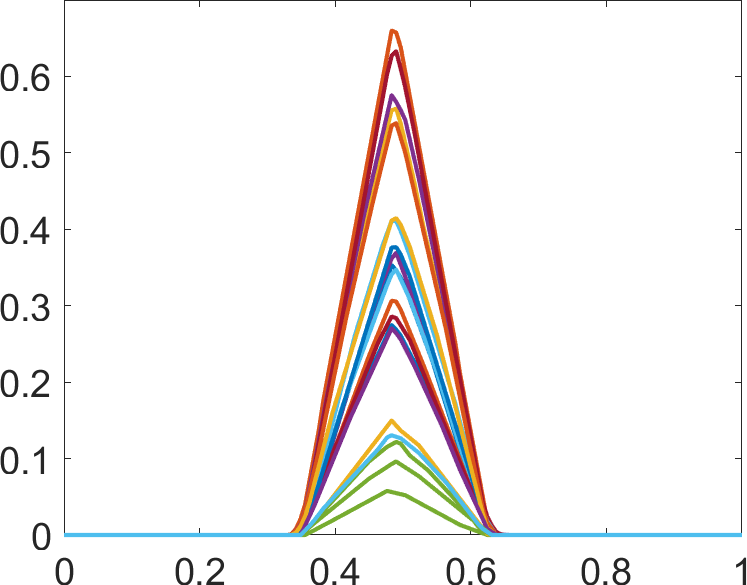}  &  \includegraphics[width = 1.1 in]{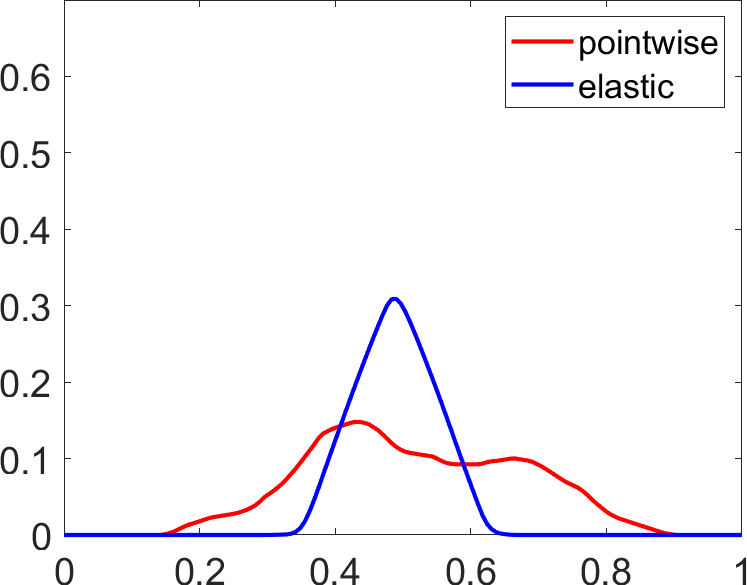} \\
    (a) & (b)& (c) \\
    \includegraphics[width = 1.1 in]{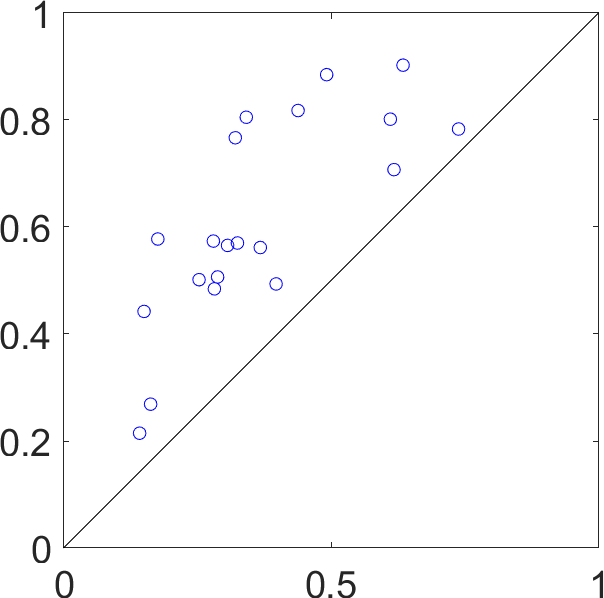} & 
    \includegraphics[width = 1.1 in]{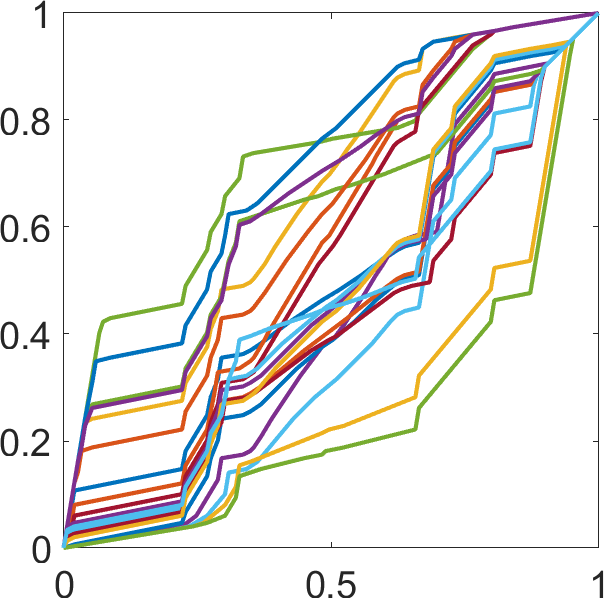}  &  \includegraphics[width = 1.1 in]{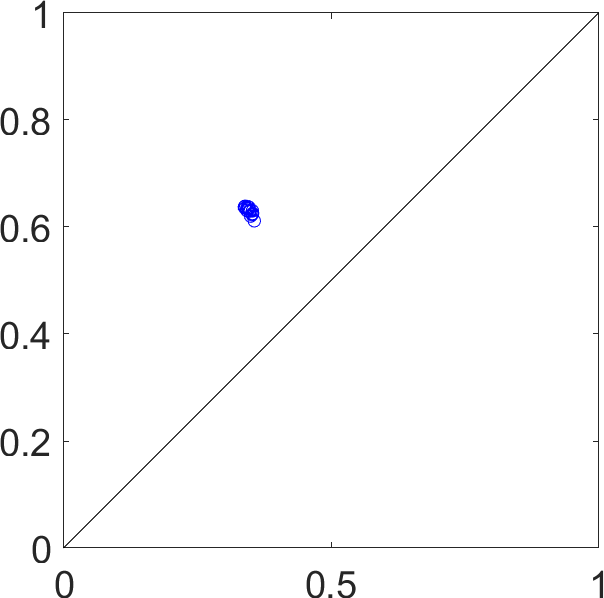} \\
     (d) & (e) & (f) \\
    \end{tabular}
    \caption{\emph{Same topology with scale and sampling variabilities}: (a) Persistence landscapes $\{\Lambda_i\}_{i=1}^{20}$ of 20 point clouds of the type in Figure \ref{pipeline}. (b) Aligned persistence landscapes $\{\Lambda_i(\gamma_i)\}_{i=1}^{20}$. (c) Mean landscape after (blue) and without (red) alignment. (d) Noisy persistence diagrams $\{(b_{i},d_{i})\}_{i=1}^{20}$ from 20 point clouds. (e) Estimated reparameterizations $\{\gamma_i\}_{i=1}^{20}$. (f) Denoised persistence diagrams $\{(\gamma_i^{-1}(b_{i}),\gamma_i^{-1}(d_{i}))\}_{i=1}^{20}$.}
    \label{meanEstimationEg}
    \end{center}
\end{figure}

\textbf{Example 1.} We consider 20 point clouds, where each point cloud is generated by (i) sampling $M$ from a $\text{Discrete-Uniform}(10,30)$, (ii) sampling $r$ from $|N(1,0.3^2)|$, and (iii) generating $M$ points uniformly on a circle with radius $r$. Figure \ref{pipeline} shows three (from 20) point clouds along with the corresponding degree $p=1$, $K=1$-dimensional landscapes. Figure \ref{meanEstimationEg}(a) shows all 20 landscapes $\{\Lambda_i\}_{i=1}^{20}$. The amplitude and phase variations in the landscapes are related to variability in the radii, sample size and dispersion. Panels (b)-(f) demonstrate the benefit of alignment of landscapes $\{\Lambda_i\}$: a visually better mean (c) is obtained by using the aligned landscapes $\{\Lambda_i(\gamma_i)\}$ (b); transforming points in the persistence diagrams (d) using reparameterizations $\{\gamma_i\}$ (e) results in denoising (f) by collapsing all points to a single one, since the topology of the 20 point clouds is the same. 


\begin{figure}[!t]
\begin{center}
\begin{tabular}{cccc}
      \includegraphics[width = .6 in]{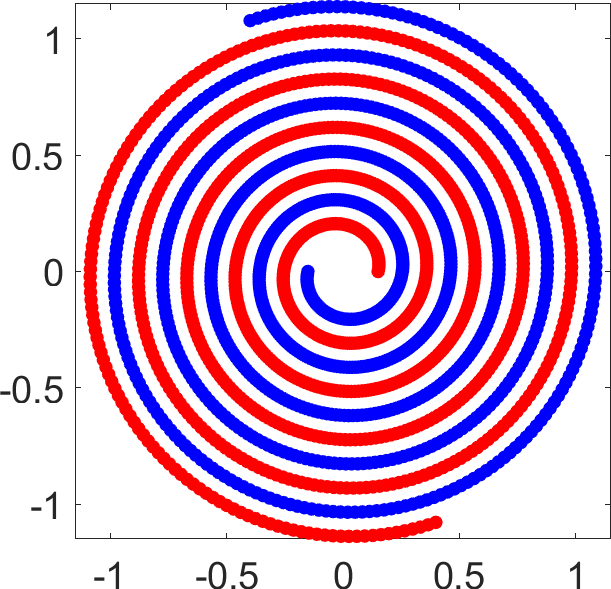} & \includegraphics[width = .8 in]{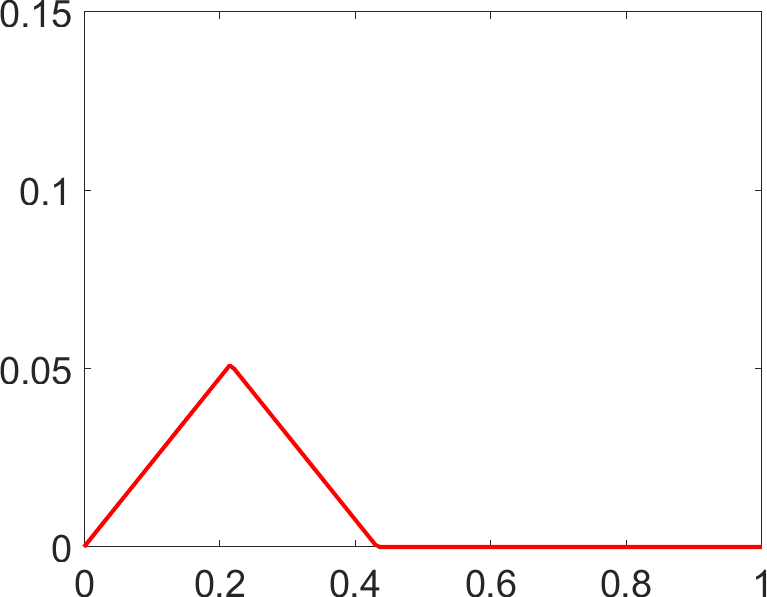}  & \includegraphics[width = .6 in]{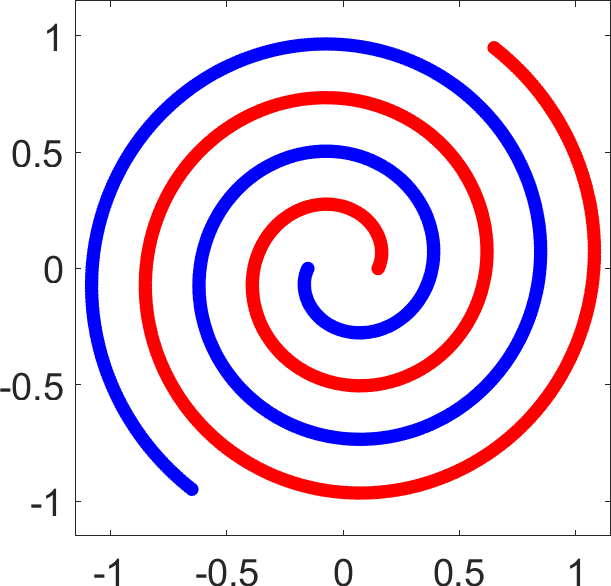} & \includegraphics[width = .8 in]{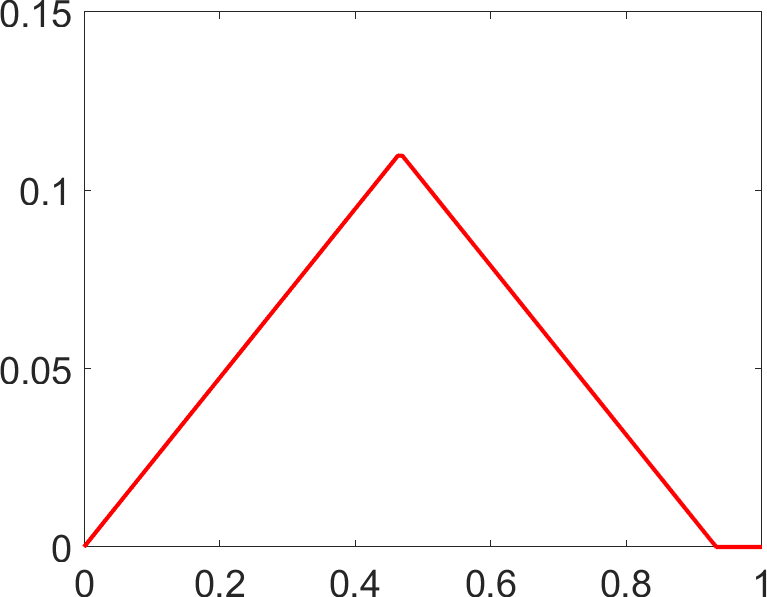}\\
      (a) & (b) & (c) & (d) \\
      \end{tabular}
      
      \begin{tabular}{ccc}
        \includegraphics[width = .8 in]{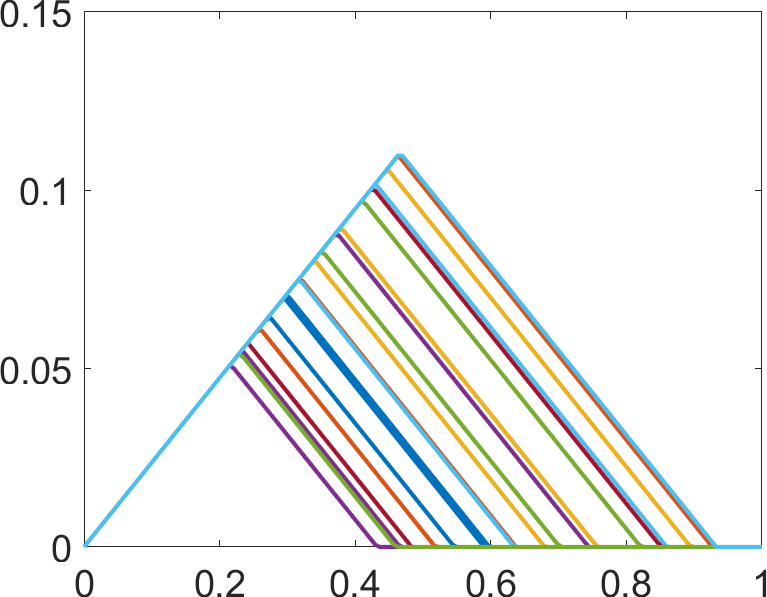} & \includegraphics[width = .8 in]{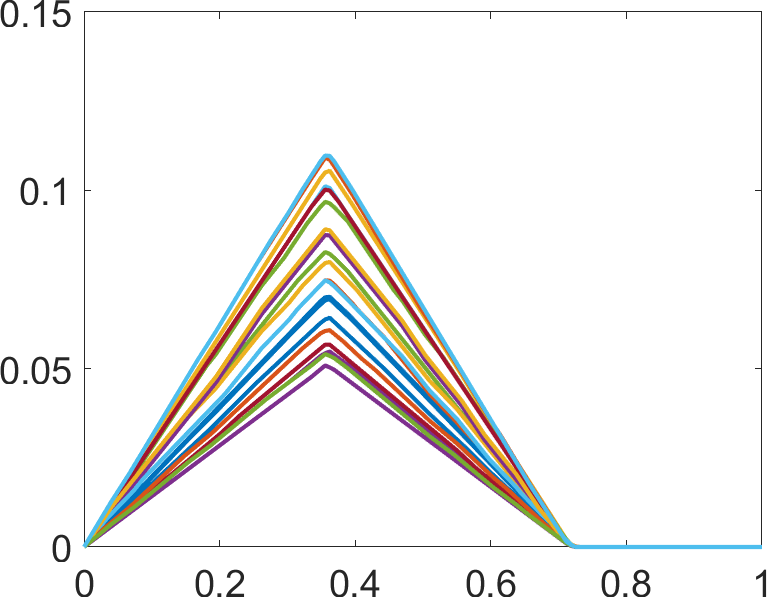}  &  \includegraphics[width = .8 in]{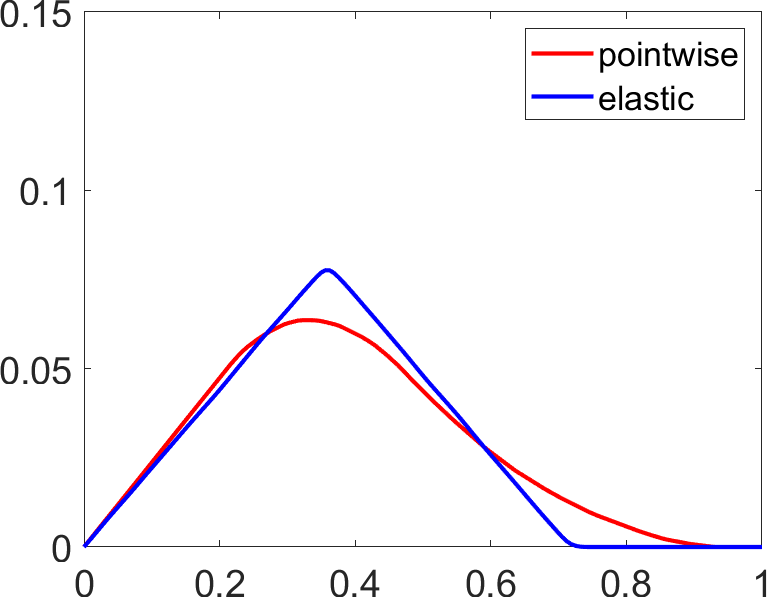} \\
         (e) & (f) & (g) \\
        \includegraphics[width = .8 in]{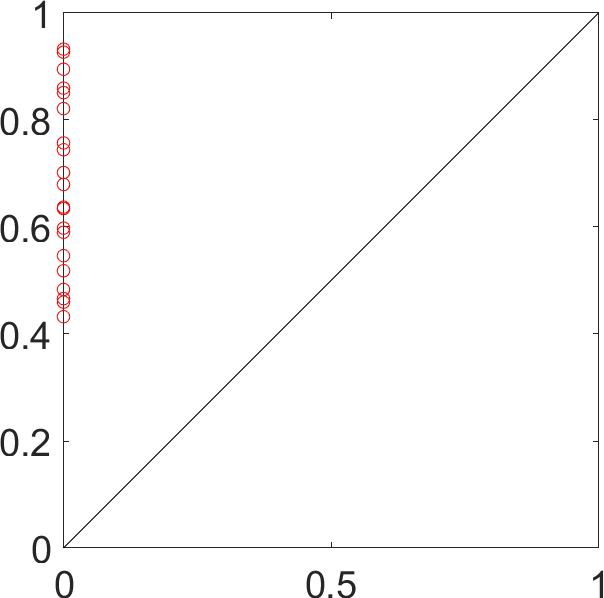} & \includegraphics[width = .8 in]{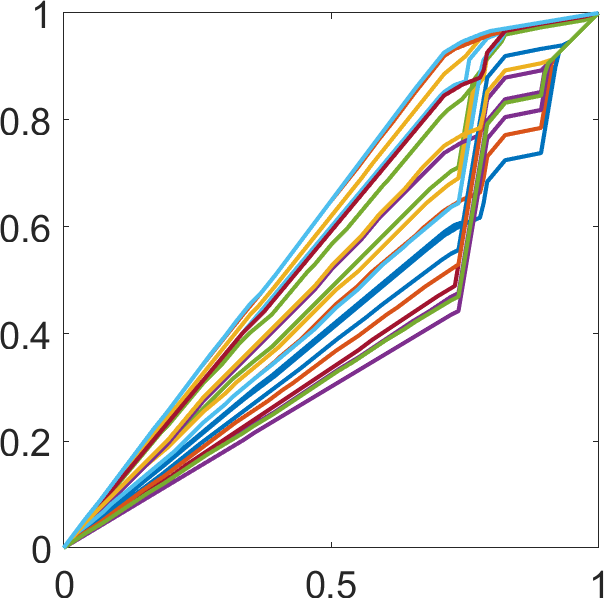}  &  \includegraphics[width = .8 in]{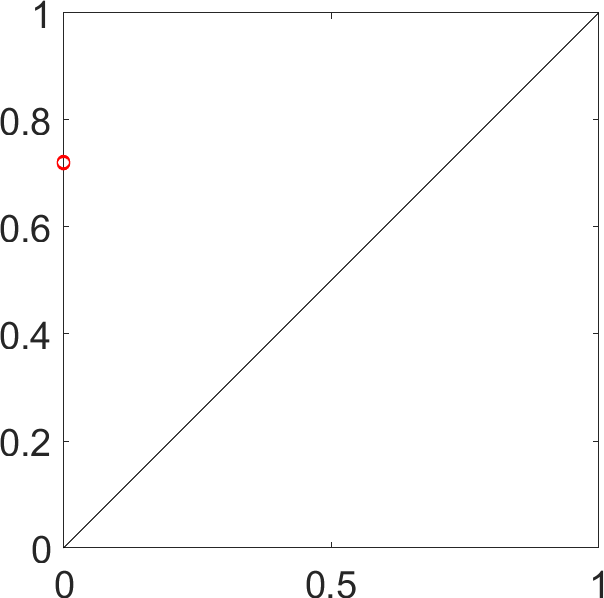} \\
        (h) & (i) & (j) \\
    \end{tabular}
    \end{center}
    \caption{\emph{Same topology with scale variability:} (a)\&(c) Two examples, from 20, of randomly generated point clouds. (b)\&(d) Corresponding persistence landscapes. (e) Persistence landscapes $\{\Lambda_i\}_{i=1}^{20}$ of 20 point clouds. (f) Aligned persistence landscapes $\{\Lambda_i(\gamma_i)\}_{i=1}^{20}$. (g) Mean landscape after (blue) and without (red) alignment. (h) Noisy persistence diagrams $\{(b_{i},d_{i})\}_{i=1}^{20}$ from 20 point clouds. (i) Estimated reparameterizations $\{\gamma_i\}_{i=1}^{20}$. (j) Denoised persistence diagrams $\{(\gamma_i^{-1}(b_{i}),\gamma_i^{-1}(d_{i}))\}_{i=1}^{20}$.}\label{meanEst:spiral}
\end{figure}

\noindent\textbf{Example 2.} In Figure \ref{meanEst:spiral}, we consider mean estimation based on degree $p=0$, $K=1$-dimensional persistence landscapes computed from 20 point clouds that consist of 2000 points uniformly sampled along two interwoven spirals. The tightness of the spirals is random, so that the spirals complete $\text{Uniform}(2,5)$ revolutions. Panels (a) and (c) show two examples of point clouds generated in such a manner with the corresponding landscapes shown in panels (b) and (d). The tighter spirals in (a) have points closer together, and the resulting landscape is smaller and shifted to the left as compared to the spirals in (c). When computing landscapes, we only considered the point in persistence diagrams that corresponded to the death time that coincided with the intersection of the two spirals present in each point cloud. Panels (e)-(g) show landscapes for all 20 point clouds, their alignment, and a comparison of the mean before (red) and after (blue) alignment. The mean based on aligned landscapes appears to have sharper features that are consistent with the observed landscapes. Based on the denoised persistence diagrams in (j), in contrast to the noisy persistence diagrams in (h), it is evident that reparameterization of landscapes completely accounts for the scale variability associated with the tightness of the spirals, i.e., all points collapse to a single point in the denoised diagrams.

\subsection{Example 3: PCA on aligned landscapes}\label{sec:sim2}

\begin{figure}[!t]
\begin{center}
\begin{tabular}{cccc}
      \includegraphics[width = 0.6 in]{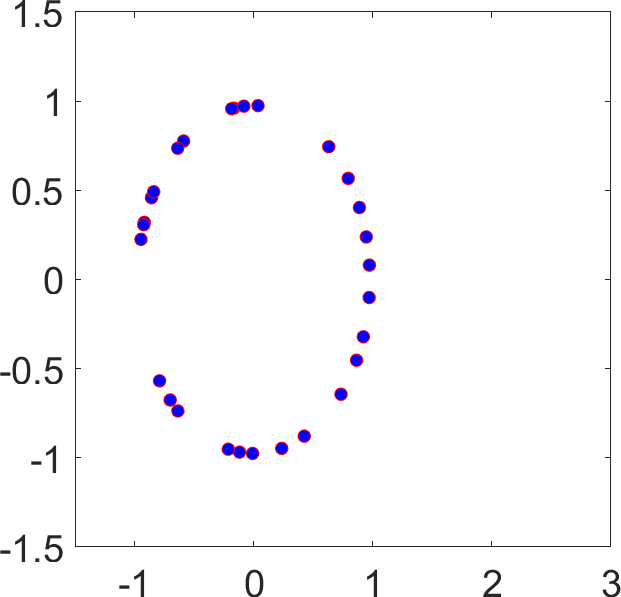} & \includegraphics[width = .7 in]{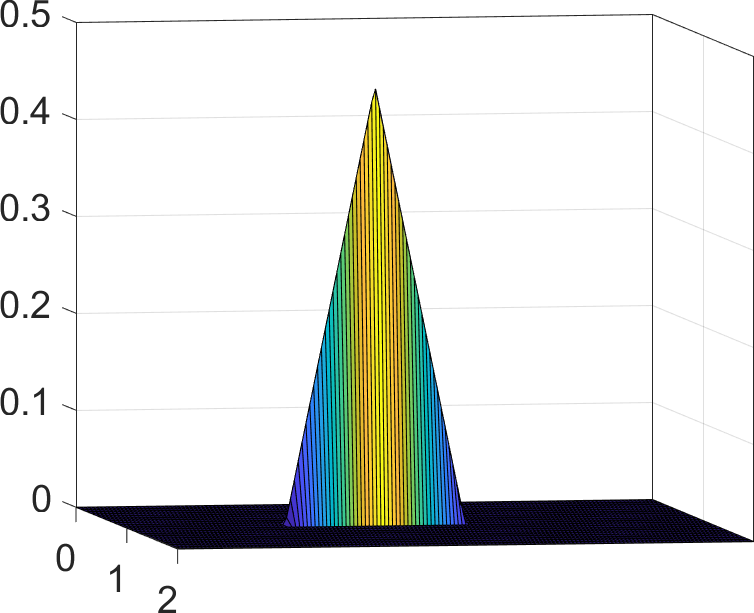}  & \includegraphics[width = 0.6 in]{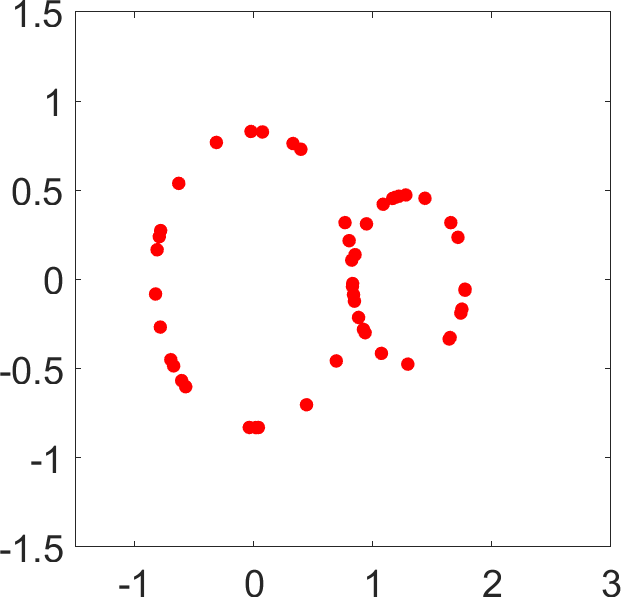} & \includegraphics[width = .7 in]{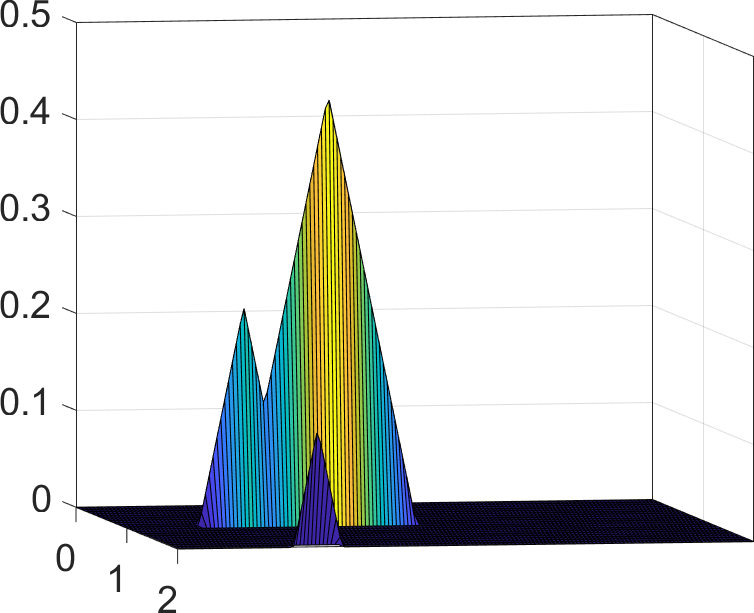}\\
     (a) & (b) & (c) & (d)\\
      
      \includegraphics[width = .7 in]{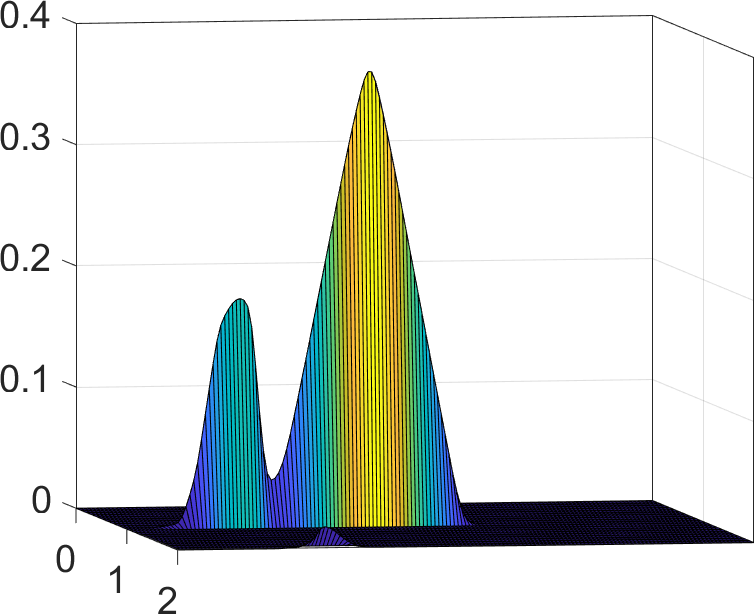} & \includegraphics[width = .7 in]{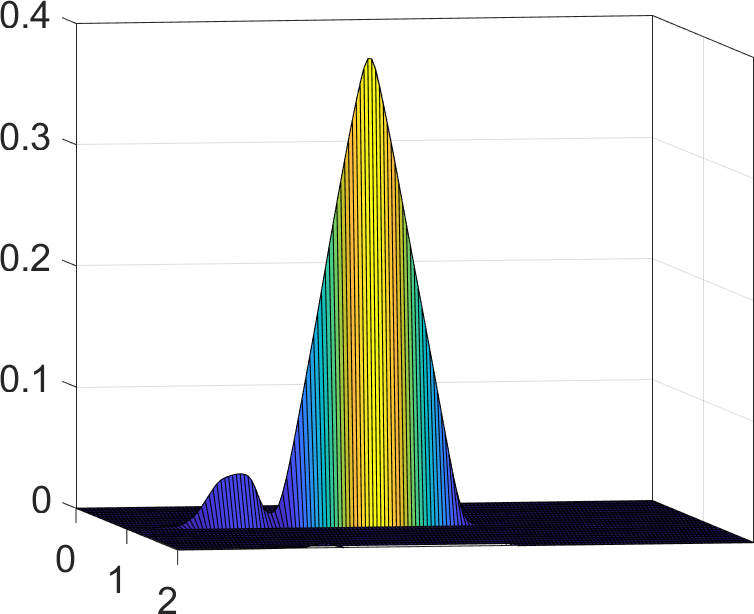}  & \includegraphics[width = .7 in]{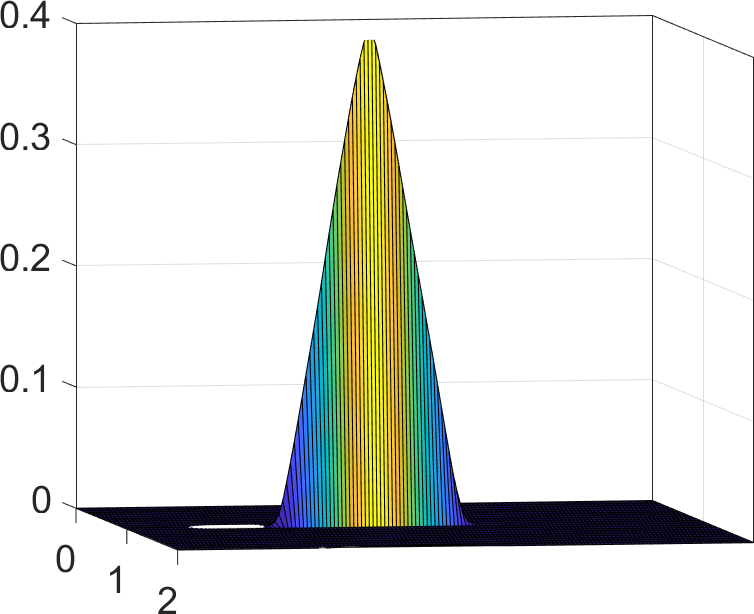} & \includegraphics[width = .7 in]{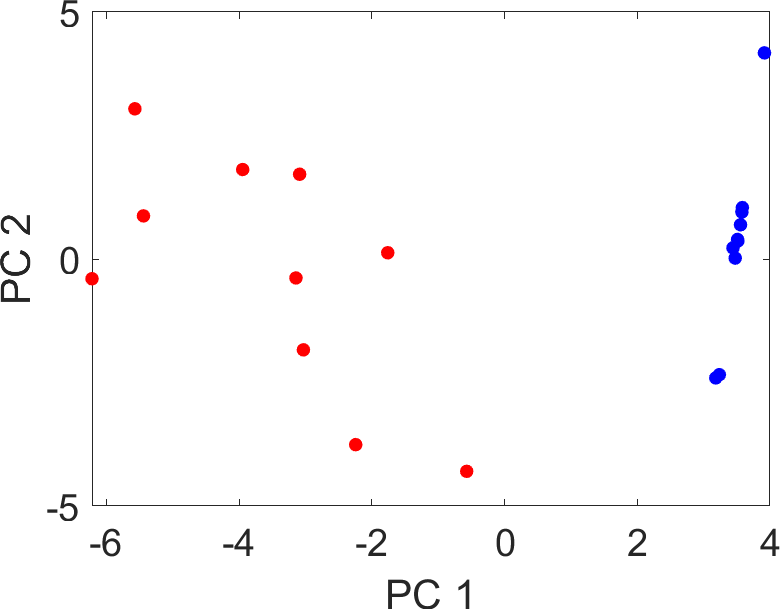}\\
    
    \includegraphics[width = .7 in]{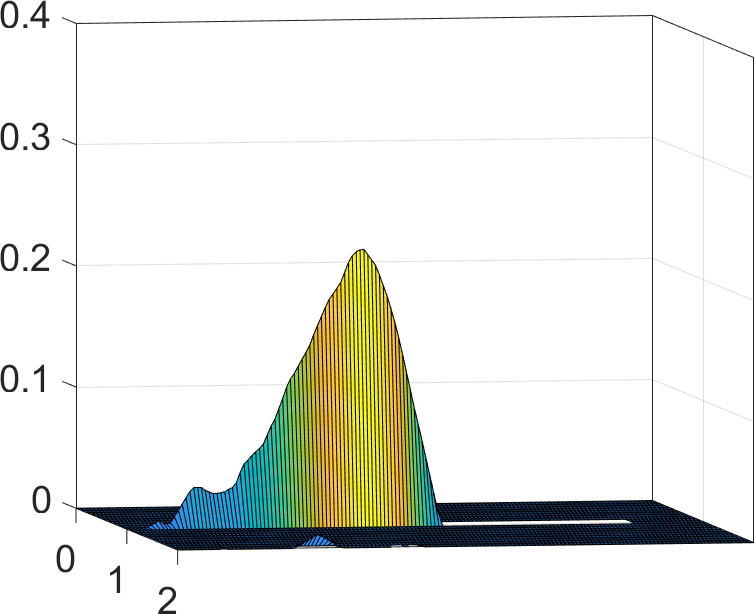} & \includegraphics[width = .7 in]{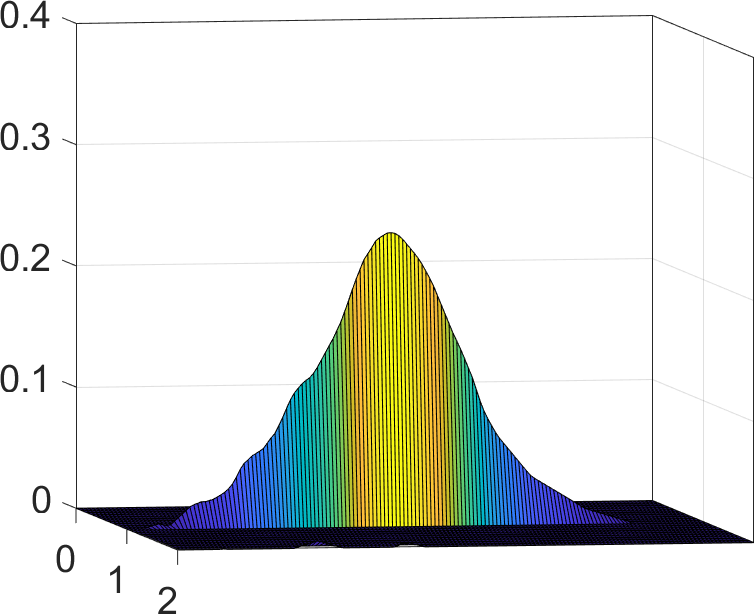}  & \includegraphics[width = .7 in]{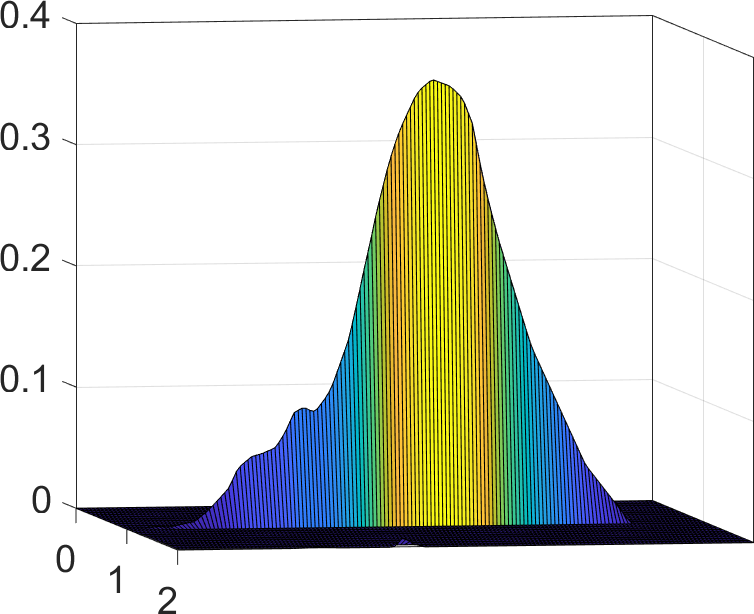} & \includegraphics[width = .7 in]{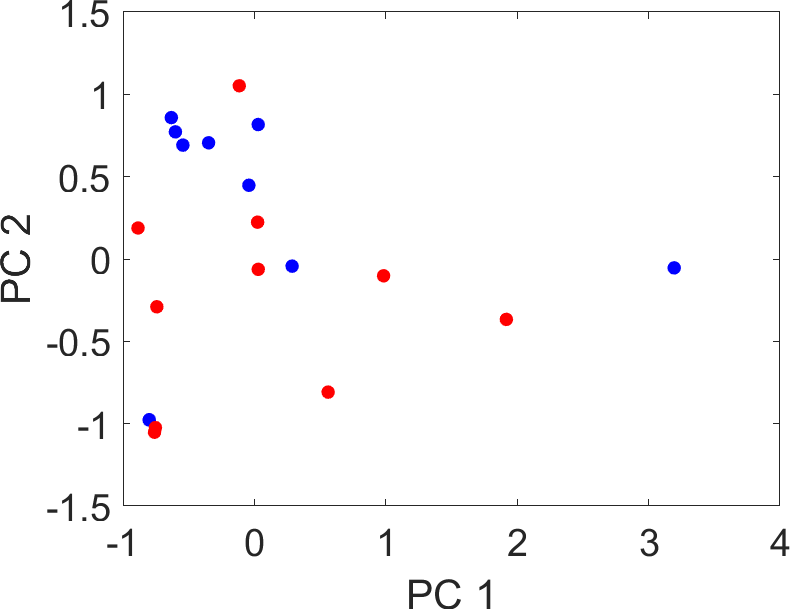}\\
     (e)  & (f) & (g) & (h)\\
      \end{tabular}
      \begin{tabular}{ccc}
        \includegraphics[width = .8 in]{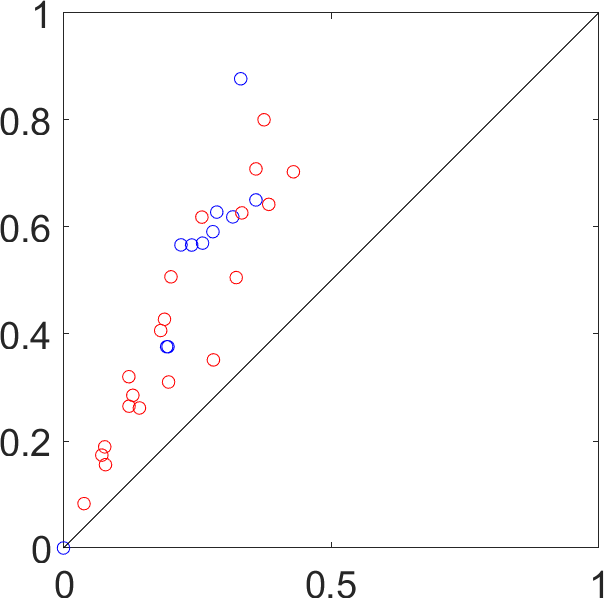} & \includegraphics[width = .8 in]{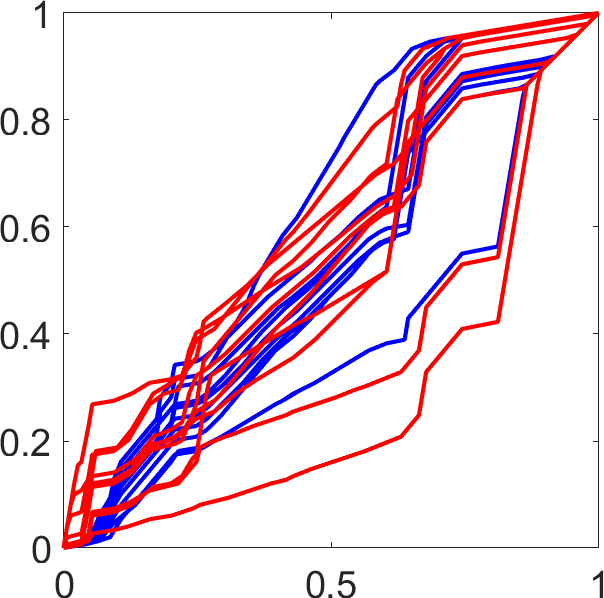}  &  \includegraphics[width = .8 in]{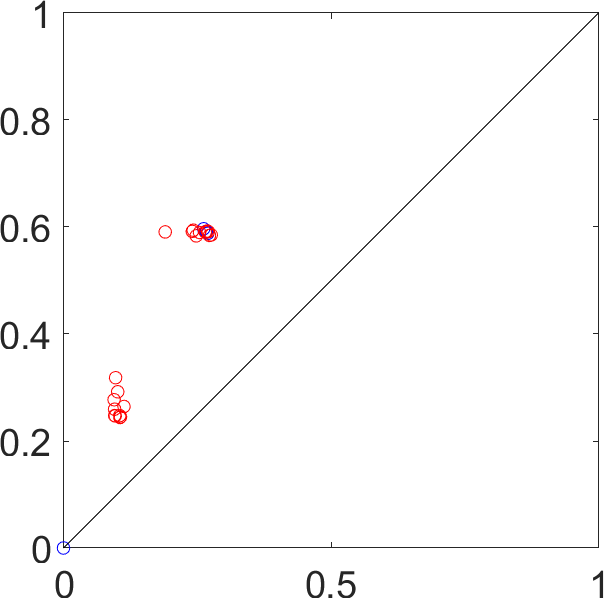} \\
     (i) & (j) & (k)
    \end{tabular}
    \caption{\emph{Different topology with scale and sampling variabilities}: (a)\&(c) Two examples, from 20, of randomly generated point clouds from topologically different spaces (blue and red, respectively, in all relevant panels). (b)\&(d) Corresponding degree $p=1$, $K=2$-dimensional persistence landscapes. (e) -1, (f) 0, (g) +1 standard deviation from the mean landscape in the first PC direction, and (h) projection of landscapes onto the first two PC directions: following alignment (top) and without alignment (bottom). (i) Noisy and (k) denoised persistence diagrams. (j) Estimated reparameterizations.}
    \label{pcaEg}
    \end{center}
\end{figure}
We consider a more involved setting involving 20 point clouds from two topologically different spaces: (i) one circle, and (ii) two connected circles. Point clouds from (i) are drawn in the same manner as in Example 1, but for the fact that the sample size $M$ is drawn from a $\text{Discrete-Uniform}(20,30)$. For point clouds from (ii), the radius of the larger circle is drawn from a $|N(1,0.3^2)|$, while the radius of the smaller circle is a random proportion of the larger circle, drawn from a $\text{Beta}(10,10)$. Panels (a) and (c) in Figure \ref{pcaEg} show one point cloud each from (i) and (ii). We consider degree $p=1$, $K=2$-dimensional landscapes $\{\Lambda_i=(\lambda_{i1},\lambda_{i2})\}_{i=1}^{20}$. For point clouds from (i), $\lambda_{i1}$ will have one peak and $\lambda_{i2}=0$ for all $t$; for point clouds from (ii), $\lambda_{i1}$ will have two peaks and $\lambda_{i2}$ will have a single peak. 

The top and bottom rows in Figure \ref{pcaEg}(e)-(g) show the primary PC direction of variability within one standard deviation of the mean following alignment and without alignment, respectively. Panel (h) highlights the benefits of alignment of landscapes through better separation of the two settings, (i) and (ii), when projected along the first two PC directions. Specifically, in the top row, when PCA is carried out on aligned landscapes, all of the point clouds that have two loops have a negative first PC score, while all of the point clouds with only one loop have a positive first PC score. There is no such clear separation of the two groups when PCA is performed on unaligned landscapes, as seen in the bottom row of (h). The separation in PC scores after alignment is directly related to the improved interpretability of the primary PC direction shown in the top row in (e)-(g): the landscape in (e) corresponding to -1 standard deviation from the mean in (f) exhibits features of a landscape for a point cloud with two loops, while the landscape in (g) corresponding to +1 standard deviation exhibits features of a landscape for a point cloud with one loop. 

Following results from previous simulations, we expect to see two clear clusters in the denoised persistence diagrams, using reparameterizations $\{\gamma_i\}$ shown in (j), corresponding to two distinct topological features; this is indeed the case as seen in (k). This is explained as follows: points concentrated around $(b,d) \approx (0.25,0.6)$ correspond to the single circle in the blue point clouds and the large circle in the red ones. This is consistent with the data generating process where the large circles across the two groups correspond to each other. The points associated with the second feature for the red point clouds are concentrated around $(b,d) \approx (0.1,0.25)$ and correspond to the additional significant homological feature (smaller circle) that generally has smaller persistence than the larger circle. It is very difficult to discern such topological information from the noisy diagrams in (i).

\section{Analysis of Brain Artery Trees}\label{sec:brainEg}
We now demonstrate the utility of the proposed approach on 3D point clouds representing brain artery trees. These data were collected to understand population attributes of brain arteries and how these attributes vary with demographic covariates. For a description of the experiment and data generation, see \cite{bullit_2005}. Information regarding human subjects in the experiment is available at \url{http://insight-journal.org/midas/community/view/21}. Chapter 10.1 in \cite{marron_2021} provides an overview and comparison of past approaches used to analyze this data.

The approach of Bendich et al. \cite{bendich_2016} computed persistence diagrams from artery trees for 98 healthy human subjects, and used persistence diagrams to extract the 100 largest birth-death differences, $\{d_{i,j} - b_{i,j}\}_{i=1,j=1}^{98,100}$, for each subject. Restricting focus to the largest differences serves as a denoising step, since points close to the line $b = d$ in a persistence diagram can be thought of as noise \citep{fasy_2014}. 

This dataset is apt to demonstrate our approach for three reasons, one \emph{a priori} and the other two \emph{a posteriori}:
(i) since each point cloud contains a large number (order of $10^5$) of points, in order to be able to compute the diagrams, \cite{bendich_2016} subsampled 3000 points from each point cloud, thus creating large sampling variability; (ii) we uncover a significant scale effect between the two sex groups of subjects (males versus females); and (iii) we confirm the finding of \cite{bendich_2016} that there exists a significant correlation between the topological structure of the brain artery trees (as captured by PCs) and age. We note that these findings are exploratory and serve as a proof of concept for the proposed approach.

\subsection{Exploration of sex differences among subjects}

The starting point for our analysis are the persistence diagrams available at \url{https://marron.web.unc.edu/brain-artery-tree-data/}, and not the original 3D point clouds. From these, we compute degree $p=1$, $K=100$-dimensional persistence landscapes. Information on the sex of each subject is also available along with the tree data. We investigate differences between mean persistence landscapes grouped by sex. One major finding of \cite{bendich_2016} is the existence of sex differences in their mean degree $p=1$ feature vectors $\{d_{i,j} - b_{i,j}\}_{i=1,j=1}^{98,100}$. For convenience, we denote the mean landscape for the male (female) group following alignment (within each group) by $\hat \mu^m_{\text{a}}\ (\hat \mu^f_{\text{a}})$, and pointwise mean computed without alignment for the males (females) by $\hat \mu^m\ (\hat \mu^f)$.

\begin{figure}[!t]
\begin{center}
\begin{tabular}{ccc}
        \includegraphics[width = 1 in]{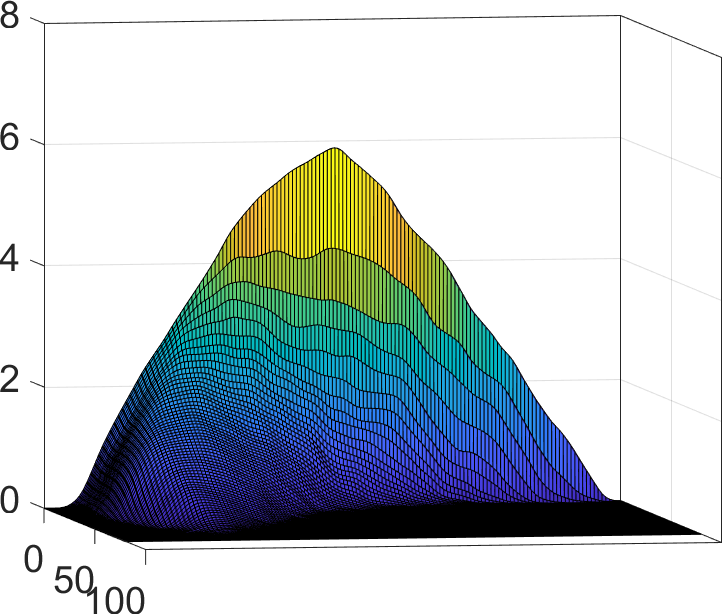} &  \includegraphics[width = 1 in]{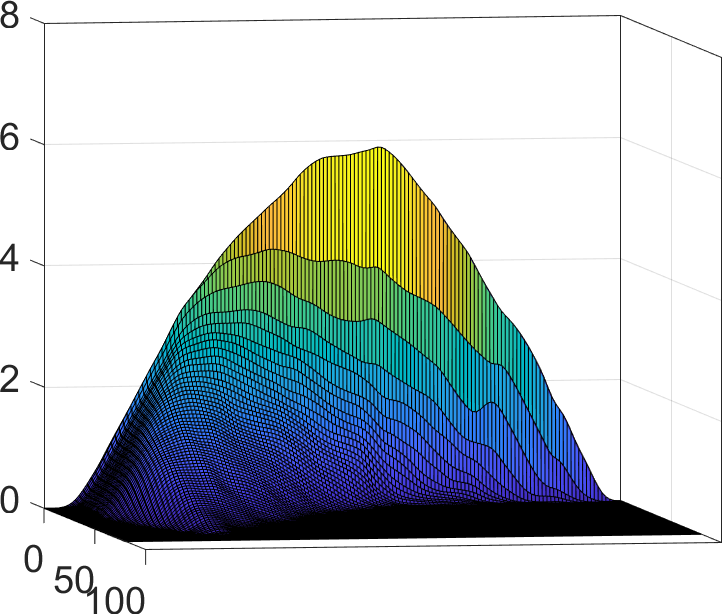}  & \includegraphics[width = 1 in]{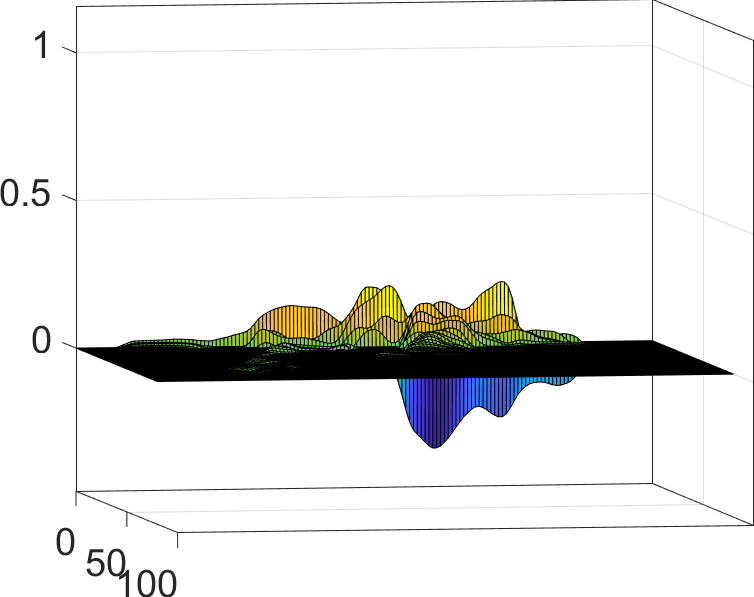} \\
       \includegraphics[width = 1 in]{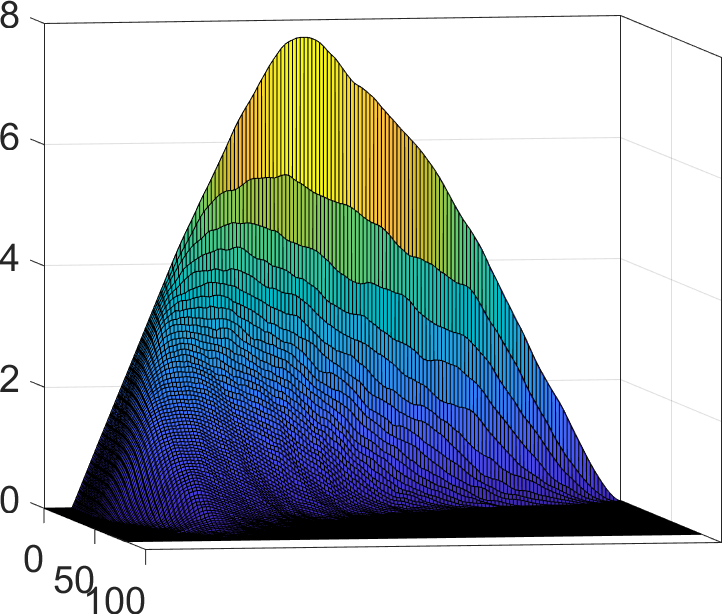} &  \includegraphics[width = 1 in]{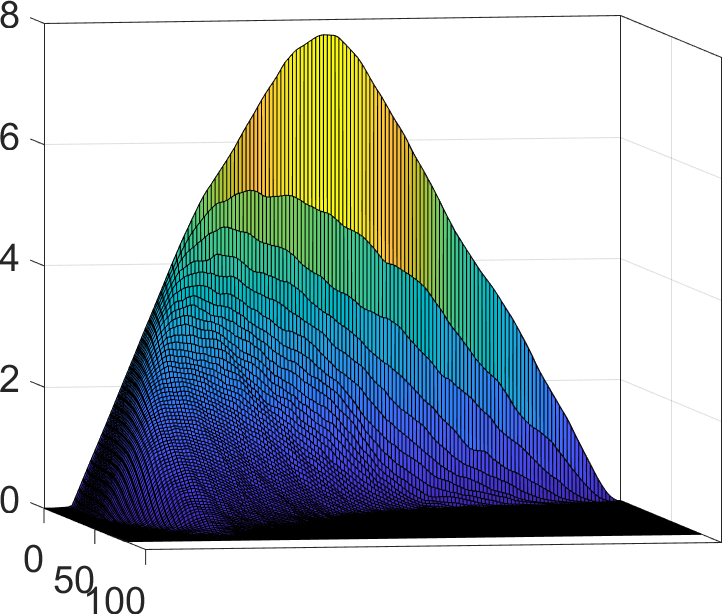}  & \includegraphics[width = 1 in]{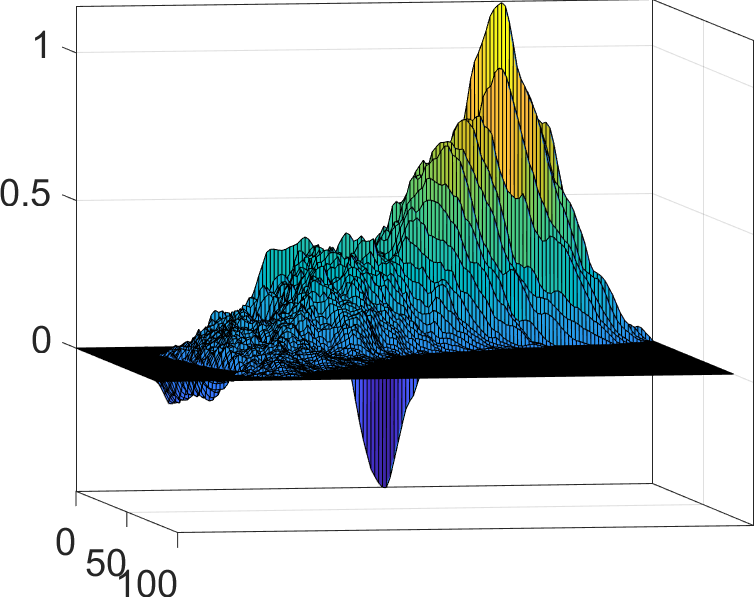} \\
       (a) & (b) & (c) \\
\end{tabular}
\begin{tabular}{cc}
        \includegraphics[width = 1.1 in]{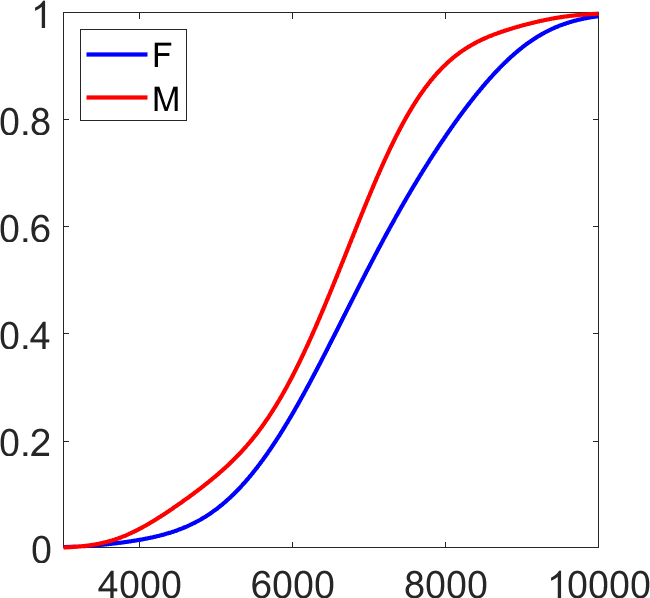}
       &  \includegraphics[width = 1 in]{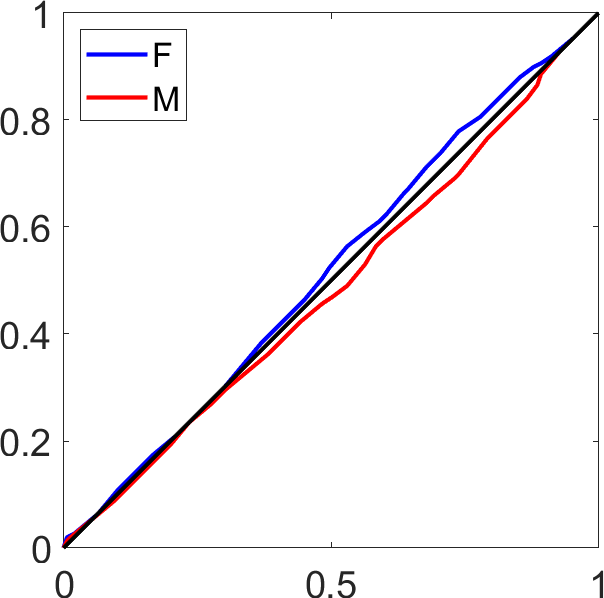}\\
       (d) & (e) \\
\end{tabular}
    \caption{\emph{Top. Mean persistence landscapes and their differences for males and females in the brain artery example with (top row) and without (bottom row) alignment:} (a) Male mean, (b) female mean, and (c) difference between (a) and (b). \emph{Bottom. Relationship between groupwise total artery length and relative phase of groupwise means following alignment, across sexes:} (d) Estimated groupwise CDFs of total artery length, and (e) reparameterizations that align groupwise aligned means to a mean computed from alignment of the pooled data. The identity parameterization (black) is shown for reference.} 
    \label{brainMean}
    \end{center}
\end{figure}

The means $\hat \mu^m_{\text{a}}$ and $\hat \mu^f_{\text{a}}$ are shown in the top row of Figure \ref{brainMean}(a)-(b), respectively. The difference between the means is obtained by first aligning the group means to the common pooled mean (computed with alignment) and then taking their difference, where all operations are carried out under the SRVF representation; this difference is shown in the top row of panel (c). The bottom row of Figure \ref{brainMean}(a)-(c) shows the pointwise means $\hat \mu^m$ and $\hat \mu^f$, and the corresponding $(\hat \mu^m-\hat \mu^f)$, when no alignment is carried out. The difference between the pointwise means has very large features indicating topological and geometric structural differences across males and females. However, these features are essentially non-existent in the difference of the aligned means. \emph{This indicates that the large difference in the pointwise means is potentially due to misalignment, and can be construed as topological noise}. 

We surmise that global scale differences and sampling variability in observed data between the male and female groups may be responsible for this phenomenon. To confirm this, we use the total artery length for each subject as a measure of global scale, which is also available as part of the tree data \citep{bullitt_2005b}. For each group, we estimate a cumulative distribution function (CDF) of total artery length using a kernel density estimate using the \verb|ksdensity| function in \verb|MATLAB| (default bandwidth). From the estimated CDFs shown in Figure \ref{brainMean}(d), it appears that total artery length is stochastically ordered by sex, with females having stochastically longer brain artery trees. Given this global scale disparity and sampling variability, we would expect $\hat \mu_a^f$ to be shifted to the right relative to $\hat \mu_a^m$. This behavior can be extracted from the phase difference between $\hat \mu_{\text{a}}^m$ and $\hat \mu_{\text{a}}^f$ after alignment to the pooled sample mean, as shown in Figure \ref{brainMean}(e). The blue reparameterization (female) shifts $\hat \mu_{\text{a}}^f$ to the left while the red (male) shifts $\hat \mu_{\text{a}}^m$ to the right. Thus, the misalignment caused by differences in global scale and sampling between the two groups appear to explain the reason behind the large difference between the groupwise pointwise means. In summary, the above analysis suggests that the sex effect detected via pointwise analysis without alignment of the landscapes, corresponding to some of the results presented in Table 10.1 in \cite{marron_2021}, is due to global scale and/or sampling differences of the observed data rather than differences in homology, and thus makes a compelling case study of the perils in ignoring the distinction between, and confounding of, amplitude and phase in landscapes.

\subsection{Correlation between age and topological structure of brain artery trees}

Bendich et al. \cite{bendich_2016} show that age is significantly correlated with the dominant PCs estimated using degree $p=0$ persistence diagrams. One potential confounding variable for this relationship is total artery length, since this quantity is also significantly correlated with age (correlation of $-0.63$; see Figure 6(a) in Appendix C in supplement). To account for this, they first rescaled each subject's persistence diagram by their total artery length and then measured correlation between age and the dominant PCs estimated using rescaled diagrams. Even after rescaling, the relationship holds.

\begin{figure}[!t]
\begin{center}
\begin{tabular}{cc}
        \includegraphics[width = 1.3 in]{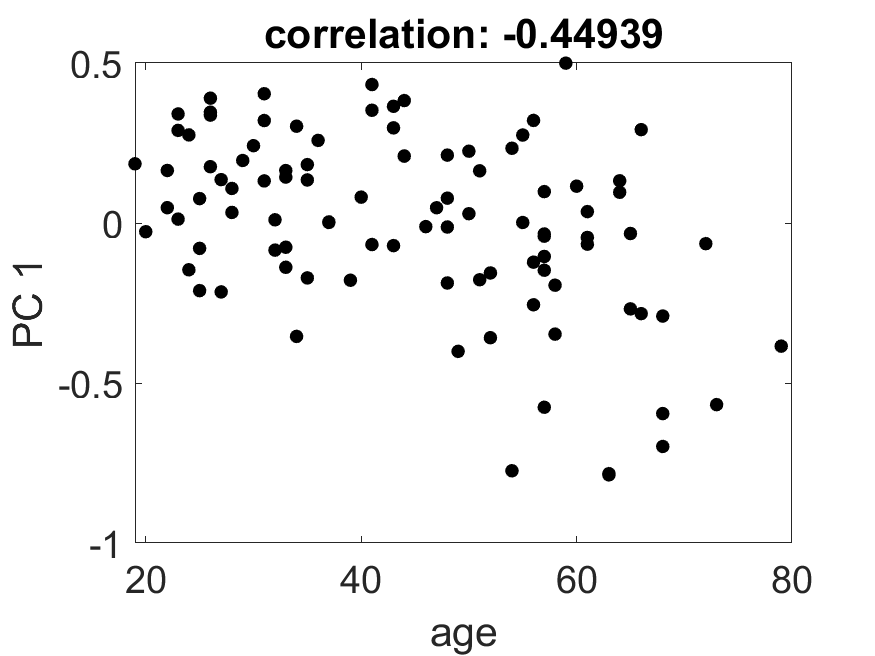} & \includegraphics[width = 1.3 in]{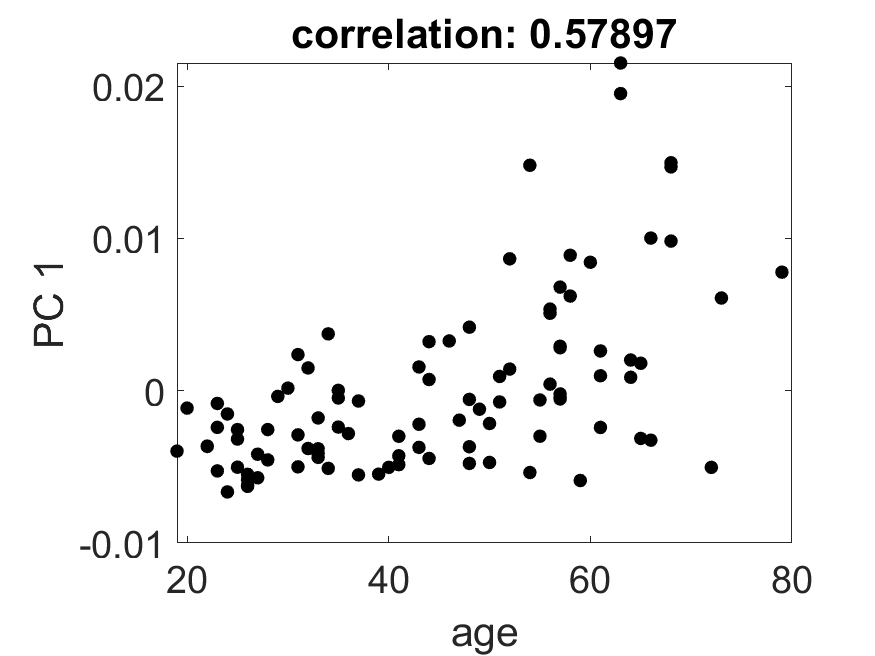}\\
        \includegraphics[width = 1.3 in]{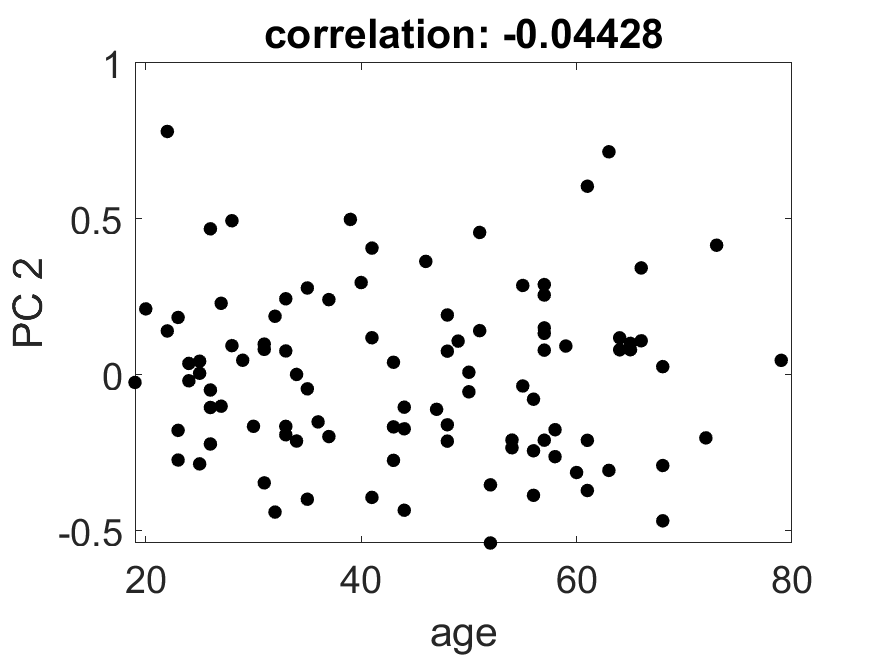}  & \includegraphics[width = 1.3 in]{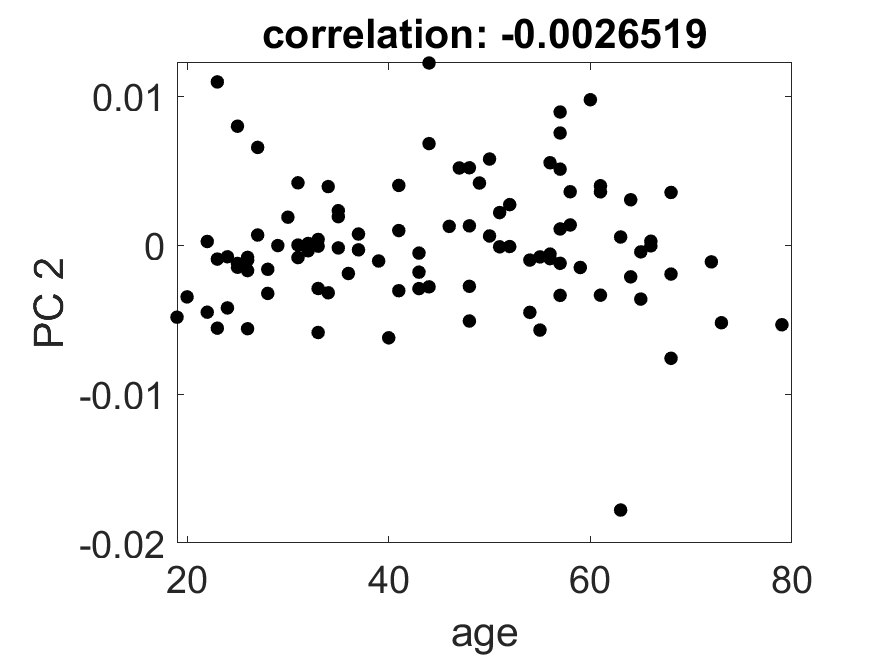} \\
       (a) & (b)\\
           \end{tabular}
    \caption{
    \emph{Correlations and scatterplots of PC 1 (top) and PC 2 (bottom) estimated using (a) aligned and (b) unaligned landscapes, computed from rescaled persistence diagrams, versus age.}}
    \label{brainH0Pcs}
    \end{center}
\end{figure}

Here, we reanalyze the data using $K=100$-dimensional persistence landscapes from rescaled degree $p=0$ persistence diagrams. We estimate PCs using aligned and unaligned landscapes, and project onto the first two PC directions to explore whether there is a correlation between either of the first two PCs and age. Figure \ref{brainH0Pcs}(a)-(b) shows the relationship between age and the first two PCs for aligned and unaligned landscapes, respectively. In both panels, age appears to be strongly associated with the first PC. \emph{Thus, our findings are consistent with \cite{bendich_2016} and the results presented in Table 10.1 in \cite{marron_2021}. We further note that alignment of the persistence landscapes computed using rescaled persistence diagrams decreases the correlation slightly from 0.58 to 0.45 (note that the sign of the correlation coefficient is irrelevant here due to lack of directionality in the PCs); this is likely due to residual scale effects (and sampling variability) after accounting for the total artery length.} Correlations computed using PCs from both unaligned and aligned persistence landscapes are similar to the correlation of 0.53 reported under TDA $H_0$ in Table 10.1 in \cite{marron_2021}. Appendix C in the supplement reports results of this analysis when original persistence diagrams (without rescaling by total artery length) are used to derive persistence landscapes; there, we show that scale is confounded with topo-geometric structure when unaligned landscapes are used to carry out PCA.

\section{Classification of Gleason Data}\label{sec:gleason}

\begin{figure}[!t]
\begin{center}
    \begin{tabular}{cccc}
    (a) & \includegraphics[width = .8 in]{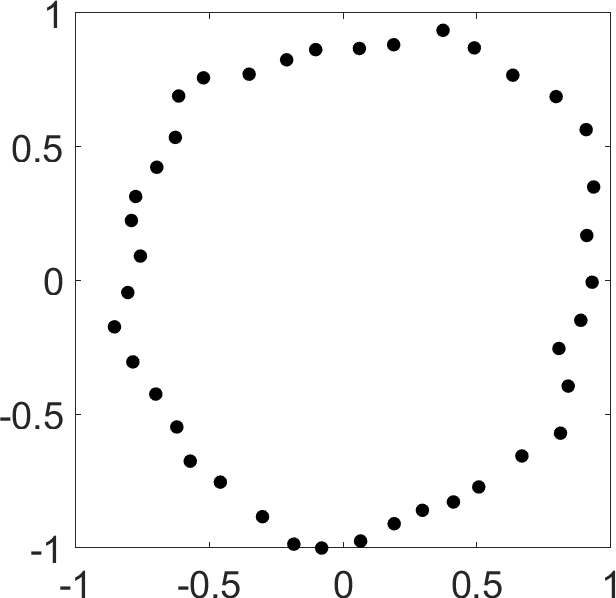} & \includegraphics[width = .8 in]{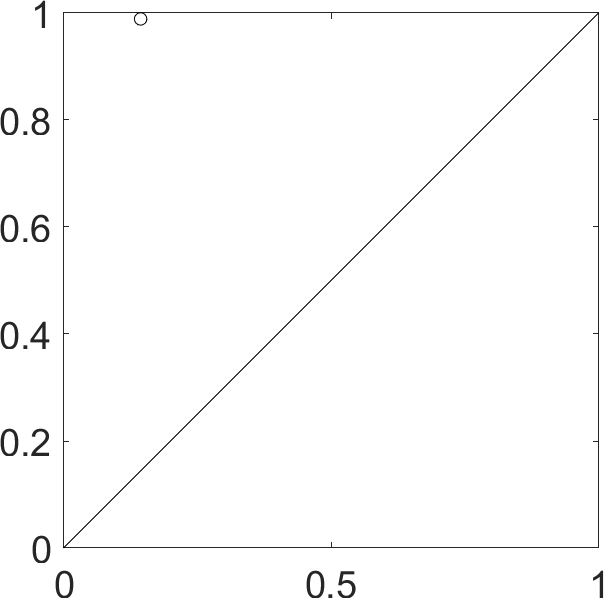} & \includegraphics[width = 1 in]{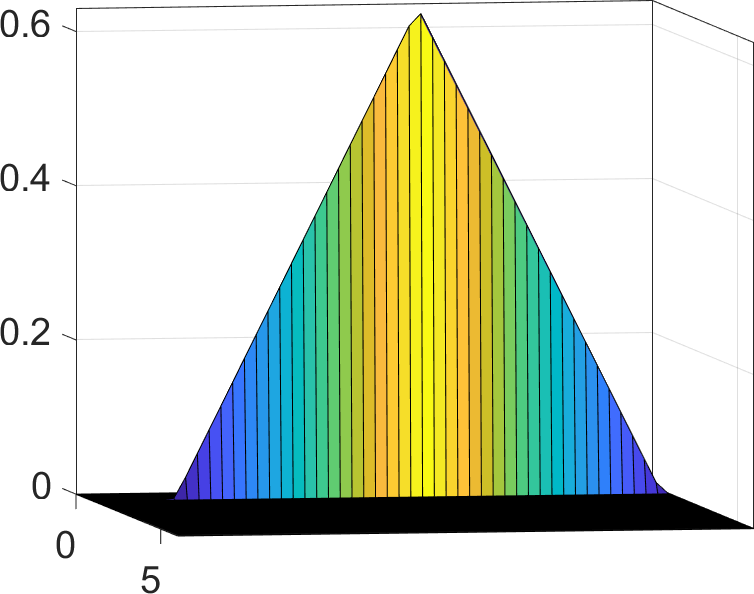}\\
    (b) & \includegraphics[width = .8 in]{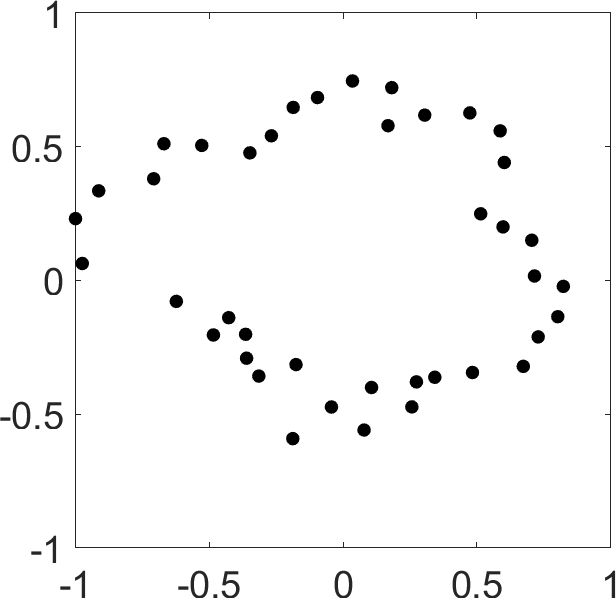} & \includegraphics[width = .8 in]{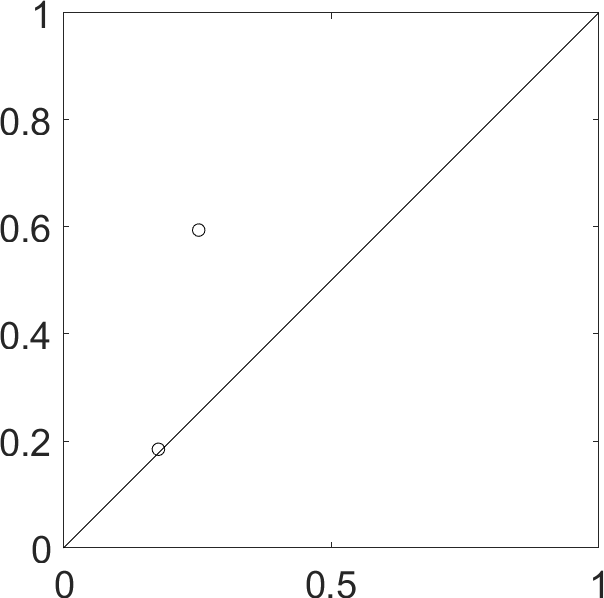} & \includegraphics[width = 1 in]{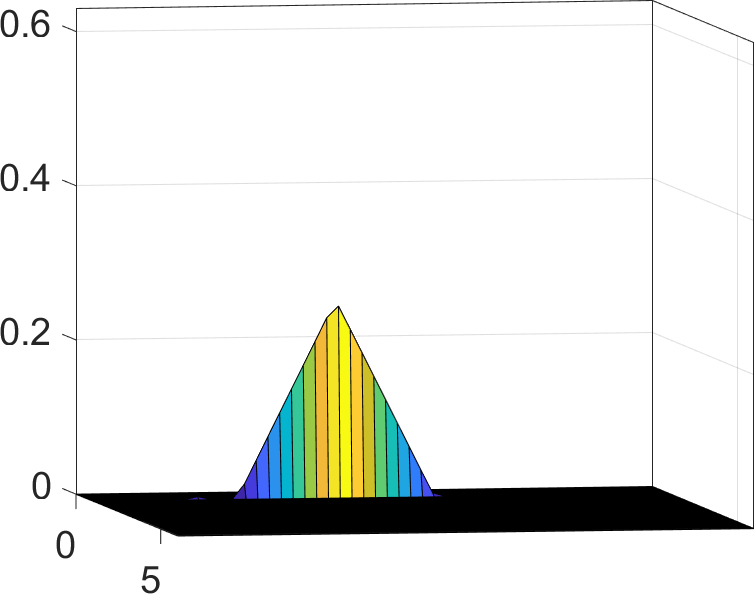}\\
    (c) & \includegraphics[width = .8 in]{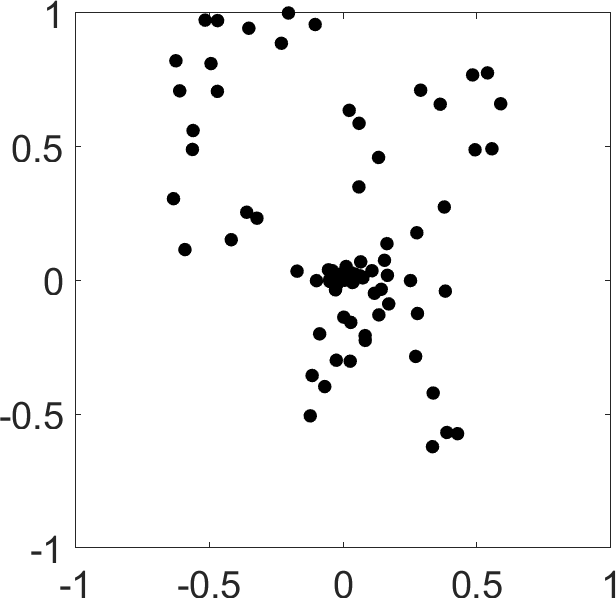} & \includegraphics[width = .8 in]{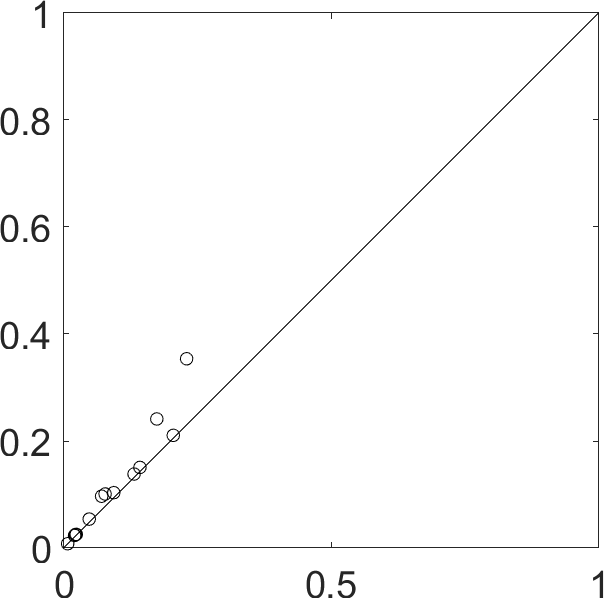} &  \includegraphics[width = 1 in]{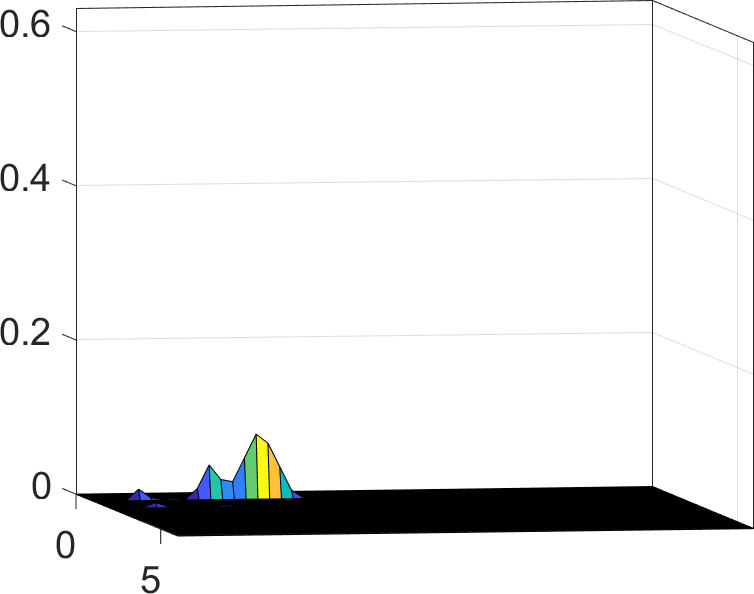}\\
    (d) & \includegraphics[width = .8 in]{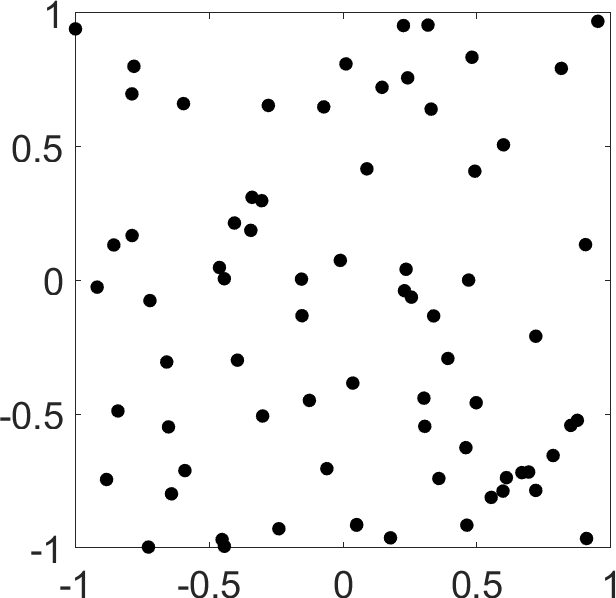} & \includegraphics[width = .8 in]{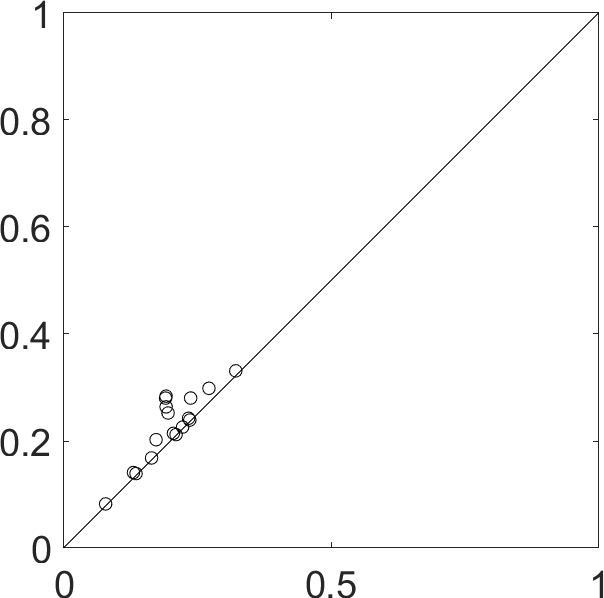} & \includegraphics[width = 1 in]{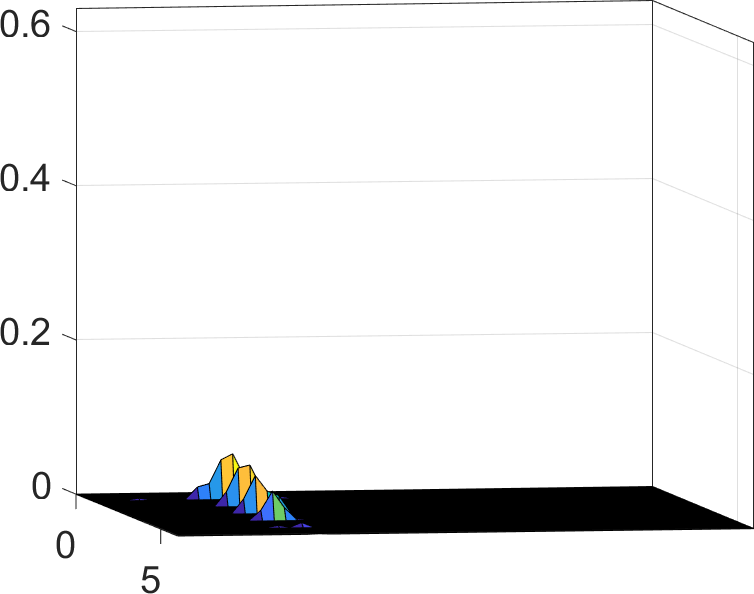} \\
    \end{tabular}
    \end{center}
    \caption{\emph{Point clouds (left), persistence diagrams (middle) and landscapes (right) from each of the four Gleason grading scales:} (a)-(d) benign to most severe.}
    \label{gleasonData}
\end{figure}

As described in \cite{berry_2020}, the Gleason grading system is a prognostic tool to help understand severity of prostate cancer. Grading groups are assigned based on features of a prostate gland biopsy. In more benign biopsies, carcinoma walls are well-defined as seen in Figure \ref{example_data}(a). On the other hand, malignant carcinoma lose their structure and have very irregular shapes as shown in Figure \ref{example_data}(d). Using a variety of functional summaries of persistence diagrams, \cite{berry_2020} classified simulated point clouds representing four different Gleason grade groups that ranged from benign to unhealthy. A representative from each of the four classes used in their study are shown in the left of Figure \ref{gleasonData}(a)-(d), from benign to most severe. The middle and left columns show the corresponding degree $p=1$ persistence diagrams and $K=5$-dimensional landscapes, respectively. As the prognostic grade worsens, the cycle in the point clouds, corresponding to the carcinoma outline, becomes less pronounced. This is accompanied by significant changes in geometry and scale. Thus, the benign class is characterized by a persistence landscape with a large maximum in the first component function, and negligible maxima in subsequent component functions. On the other hand, the most severe class is characterized by a landscape with multiple small maxima along several component functions. Clearly, the landscapes contain significant amplitude and phase variation, and our aim is to classify the point clouds (landscapes) into the four Gleason grades based on these two components. The data consists of 2400 point clouds with 600 in each of the four Gleason grades.

In particular, using this data, we study $k$-nearest neighbor (KNN) classification accuracy using the amplitude and phase features of the landscapes. We also compare to one of the approaches taken in \cite{berry_2020}, which applied KNN classification to the landscapes without alignment. KNN classification is a distance-based approach, and we compare performance based on the following three distances: (i) $\mathbb{L}^2$ distance: $d_{\mathbb{L}^2}(\Lambda_1,\Lambda_2) = \|\Lambda_1 - \Lambda_2\|_2$; (ii) amplitude distance: $d_{\text{a}}(\Lambda_1,\Lambda_2) = \underset{\gamma \in \Gamma}{\min}\|q_1 - (q_2,\gamma)\|_2$; (iii) phase distance: $d_{\text{p}}(\Lambda_1,\Lambda_2) = \arccos\left(\int_0^1\sqrt{\dot{\gamma}^*(s)}ds\right)$, where $\gamma^* = \underset{\gamma \in \Gamma}{\text{argmin}}\|q_1 - (q_2,\gamma)\|_2$. The KNN classification procedure is implemented as follows. For a test observation (unknown class), we first compute its distance from each observation in the training data (known class). Then, we find the $k$ nearest neighbors in the training set to the test observation, and predict its class as the one that is most frequent among the $k$ nearest training neighbors. In case of a tie, we use the class of the nearest training neighbor. The classification accuracy is then computed as the percentage of correctly predicted classes in the test set. Using the same training and testing split as \cite{berry_2020}, data are split into $83\%$ training (2000 landscapes, 500 in each of the four severity classes) and $17\%$ testing (400 landscapes, 100 in each of the four severity classes).

\begin{figure}[!t]
\begin{center}
      \begin{tabular}{ccc}
    \includegraphics[width = 1 in]{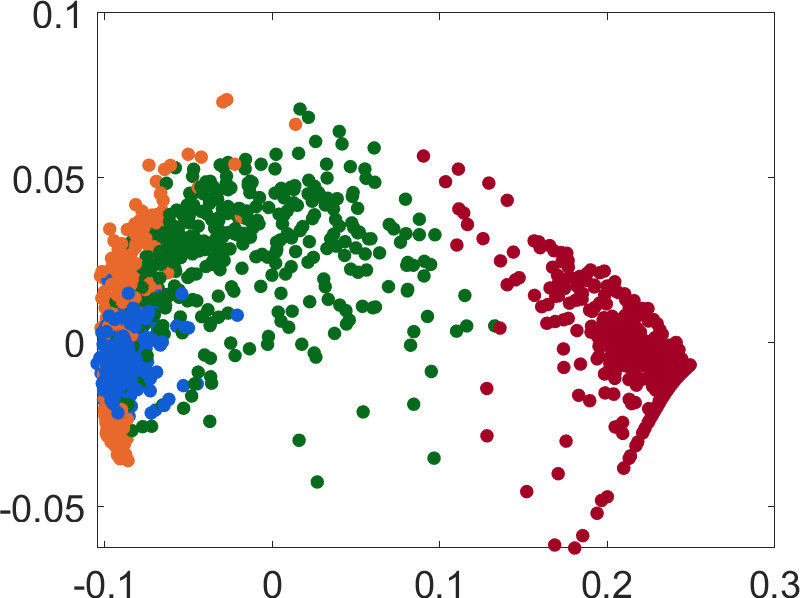} & \includegraphics[width = .95 in]{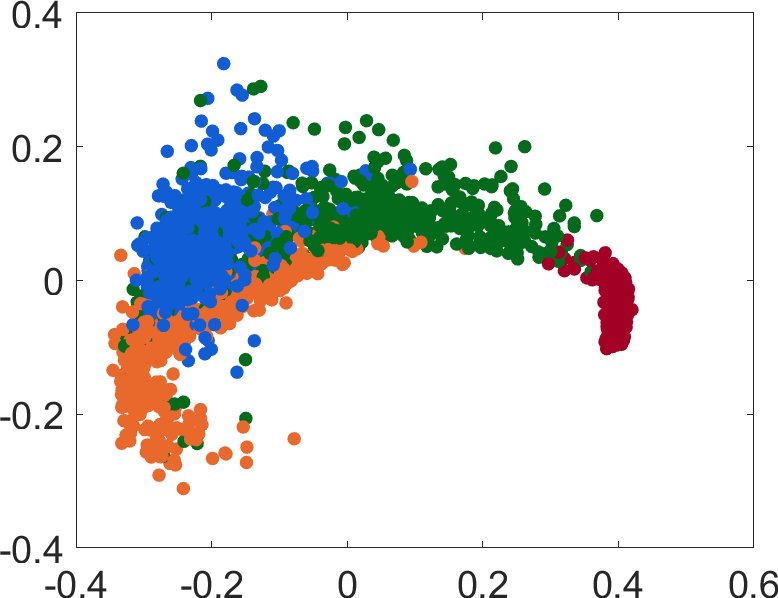} & \includegraphics[width = .9 in]{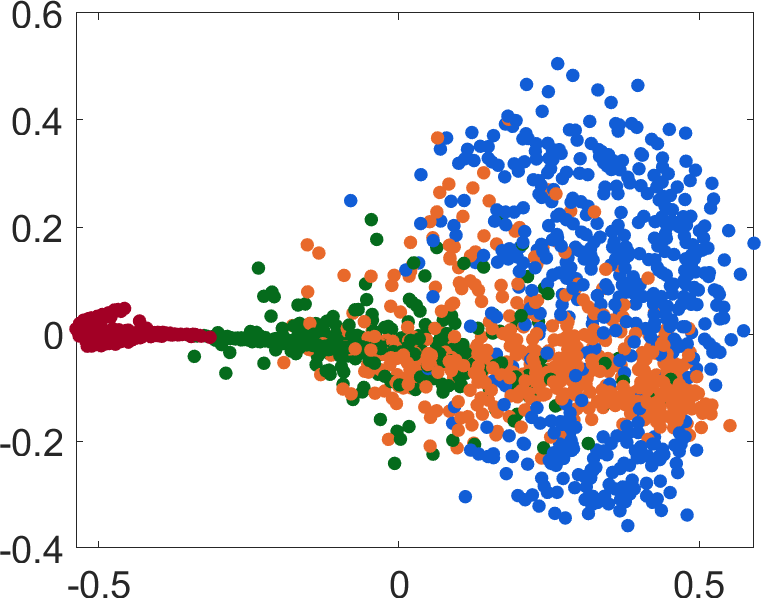} \\
    (a) & (b) & (c) \\
    \end{tabular}
    \end{center}
    \caption{\emph{Multidimensional scaling plots for the training dataset:} (a) $\mathbb{L}^2$ distance, (b) phase distance, and (c) amplitude distance. \emph{Each point is colored according to class membership with benign = red, grade 2 = green, grade 3 = orange, and grade 4 = blue.}}
    \label{gleasonMDS}
\end{figure}

We begin by displaying the 2D multidimensional scaling (MDS) plots, computed using the three different distances, for the training dataset. In short, MDS uses pairwise distances to compute lower-dimensional Euclidean coordinates of the data such that interpoint Euclidean distances based on these coordinates are as similar as possible to the original distances. It is evident that there is good separation between the benign (red) and Gleason grade 2 (green) classes for each of the three distances. The $\mathbb{L}^2$ and phase distances also provide good separation between the grade 2, grade 3 (orange) and grade 4 (blue) classes, with phase appearing more discriminative between the grade 3 and grade 4 classes. However, the amplitude distance is ineffective at separating the grade 2, grade 3 and grade 4 classes. This result is not surprising. The amplitude (shape) of persistence landscapes is effective in capturing whether and how many cycles exist in the point clouds. Thus, while it is very effective in discriminating between the benign and severe classes, it does not provide effective finer classification into the four Gleason grades. On the other hand, the signal captured in the phase component is related to the size and geometry of the homological features. Visually inspecting the four point clouds in Figure \ref{gleasonData}, it is clear that these are the most discriminative features in the data. Finally, the $\mathbb{L}^2$ distance uses both amplitude and phase information of the landscapes without explicit control of the contribution of each component. We expect that these observations will lead to very good KNN classification rate based on the phase distance. On the other hand, the amplitude distance will only be effective at classifying benign versus severe classes.

A key question that has not yet been addressed is the choice of the number of nearest neighbors $k$. While we could fix this number \emph{a priori} to some small number of neighbors, say 1 or 3, this approach will not result in optimal classification performance. Instead, we will learn an optimal $k$, for each of the three distances, based on training data and then apply the KNN classifier with the optimal $k$ to testing data. We allowed values of $k=1,\dots,20$. Based on KNN classification applied to the training data, we determined the optimal values of $k$ to be 11 for the $\mathbb{L}^2$ distance, 19 for the amplitude distance, and 9 for the phase distance, using leave-one-out cross-validation.

\renewcommand{\tabcolsep}{3pt}
\begin{table}[!t]
\begin{center}
\begin{tabular}{|cc|cccc||cccc||cccc|}
\hline
                                                 \multicolumn{2}{|c|}{\multirow{3}{*}{{(a)}}}& \multicolumn{4}{c||}{$\pmb{\mathbb{L}^2}$}                                      & \multicolumn{4}{c||}{\textbf{Phase}}                                               & \multicolumn{4}{c|}{\textbf{Amplitude}}                                           \\ \cline{3-14} 
                                                 &&\multicolumn{12}{c|}{\textit{Overall Accuracy}}\\\cline{3-14}
                                                 && \multicolumn{4}{c||}{$91\%$}                                      & \multicolumn{4}{c||}{$92.75\%$}                                               & \multicolumn{4}{c|}{$70\%$}                                           \\ \hline\hline
                                                 \multicolumn{2}{|c|}{\multirow{3}{*}{{(b)}}}&\multicolumn{12}{c|}{\textit{Confusion Tables}}\\\cline{3-14}
                                                 &   & \multicolumn{4}{c||}{True}                                                         & \multicolumn{4}{c||}{True}                                                         & \multicolumn{4}{c|}{True}                                                         \\ \cline{3-14} 
                                                 &   & \multicolumn{1}{c|}{\textbf{1}}   & \multicolumn{1}{c|}{\textbf{2}}  & \multicolumn{1}{c|}{\textbf{3}}  & \textbf{4}  & \multicolumn{1}{c|}{\textbf{1}}   & \multicolumn{1}{c|}{\textbf{2}}  & \multicolumn{1}{c|}{\textbf{3}}  & \textbf{4}  & \multicolumn{1}{c|}{\textbf{1}}   & \multicolumn{1}{c|}{\textbf{2}}  & \multicolumn{1}{c|}{\textbf{3}}  & \textbf{4}  \\ \hline 
\multicolumn{1}{|c|}{\multirow{4}{*}{\rotatebox[origin=c]{90}{Predicted}}} & \textbf{1} & \multicolumn{1}{c|}{100} & \multicolumn{1}{c|}{2}  & \multicolumn{1}{c|}{0}  & 0  & \multicolumn{1}{c|}{100} & \multicolumn{1}{c|}{3}  & \multicolumn{1}{c|}{0}  & 0  & \multicolumn{1}{c|}{100} & \multicolumn{1}{c|}{4}  & \multicolumn{1}{c|}{0}  & 0  \\ \cline{2-14} 
\multicolumn{1}{|c|}{}                           & \textbf{2} & \multicolumn{1}{c|}{0}   & \multicolumn{1}{c|}{87} & \multicolumn{1}{c|}{9}  & 5  & \multicolumn{1}{c|}{0}   & \multicolumn{1}{c|}{89} & \multicolumn{1}{c|}{5}  & 10 & \multicolumn{1}{c|}{0}   & \multicolumn{1}{c|}{91} & \multicolumn{1}{c|}{23} & 2  \\ \cline{2-14} 
\multicolumn{1}{|c|}{}                           & \textbf{3} & \multicolumn{1}{c|}{0}   & \multicolumn{1}{c|}{8}  & \multicolumn{1}{c|}{86} & 4  & \multicolumn{1}{c|}{0}   & \multicolumn{1}{c|}{4}  & \multicolumn{1}{c|}{94} & 2  & \multicolumn{1}{c|}{0}   & \multicolumn{1}{c|}{5}  & \multicolumn{1}{c|}{77} & 86 \\ \cline{2-14} 
\multicolumn{1}{|c|}{}                           & \textbf{4} & \multicolumn{1}{c|}{0}   & \multicolumn{1}{c|}{3}  & \multicolumn{1}{c|}{5}  & 91 & \multicolumn{1}{c|}{0}   & \multicolumn{1}{c|}{4}  & \multicolumn{1}{c|}{1}  & 88 & \multicolumn{1}{c|}{0}   & \multicolumn{1}{c|}{0}  & \multicolumn{1}{c|}{0}  & 12 \\ \hline 
\end{tabular}
\end{center}
\caption{\emph{(a) Test classification accuracy, based on the KNN classifier, using the $\mathbb{L}^2$ ($k=11$), phase ($k=9$) and amplitude ($k=19$) distances. (b) Corresponding confusion matrices.}}\label{conf_matrices}
\end{table}

We report the overall classification rate on the testing data in Table \ref{conf_matrices}(a). The phase-based KNN classifier provides highest classification accuracy with the $\mathbb{L}^2$ distance-based approach in close second; the phase distance results in correct classification of 7 more cases than the $\mathbb{L}^2$ distance. Finally, the amplitude distance provides the lowest classification accuracy. We further report confusion matrices in Table \ref{conf_matrices}(b). Overall, the phase distance is more effective than the other two distances in discriminating between neighboring classes, e.g., class 2 versus class 3. The amplitude distance is only effective in discriminating between the benign and severe classes. These observations are very similar to those reported earlier based on the MDS plots.

\section{Discussion}\label{sec:Discussion}

In the analysis of the brain artery tree data, alignment of persistence landscapes adds substantially to the findings of \cite{bendich_2016,marron_2021} by uncovering that the apparent differences in the unaligned mean landscapes of the two sex groups can be partially attributed to a difference in scale and sampling variability, and confirms this finding by comparing the distributions of the total artery lengths of males and females. In the analysis of the Gleason dataset \cite{berry_2020}, we show that the amplitude of landscapes (topological information) is most effective in discriminating between benign and severe cancer, while the phase (geometry and scale) is very effective in discriminating between all four grades. In particular, phase-based classification outperforms the standard $\mathbb{L}^2$-based approach. In both settings, we demonstrate the need to consider amplitude and phase variability in persistence landscapes to address the scientific questions of interest. 

The novel approach presented in this paper can be viewed as a first step toward understanding how geometry of the manifold on which point clouds are sampled influences TDA. To see this, suppose $e:M \hookrightarrow \mathbb R^D$ is an equivariant embedding of a $d$-dimensional manifold $M$ into $\mathbb R^D,\ D \geq d$. Then, a diffeomorphism $\phi:\mathbb R^D \to \mathbb R^D$ acts on the embedding as $\phi\circ e(M)$. The map $\phi$ does not change the topology of $M$, but constructing simplicial filtrations for point clouds under the embedding in $\mathbb R^D$ using balls will transform according to $\phi$ since the metric is accordingly transformed; that is, for a fixed $x \in e(M)$, $\{y\in e(M): \|x-y\|_{\mathbb R^D} < t\}$ will transform to $\phi(x)\in \phi\circ e(M)$, $\{\phi(y) \in \phi\circ e(M): \|\phi(x)-\phi(y)\|_{\mathbb R^D} < t\}$. In the special case where $\phi$ corresponds to a (constant) scale change, the radius $t$ changes nonlinearly as $t \mapsto \gamma(t)$, for a reparameterization $\gamma$, since $t$ is forced to lie within $[0,1]$. This phenomenon also relates to when points are sampled with variability on $M$, since by judiciously changing the metric depending on the locations of points, balls of different (or differently changing) radii can be used to construct the simplicial filtration, not dissimilar to the multiscale approach considered by \cite{yoon2020persistence}. Much remains to be done in this direction. 

The limitations of this work inspire directions for future work. First, using scaled persistence diagrams by rescaling to $[0,1]^2$ is also a source of topological noise, but is entirely driven by practical considerations. In principle, we could instead consider the group of diffeomorphisms of $[0,\infty)$ to align the persistence landscapes, although there would be very little phase variation for parameter values exceeding the maximum across different point clouds. A compromise would be to consider the subgroup of diffeomorphisms of $[0,\infty)$ based on scaling and translating diffeomorphisms of $[0,1]$, considered in \cite{BS}, to perform alignment that better reflects phase variability in the landscapes. Second, notwithstanding the promising results for the noisy simulations presented in Appendix A in the supplement, robustness of the alignment-based approach to measurement error will strongly depend on the geometry of the manifold $M$, sampling density and magnitude of noise in observed point clouds on $\mathbb R^d$, especially if data have been sampled from a distribution with support only on $M$. One possible approach would constitute of first estimating $M$ (and its dimension) using a manifold fitting method, and using this information to construct tailored simplicial filtrations; however, additional noise induced by the fitting procedure would have to be accounted for in downstream tasks. Another option for the large noise setting is to use explicit statistical models to align persistence landscapes that account for all sources of uncertainty. For example, Bayesian models based on shape constraints to infer the pattern and number of extrema in landscapes may be profitably used \citep{matuk_2021}.

While the focus of this paper is on persistence landscapes, we expect our approach to be fruitful with silhouettes \citep{chazal_2014}, since similar triangular functions are used in their definition. However, feasibility of the alignment method for other functional summaries mentioned in Section \ref{sec:intro} is not clear, and is worthy of further investigation.  Finally, for denoising a persistence diagram directly without using landscapes, it is possible to consider generalizations of the one-dimensional transforms $\gamma$ of points on diagrams to the group of diffeomorphisms of $\mathbb R^2$, along the lines of what is done in the Large Deformation Diffeomorphic Metric Mapping (LDDMM) framework \citep{grenander}.

\noindent\textbf{Acknowledgements:} The MR brain images from healthy volunteers used in this paper were collected and made available by the CASILab at The University of North Carolina at Chapel Hill and were distributed by the MIDAS Data Server at Kitware, Inc. The authors would like to thank Jessi Cisewski Kehe for sharing the Gleason dataset. This research was partially funded by NIH R37-CA214955 (to SK and KB), NSF DMS-2015374 (to KB), NSF CCF-1740761, NSF DMS-2015226 and NSF CCF-1839252 (to SK), and NIH R01-ES028804 (to JM). We also thank the two anonymous reviewers for providing constructive comments that have improved this manuscript.

\bibliographystyle{agsm}
\bibliography{bibliography}

\newpage

\appendix

\setcounter{figure}{0} 

\section{Simulated Examples with Additive Noise}

\begin{figure}[!t]
\begin{center}
\begin{tabular}{cccc}
      \includegraphics[width = 0.65 in]{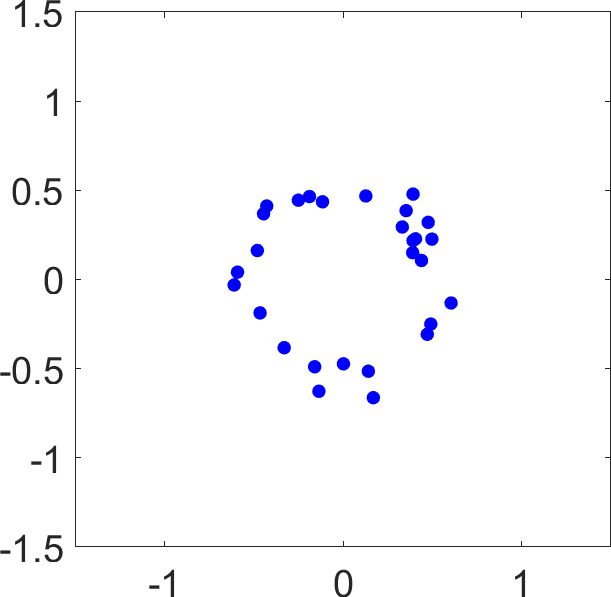} & \includegraphics[width = 0.8 in]{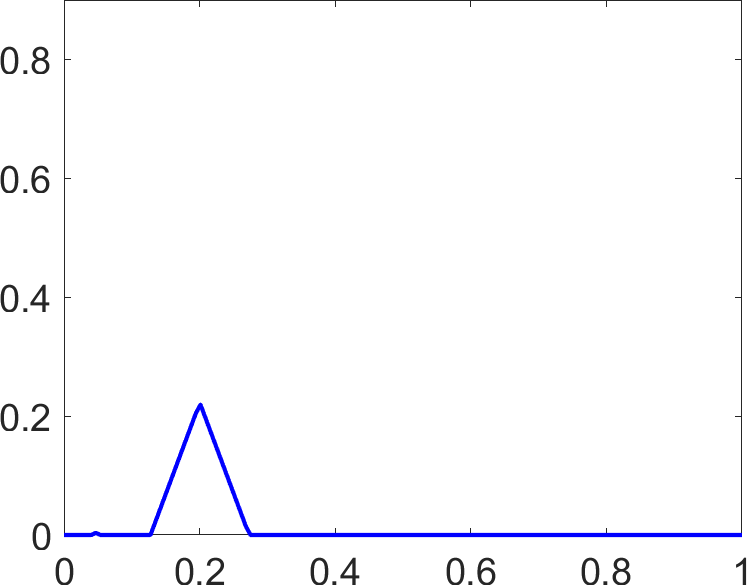}  & \includegraphics[width = 0.65 in]{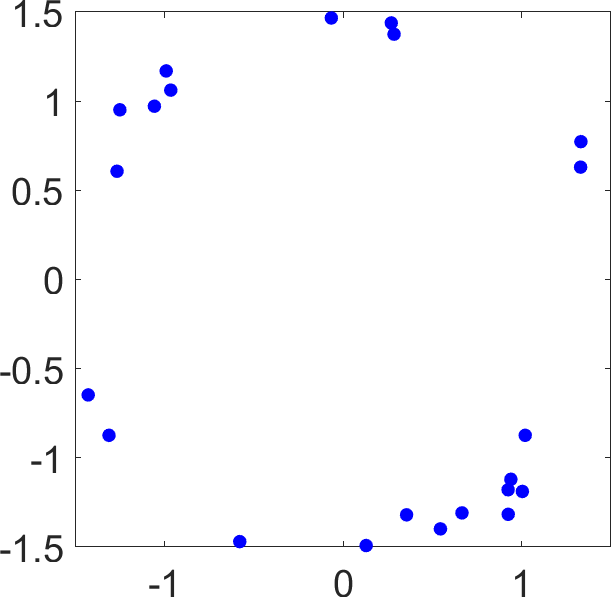} & \includegraphics[width = 0.8 in]{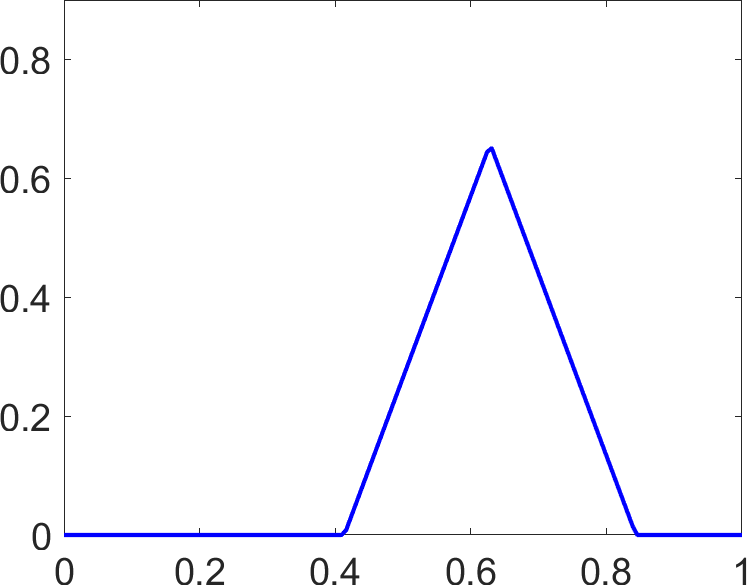}\\
      (a) & (b) & (c)& (d)  \\
      \end{tabular}
      \begin{tabular}{ccc}
    \includegraphics[width = 1 in]{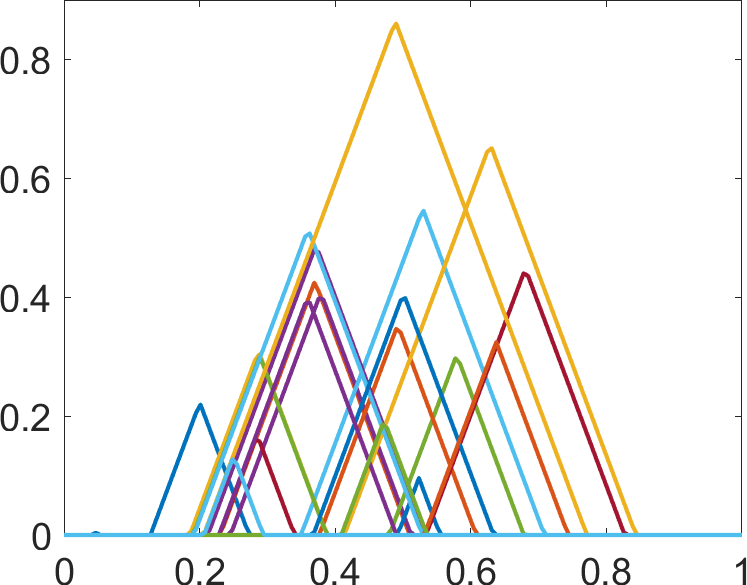} & \includegraphics[width = 1 in]{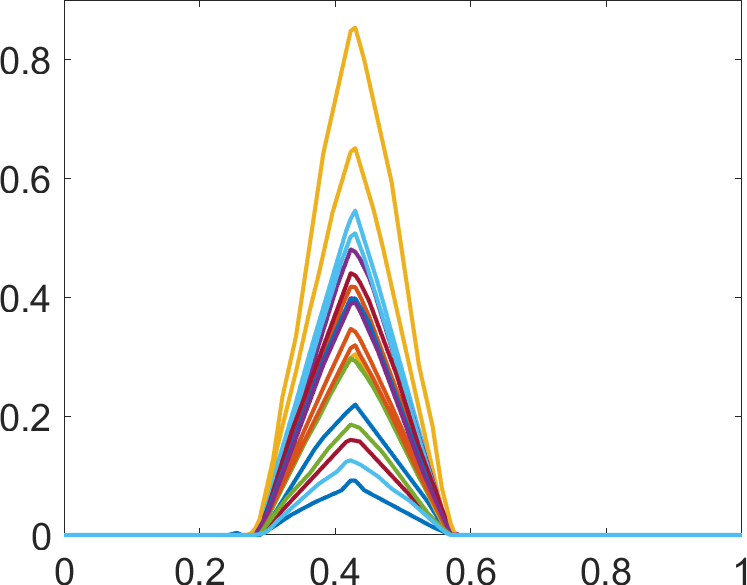}  &  \includegraphics[width = 1 in]{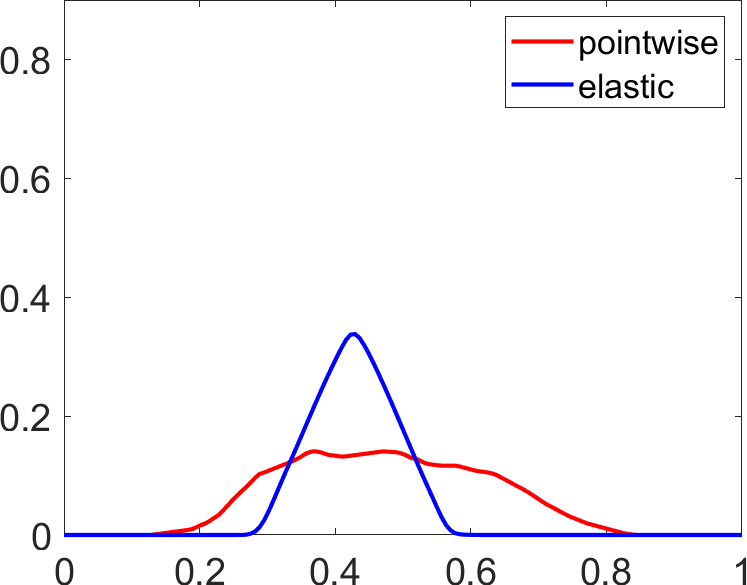} \\ 
    (e) & (f)& (g)  \\
        \includegraphics[width = 1 in]{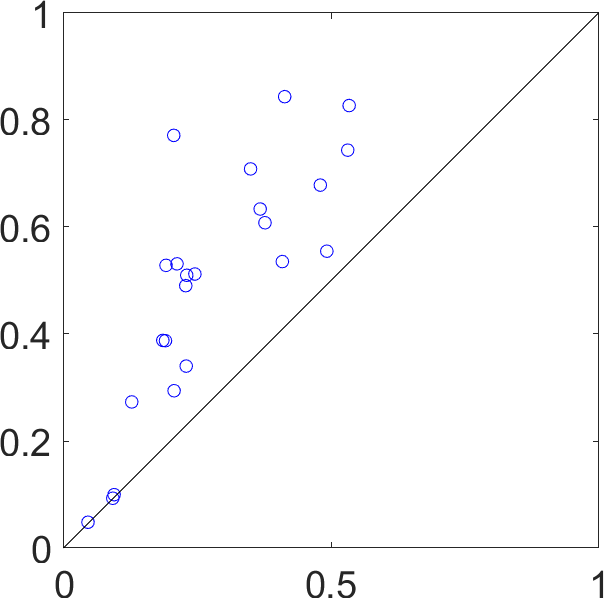} & \includegraphics[width = 1 in]{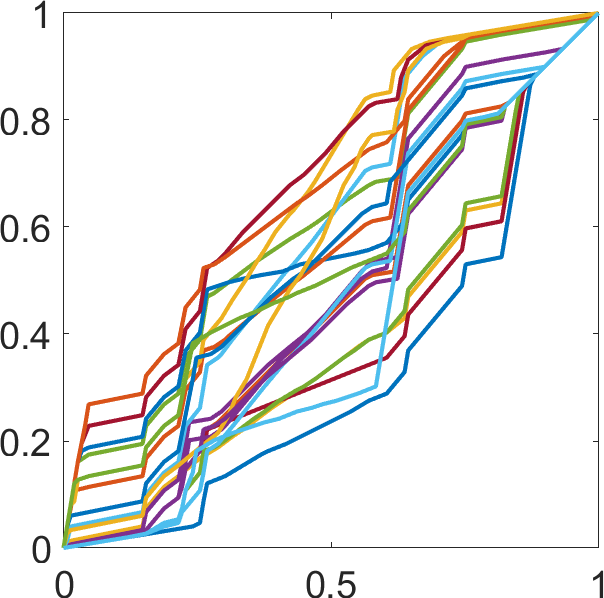}  &  \includegraphics[width = 1 in]{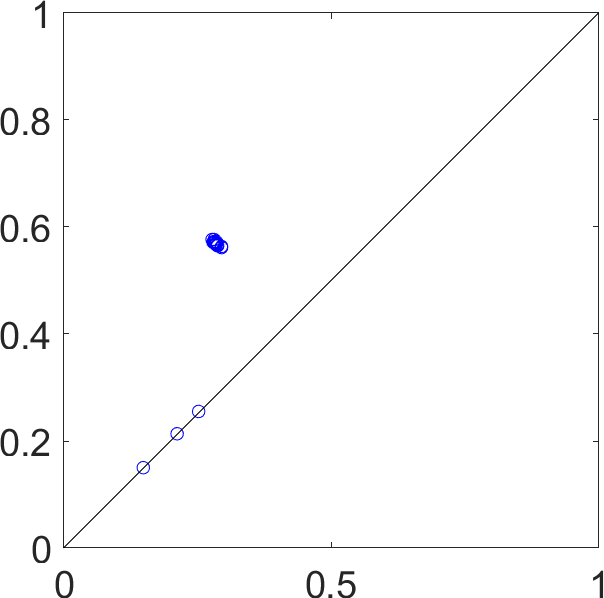} \\
        (h) & (i) & (j)  \\
    \end{tabular}
    \caption{\emph{Same topology with scale and sampling variabilities as well as additive noise:} (a)\&(c): Two examples, from 20, of randomly generated point clouds on circles with randomly chosen radii, random sample sizes and additive noise. (b)\&(d): Corresponding persistence landscapes. (e) Persistence landscapes $\{\Lambda_i\}_{i=1}^{20}$ of 20 point clouds. (f) Aligned persistence landscapes $\{\Lambda_i(\gamma_i)\}_{i=1}^{20}$. (g) Mean landscape after (blue) and without (red) alignment. (h) Noisy persistence diagrams $\{(b_{ij},d_{ij})\}_{i=1}^{20}$ from 20 point clouds. (i) Estimated reparameterizations $\{\gamma_i\}_{i=1}^{20}$. (j) Denoised/transformed persistence diagrams $\{(\gamma_i^{-1}(b_{ij}),\gamma_i^{-1}(d_{ij}))\}_{i=1}^{20}$.}
    \label{meanEstimationEgNoise}
    \end{center}
\end{figure}

\begin{figure}[!t]
\begin{center}
\begin{tabular}{cccc}
      \includegraphics[width = 0.6 in]{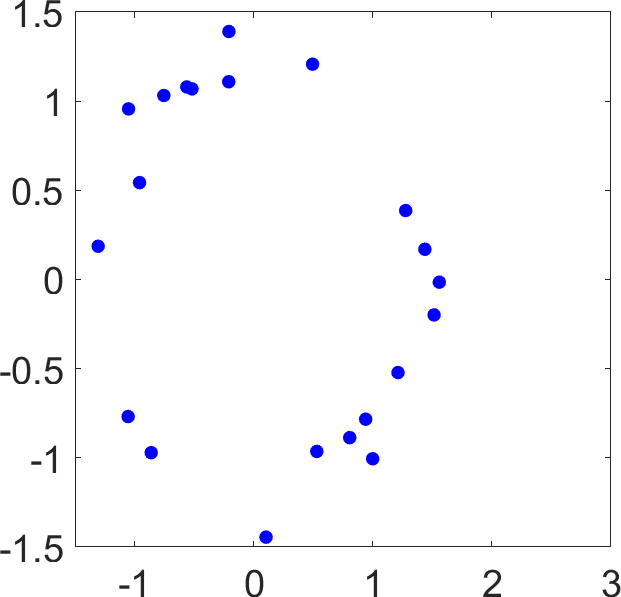} & \includegraphics[width = .7 in]{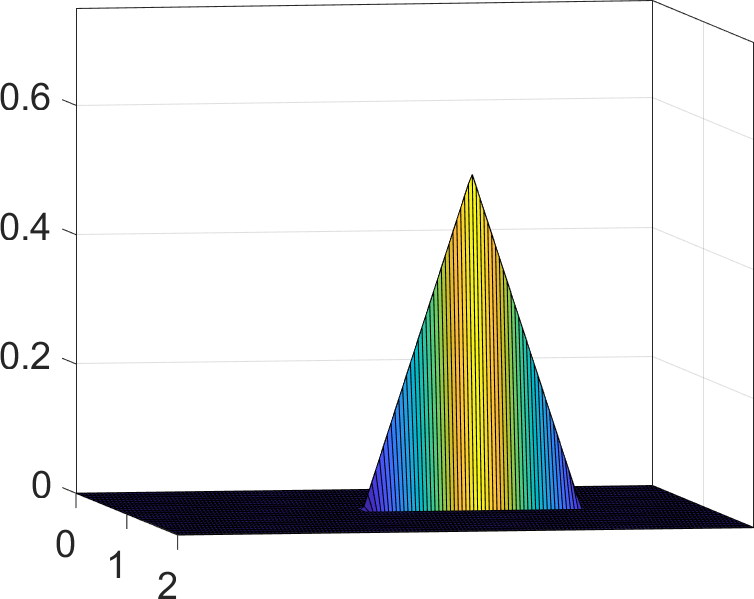}  & \includegraphics[width = 0.6 in]{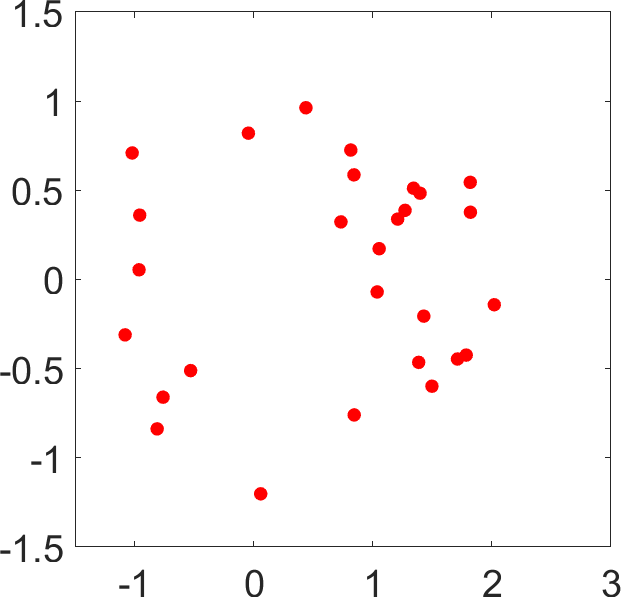} & \includegraphics[width = .7 in]{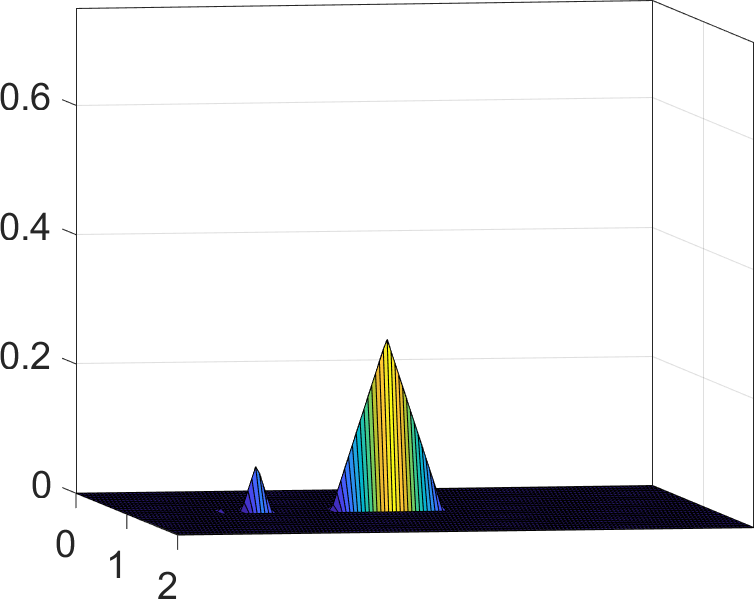}\\
     (a) & (b) & (c) & (d)\\
      
      \includegraphics[width = .7 in]{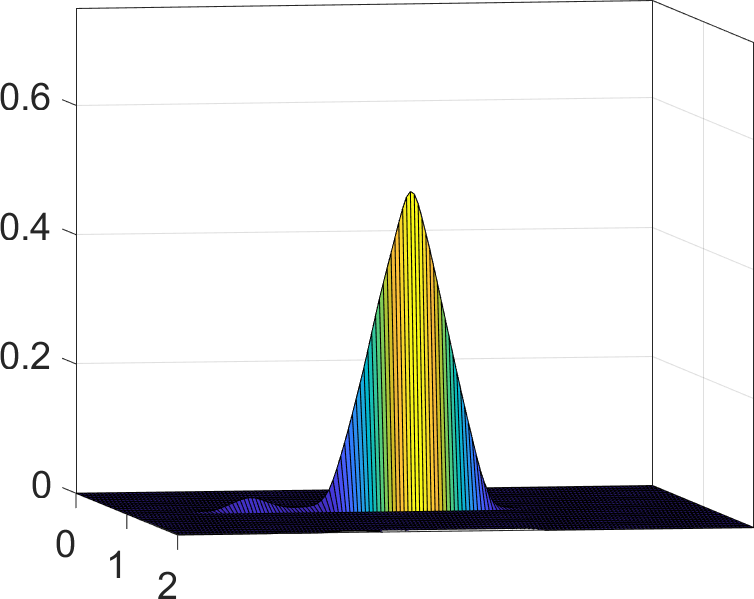} & \includegraphics[width = .7 in]{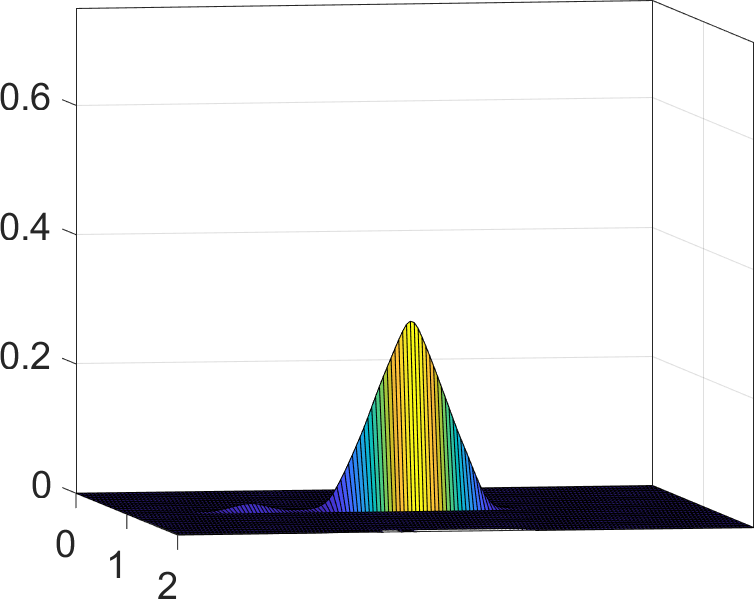}  & \includegraphics[width = .7 in]{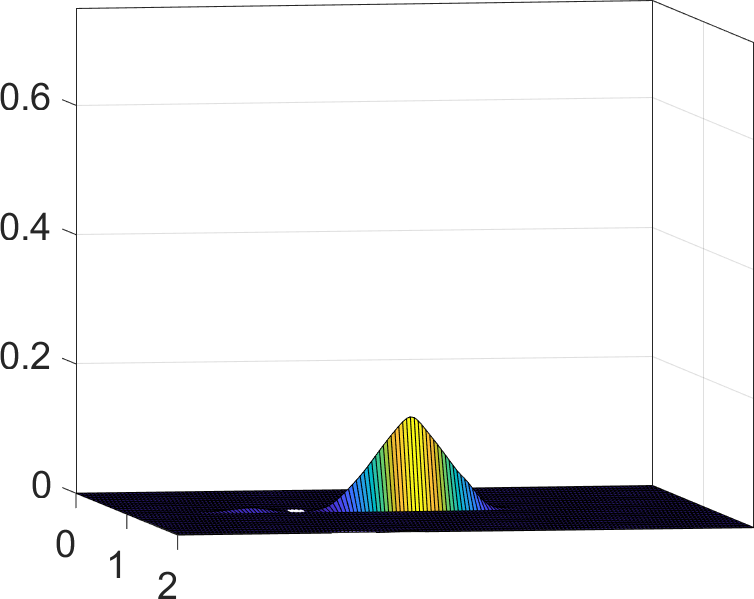} & \includegraphics[width = .7 in]{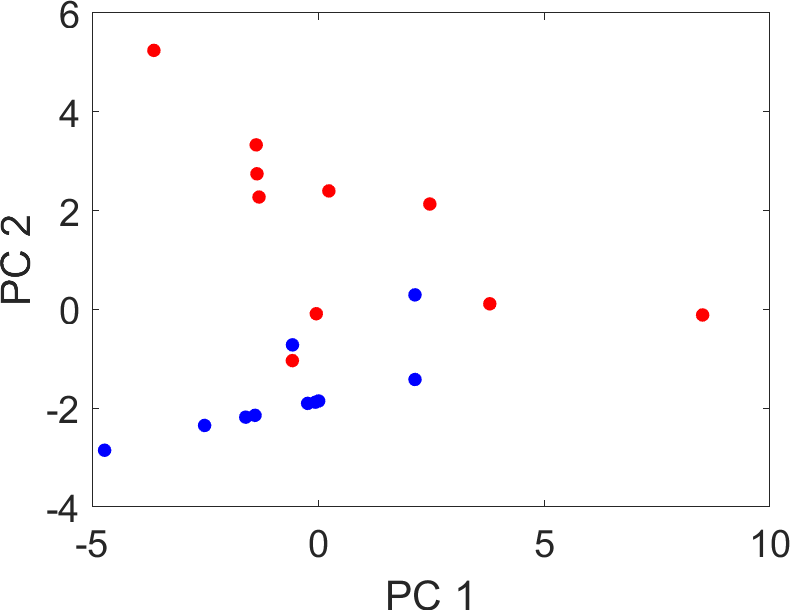}\\
    
    \includegraphics[width = .7 in]{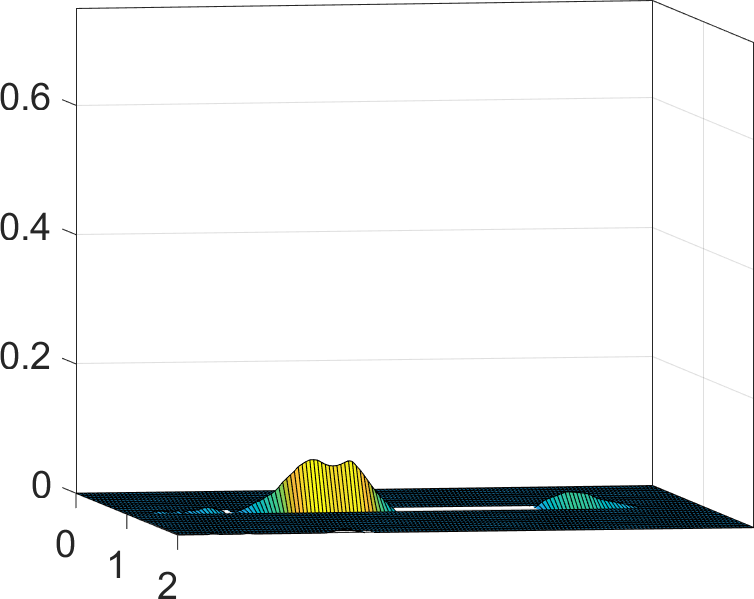} & \includegraphics[width = .7 in]{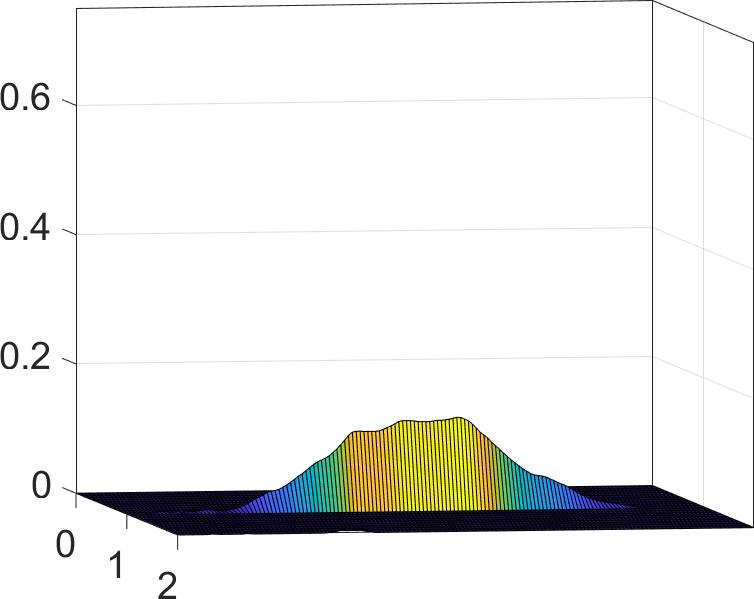}  & \includegraphics[width = .7 in]{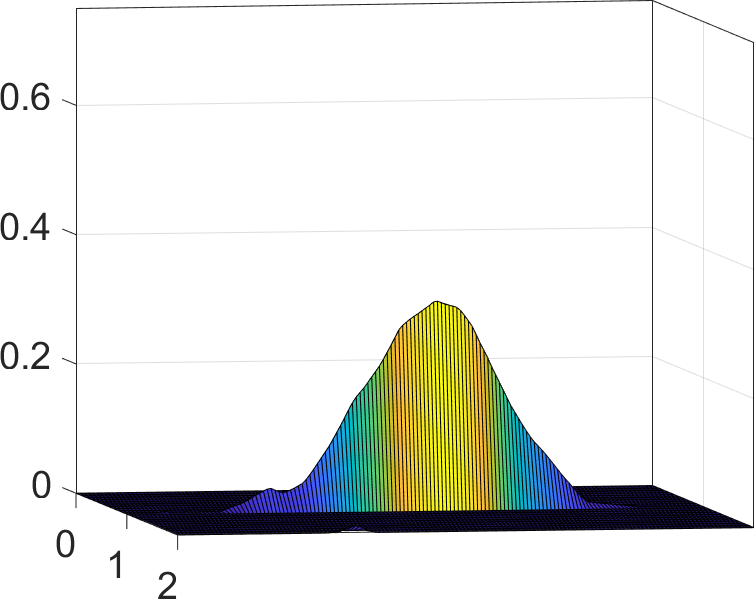} & \includegraphics[width = .7 in]{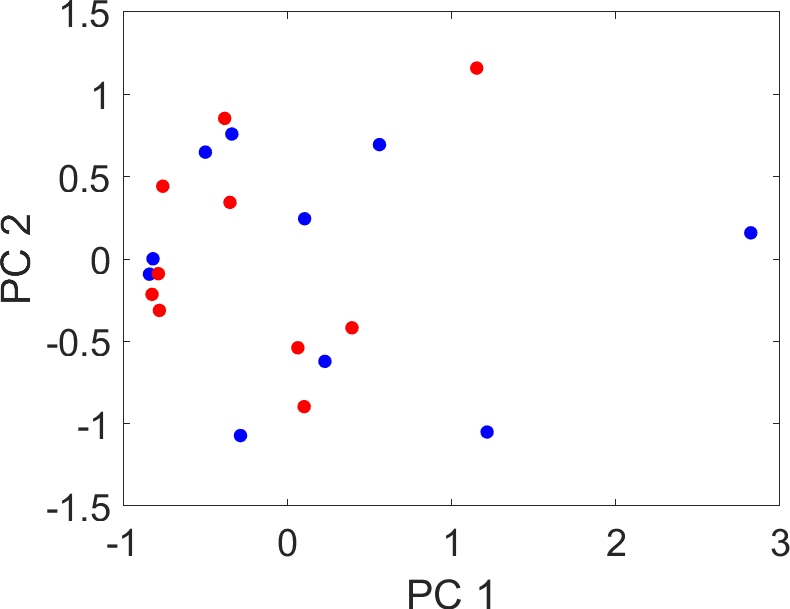}\\
     (e)  & (f) & (g) & (h)\\
      \end{tabular}
      \begin{tabular}{ccc}
        \includegraphics[width = 1 in]{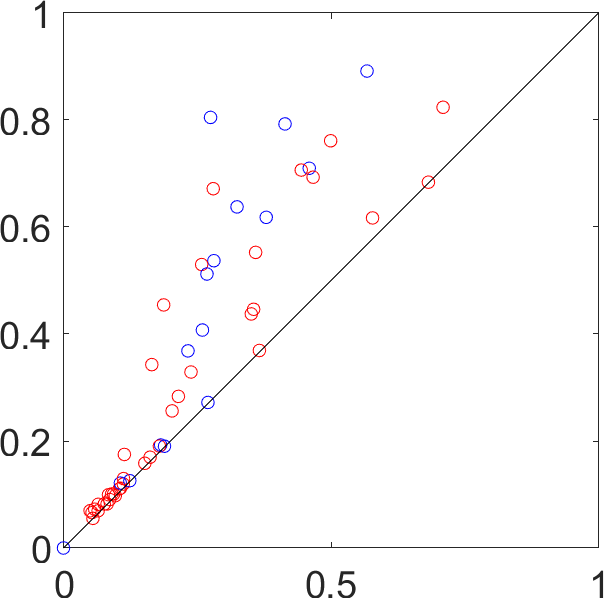} & \includegraphics[width = 1 in]{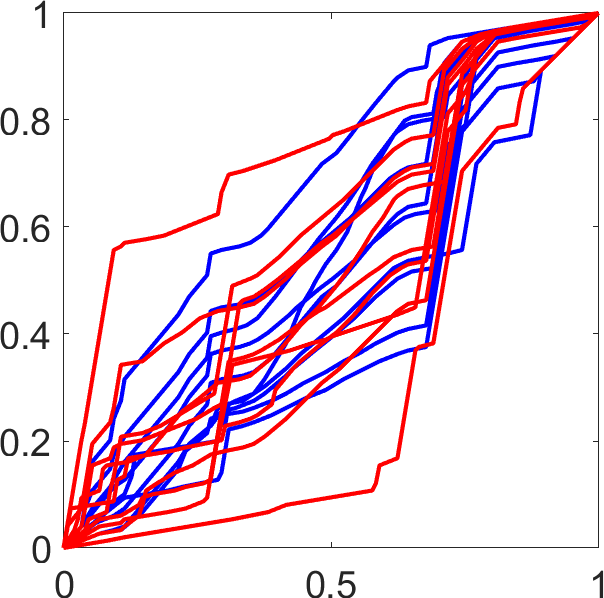}  &  \includegraphics[width = 1 in]{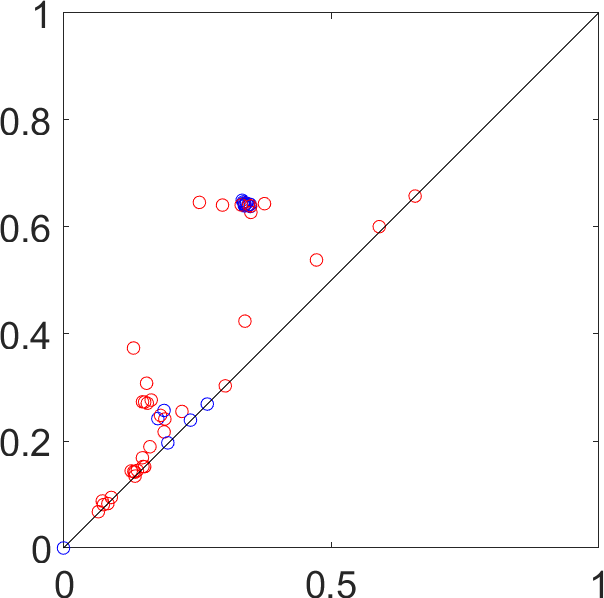} \\
     (i) & (j) & (k)
    \end{tabular}
    \caption{\emph{Different topology with scale and sampling variabilities as well as additive noise}: (a)\&(c) Two examples, from 20, of randomly generated point clouds from topologically different spaces (blue and red, respectively, in all relevant panels). (b)\&(d) Corresponding degree $p=1$, $K=2$-dimensional persistence landscapes. (e) -1, (f) 0, (g) +1 standard deviation from the mean landscape in the first PC direction, and (h) projection of landscapes onto the first two PC directions: following alignment (top) and without alignment (bottom). (i) Noisy and (k) denoised persistence diagrams. (j) Estimated reparameterizations.}
    \label{pcaEgNoise}
    \end{center}
\end{figure}

We repeat Simulated Example 1 from the main article with additive pointwise noise, i.e., for each point cloud, we independently generate additive noise from a zero-mean bivariate Gaussian distribution with covariance $r(0.1)^2I_2$, where $r$ is the radius of the circle that underlies the `noiseless' point cloud. Two examples of point clouds and their corresponding degree $p=1$, $K=1$-dimensional landscape functions are shown in panels (a) and (c), and (b) and (d) in Figure \ref{meanEstimationEgNoise}, respectively. The landscape functions for all 20 point clouds are shown in panel (e). Most of the landscapes generated from the noisy point clouds are similar in shape to those from the `noiseless' setting. However, sometimes there is an extremely small peak, corresponding to noise, in some of the landscape functions. This is most visually apparent in panel (f), where there is an extremely small peak at $t \approx 0.25$ prior to the major peak in many of the landscape functions. Due to the small magnitude of this noise-induced feature, additive noise appears to have little effect on landscape alignment and mean computation. The mean based on aligned landscapes, visualized in panel (g), is consistent with that of a circle. There is considerable variance reduction in the denoised/transformed persistence diagrams in panel (j), via reparameterizations shown in (i). Importantly, points corresponding to the main feature of the point clouds are collapsed to a single point, while points near the diagonal, corresponding to features created by the additive noise, remain near the diagonal. Such clarity is absent in the noisy persistence diagrams in (h). 
Next, we repeat Simulated Example 2 from the main article with additive noise. For point clouds in the blue group (single circle), we independently generate additive noise from a zero-mean bivariate Gaussian distribution with covariance $r(0.1)^2I_2$, where $r$ is the radius of the circle that underlies the `noiseless' point cloud. We repeat this procedure for point clouds in the red group, but noise is generated such that the covariance depends on the radius of one of the two circles that it belongs to. A single example of a point cloud in the blue and red groups is shown in Figure \ref{pcaEgNoise}(a)\&(c); the corresponding degree $p=1$, $K=2$-dimensional landscapes are shown in panels (b) and (d). The shapes of the landscapes in each group are similar to those displayed in the `noiseless' setting. However, here, we notice some differences in PCA carried out on aligned landscapes. The first direction of variability, viewed in the top row of panels (e)-(g) appears to be associated with scale variability in the point cloud data; the results here are displayed in the same manner as in Figure 6 in the main article. On the other hand, as confirmed in the top row of panel (h), the second direction of variability appears to be associated with the homology of the point clouds, i.e., most of the red point clouds, generated from two connected circles, have a positive second PC score, while most of the blue point clouds, generated from a single circle, have a negative second PC score. Since noise can distort topological features, there does appear to be some overlap between the two classes based on the first two PC scores. This observation is in contrast to the `noiseless' setting, where the two classes are clearly separated based on the first PC score alone. The corresponding PC scores computed based on unaligned landscapes, shown in the bottom of panel (h), provide no such distinction between the two classes. A comparison of the denoised persistence diagrams presented in Figure \ref{pcaEgNoise}(k) to their noisy counterparts in panel (i) shows the benefits of our approach. While the clustering of features in panel (k) is not as clear as in the `noiseless' setting, one can still extract useful homological information from the denoised persistence diagrams. On the other hand, this is not possible based on the noisy diagrams in panel (i).

\begin{figure}[!t]
\begin{center}
\begin{tabular}{ccc}
        \includegraphics[width = 1 in]{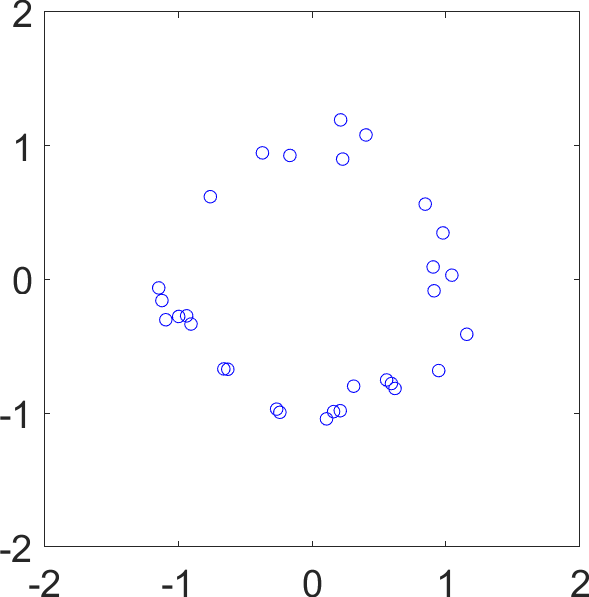} & \includegraphics[width = 1 in]{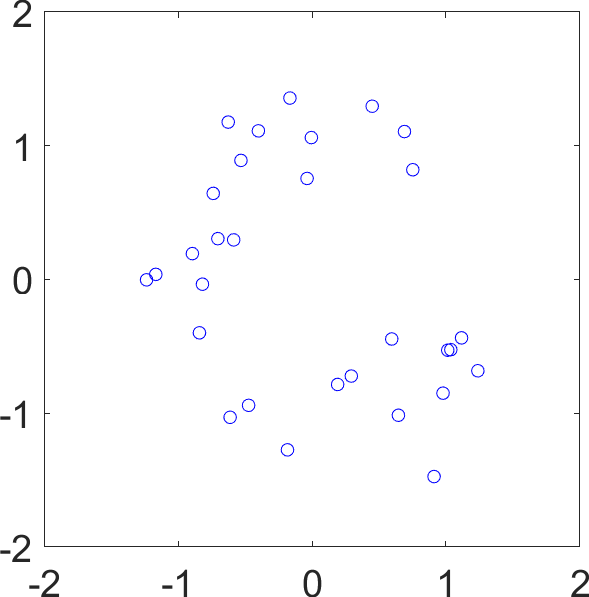}  &  \includegraphics[width = 1 in]{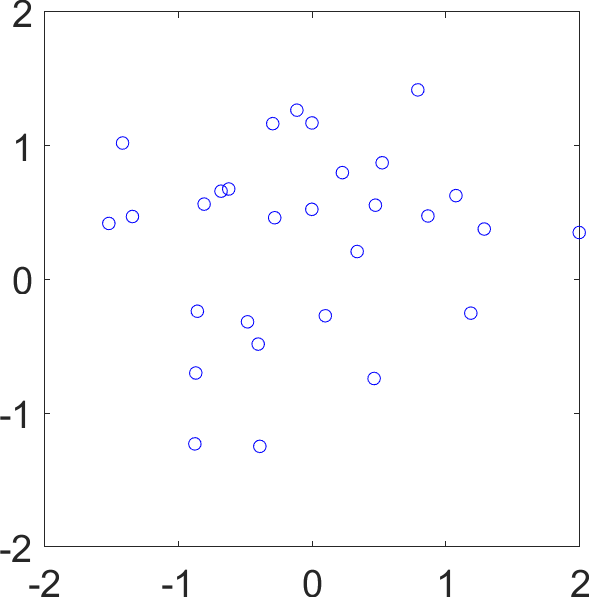}\\
        (a) & (b) & (c) \\
        \includegraphics[width = 1 in]{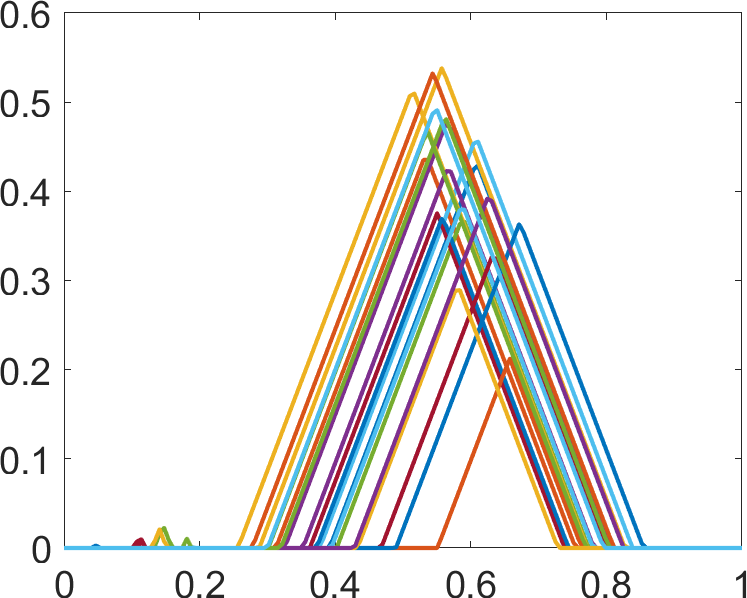} & \includegraphics[width = 1 in]{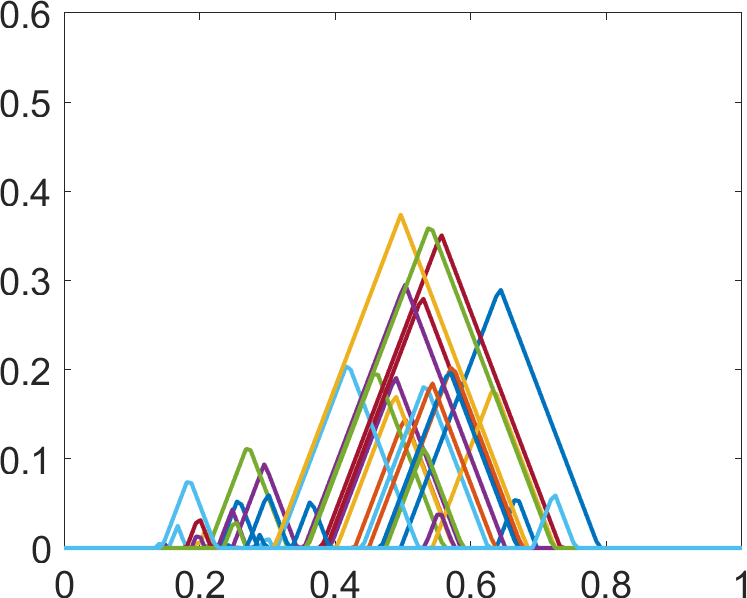}  &  \includegraphics[width = 1 in]{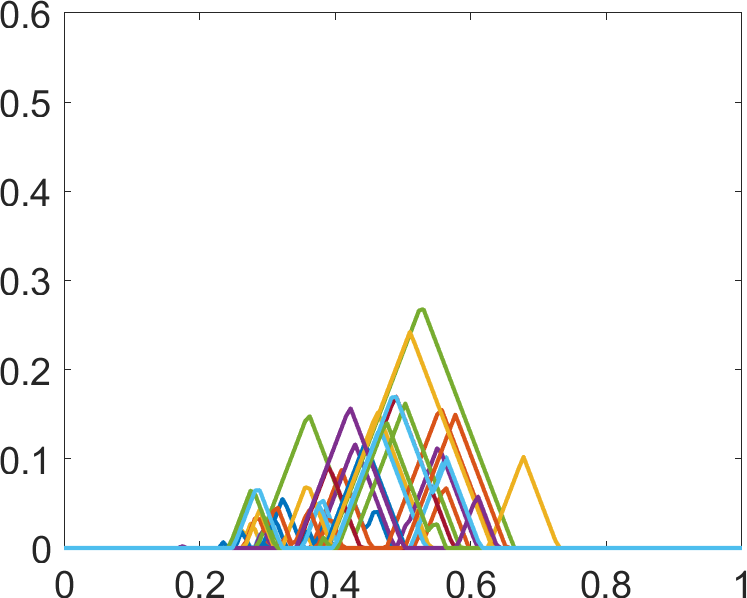}\\
        (d) & (e) & (f) \\
        \includegraphics[width = 1 in]{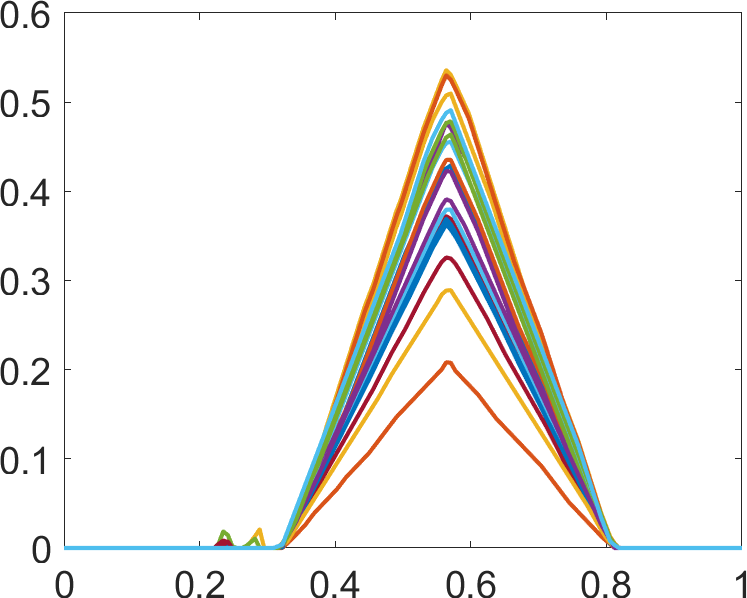} & \includegraphics[width = 1 in]{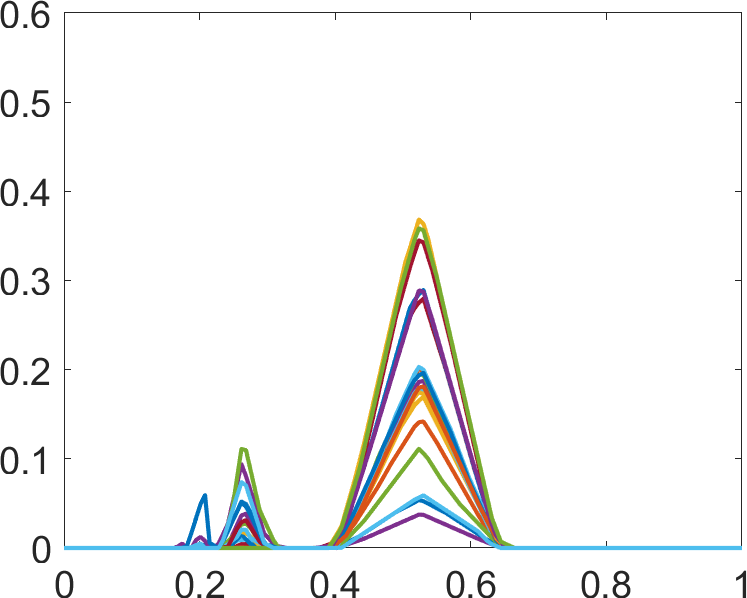}  &  \includegraphics[width = 1 in]{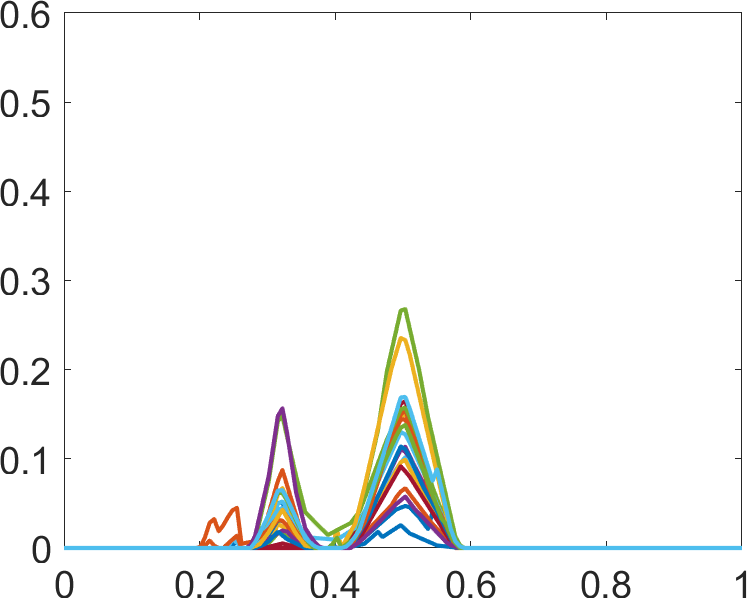}\\
        (g) & (h) & (i) \\
        \includegraphics[width = 1 in]{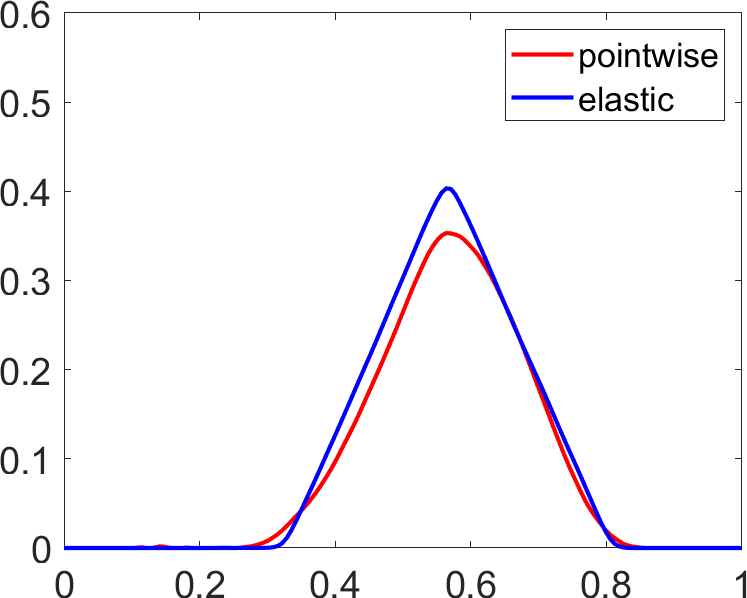} & \includegraphics[width = 1 in]{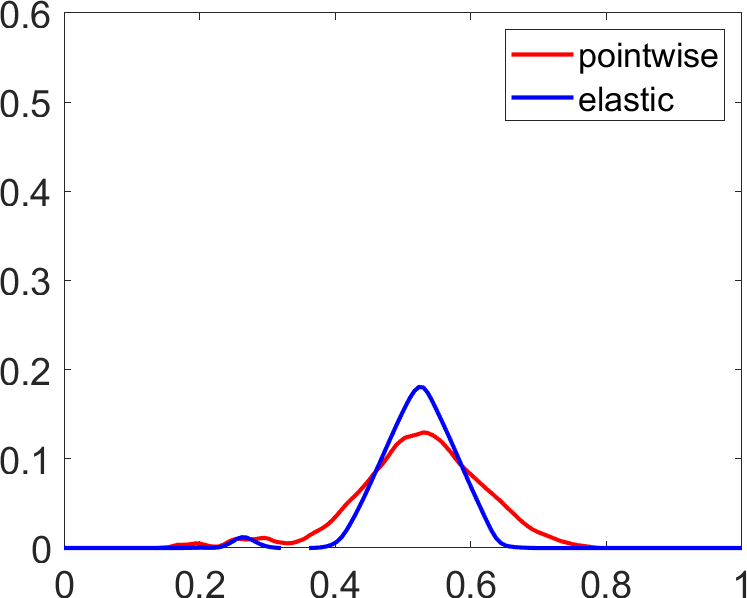}  &  \includegraphics[width = 1 in]{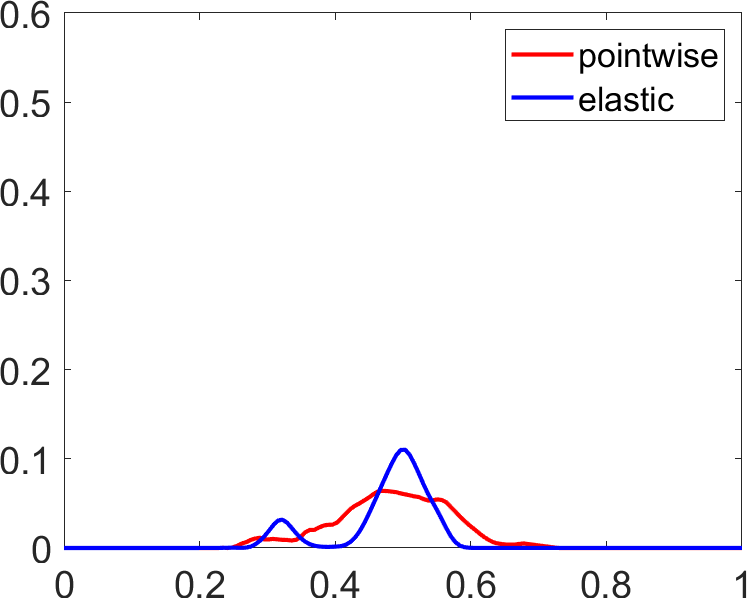}\\
        (j) & (k) & (l) \\
        \includegraphics[width = 1 in]{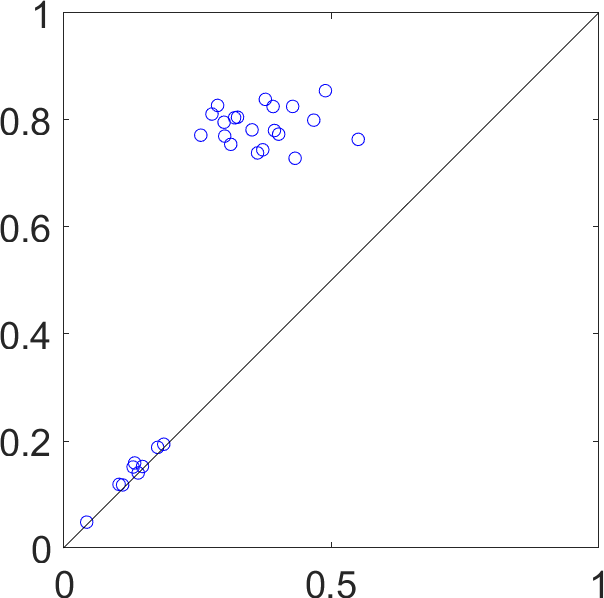} & \includegraphics[width = 1 in]{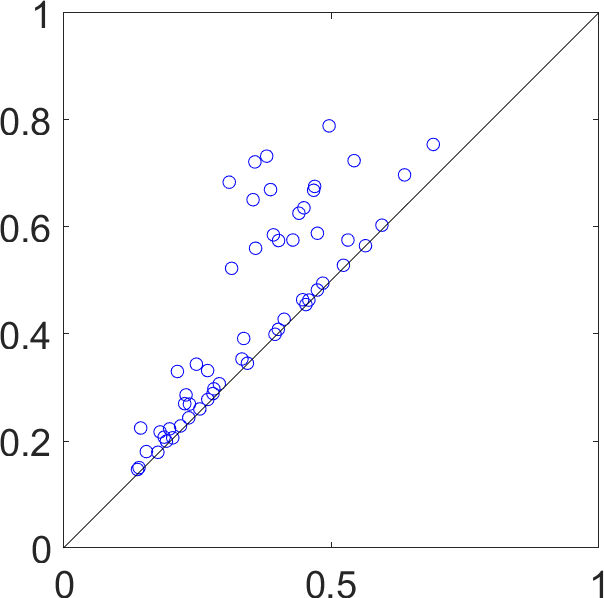}  &  \includegraphics[width = 1 in]{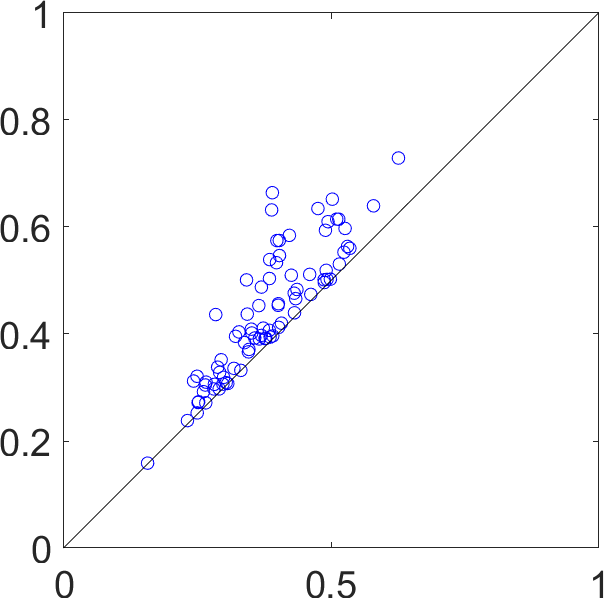}\\
        (m) & (n) & (o) \\
           \includegraphics[width = 1 in]{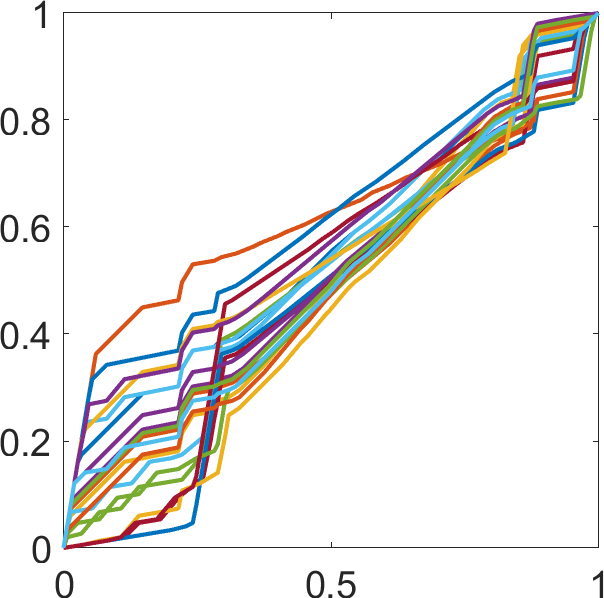} & \includegraphics[width = 1 in]{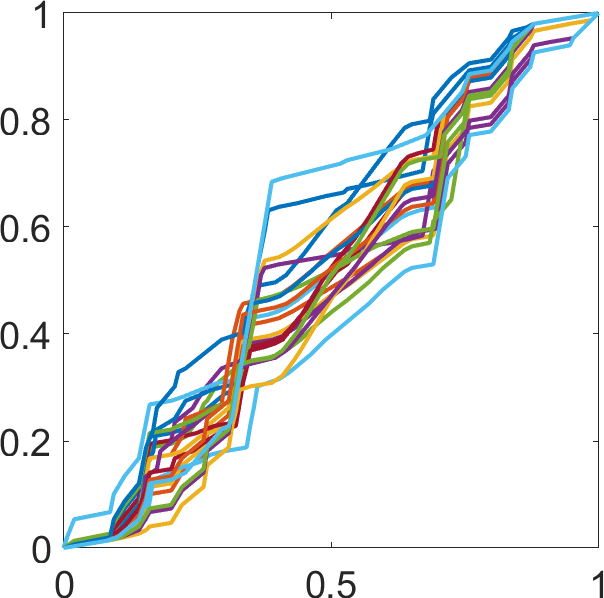}  &  \includegraphics[width = 1 in]{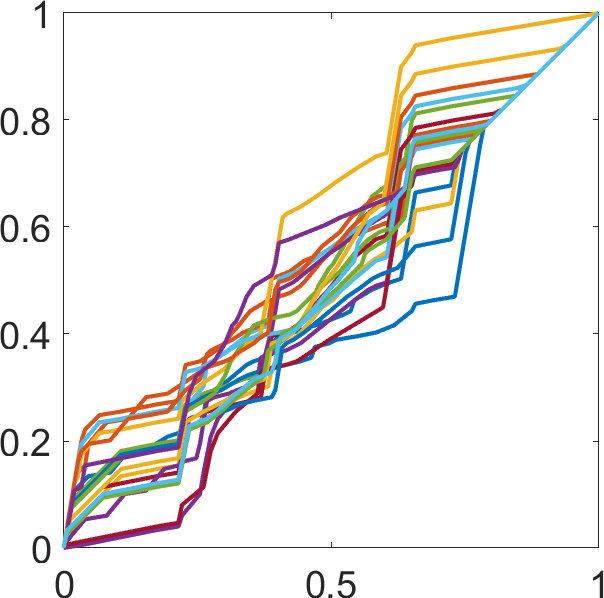}\\
     (p) & (q) & (r) \\
           \includegraphics[width = 1 in]{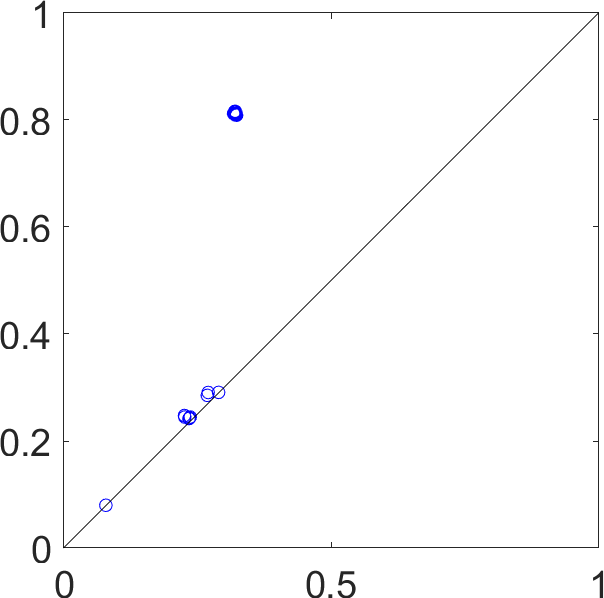} & \includegraphics[width = 1 in]{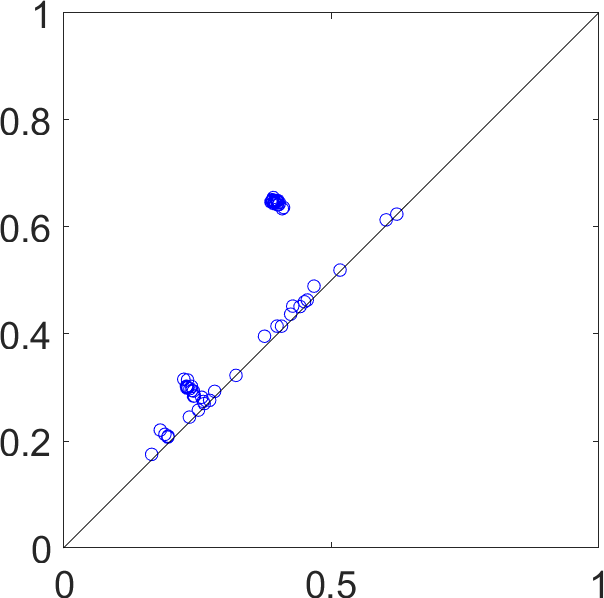}  &  \includegraphics[width = 1 in]{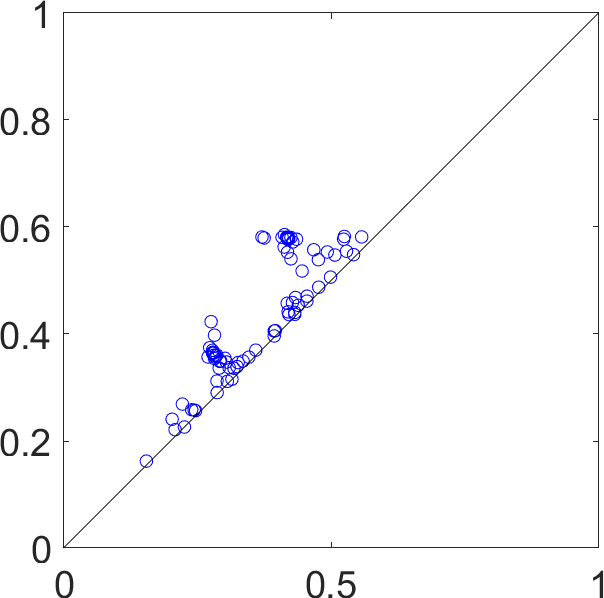}\\
     (s) & (t) & (u) \\
     \end{tabular}
    \caption{\emph{Same topology and scale with increasing pointwise noise}: Point clouds in the left, middle and right columns were generated with noise variance of $(0.1)^2$, $(0.25)^2$ and $(0.5)^2$, respectively. (a)-(c) Example of one of the 20 point clouds with additive pointwise noise. (d)-(f) Degree $p=1$, $K=1$-dimensional persistence landscapes $\{\Lambda_i\}_{i=1}^{20}$. (g)-(i) Aligned persistence landscapes $\{\Lambda_i(\gamma_i)\}_{i=1}^{20}$. (j)-(l) Mean landscape after (blue) and without (red) alignment. (m)-(o) Noisy persistence diagrams. (p)-(r) Estimated reparameterizations $\{\gamma_i\}_{i=1}^{20}$. (s)-(u) Denoised persistence diagrams $\{(\gamma_i^{-1}(b_{ij}),\gamma_i^{-1}(d_{ij}))\}_{i=1}^{20}$.}
    \label{increasingNoise}
    \end{center}
\end{figure}

In Figure \ref{increasingNoise}, we illustrate the effects of increasing pointwise noise on alignment of persistence landscapes and denoising of the corresponding persistence diagrams. The point cloud data considered in the left (low noise), middle (medium noise) and right (high noise) columns in this figure were generated by first uniformly sampling 30 points on a circle with radius $r=1$ and then adding pointwise noise sampled from a zero-mean Gaussian distribution with covariance $(0.1)^2I_2$, $(0.25)^2I_2$ and $(0.5)^2I_2$, respectively. For each of the three examples, we generate 20 point clouds in this manner, and consider their alignment and averaging using degree $p=1$, $K=1$-dimensional persistence landscapes. Panels (a)-(c) display one example point cloud for each of the noise settings. The corresponding landscapes for all 20 observations are shown in panels (d)-(f) with corresponding noisy persistence diagrams in panels (m)-(o). While the underlying circle is visibly discernible in panels (a) and (b), the noise overwhelms the structure of the data in panel (c). This is further reflected in the corresponding persistence landscapes, which have one prominent peak in panels (g) and (h), but a relatively smaller prominent peak (and another peak with similar magnitude) in panel (i). Reparameterizations used to align the persistence landscapes in panels (d)-(f), resulting in the aligned landscapes in panels (g)-(i), are shown in (p)-(r). It is evident that alignment is effective in the low and medium noise settings resulting in a mean that is consistent with that of a circle, as seen in panels (j) and (k). Also, as expected, there is considerable variance reduction in the denoised persistence diagrams in panels (s) and (t) as compared to their noisy counterparts shown in panels (m) and (n). As before, points corresponding to the main feature of the point clouds are collapsed to a single point, while points near the diagonal, corresponding to features created by the additive noise, remain near the diagonal. This structure is less clear in the high noise setting. First, the mean shown in (l) contains two clear peaks: a small peak near $t=0.3$ and a much larger peak near $t=0.5$. Second, while some variance reduction is observed when comparing the noisy diagrams in (o) to their denoised versions in (u), alignment in this case does not appear to be as effective in distinguishing the underlying structure in the point clouds from noise.

\clearpage

\section{Additional Mean Estimation Examples}\label{sec:simAdditionalMeans}

\begin{figure}[!t]
\begin{center}
\begin{tabular}{cccc}
      \includegraphics[width = .6 in]{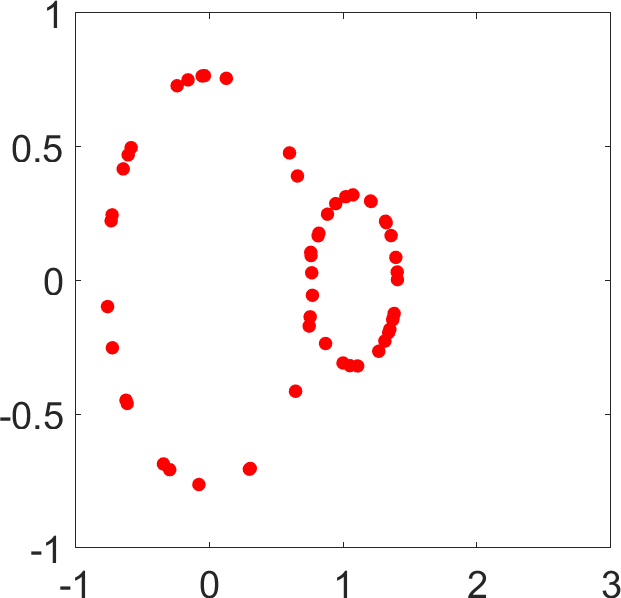} & \includegraphics[width = .7 in]{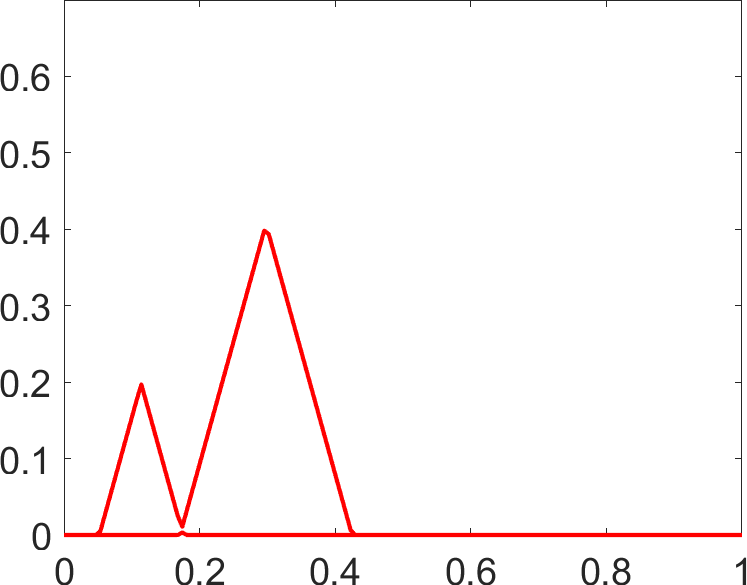}  & \includegraphics[width = .6 in]{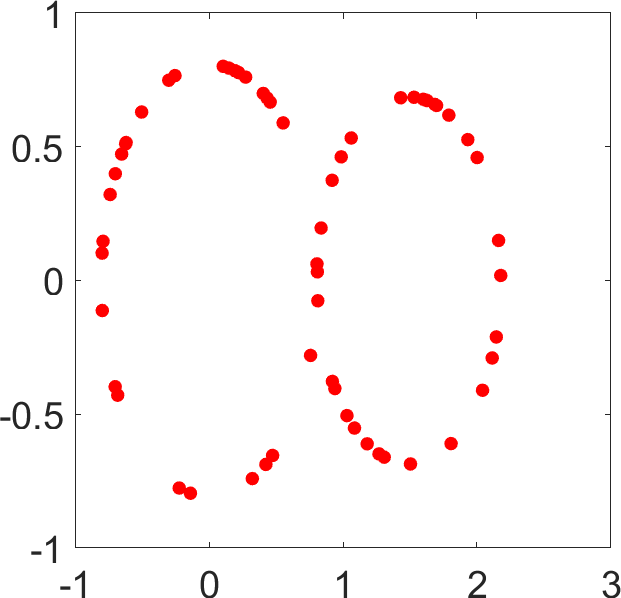} & \includegraphics[width = .7 in]{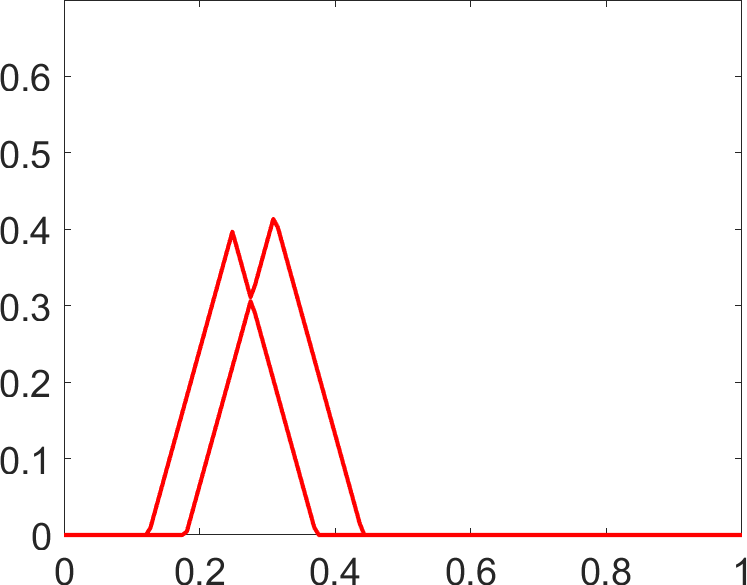}\\
     (a) & (b) & (c) & (d) \\
      \end{tabular}
      
      \begin{tabular}{ccc}
    \includegraphics[width = 1 in]{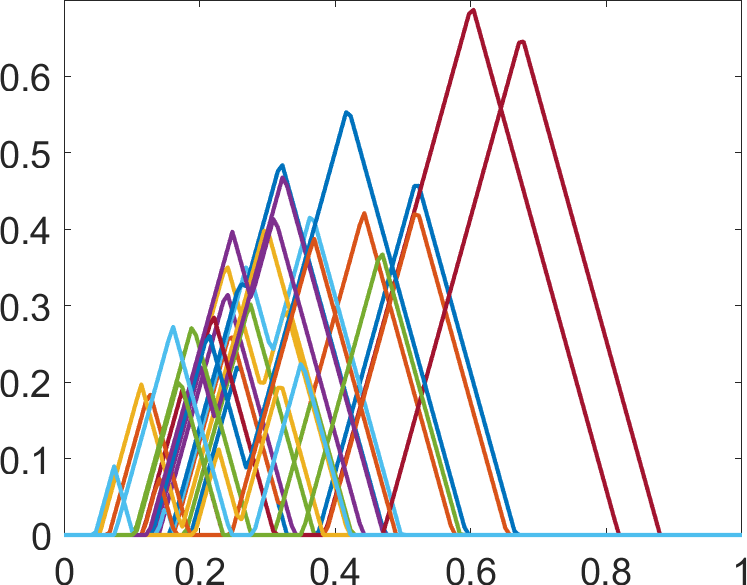} & \includegraphics[width = 1 in]{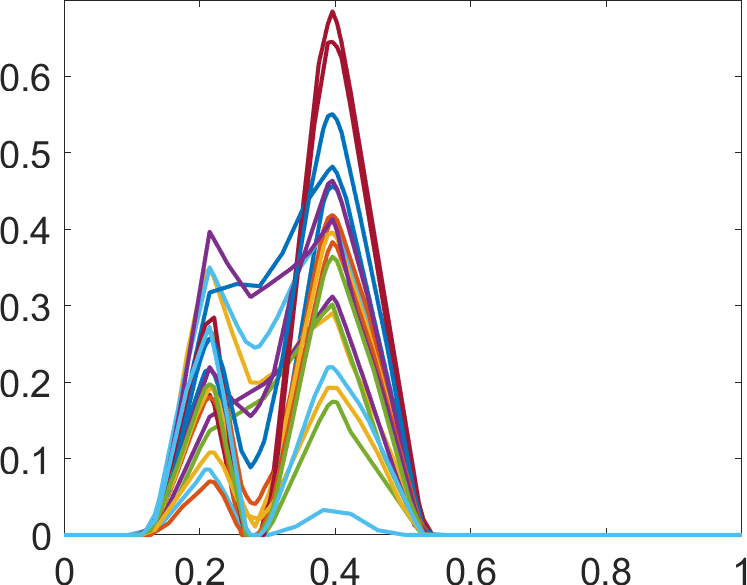}  &  \includegraphics[width = 1 in]{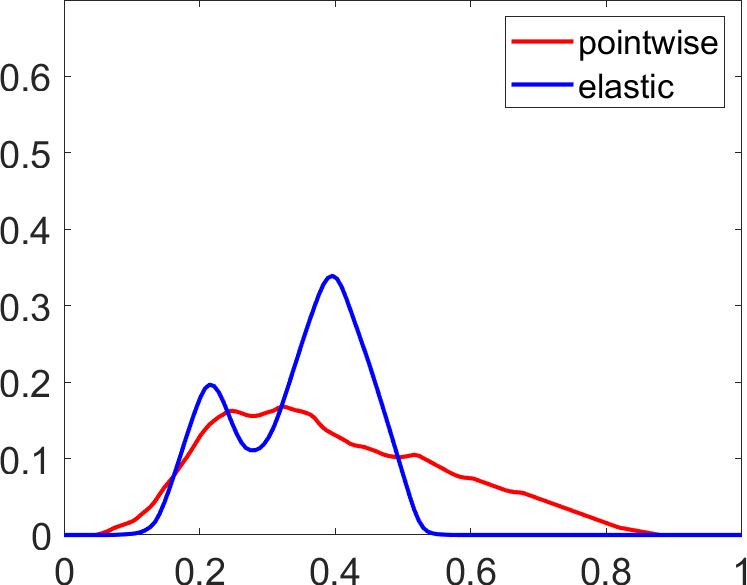} \\
        \includegraphics[width = 1 in]{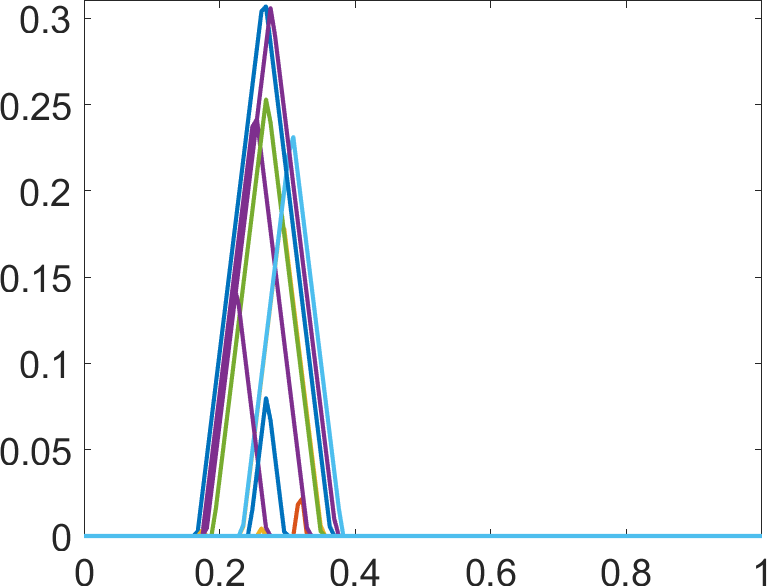} & \includegraphics[width = 1 in]{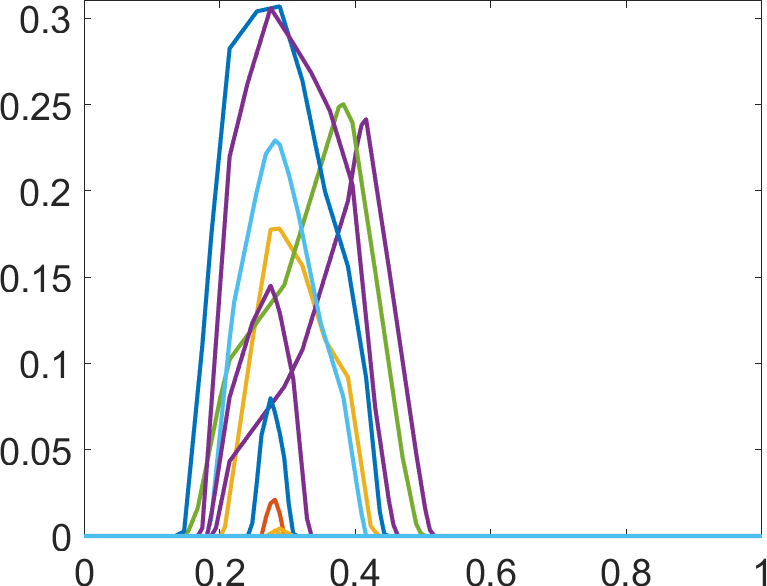}  &  \includegraphics[width = 1 in]{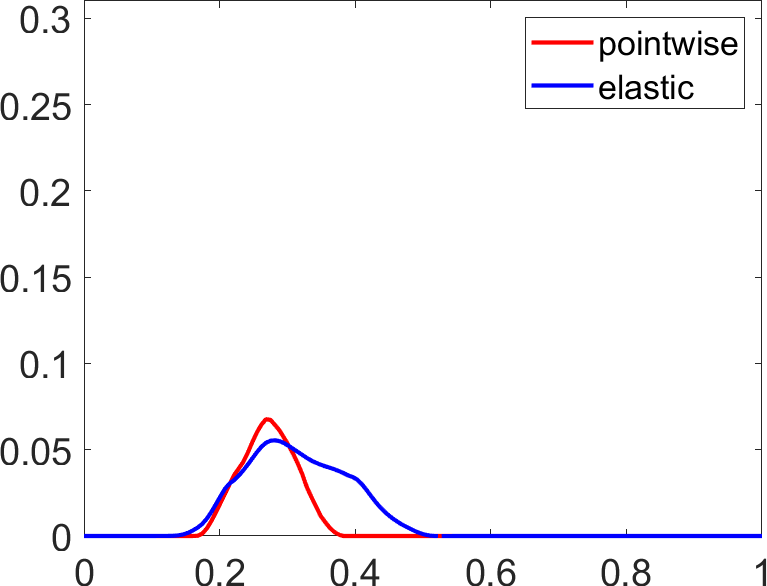} \\
        (e) & (f) & (g) \\
        \includegraphics[width = 1 in]{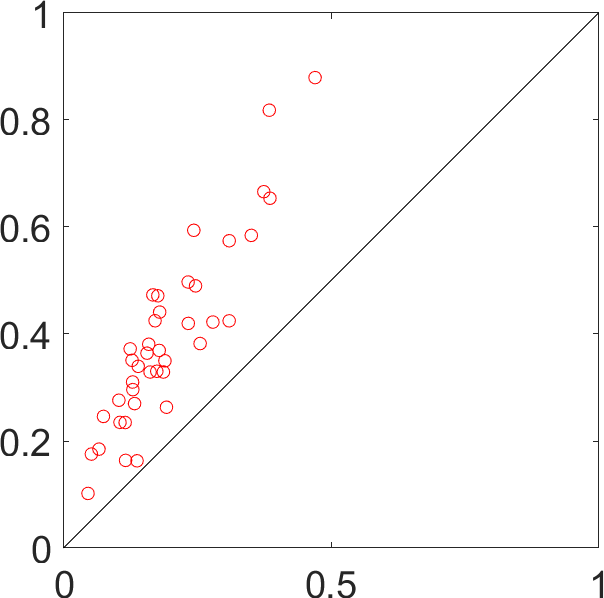} & \includegraphics[width = 1 in]{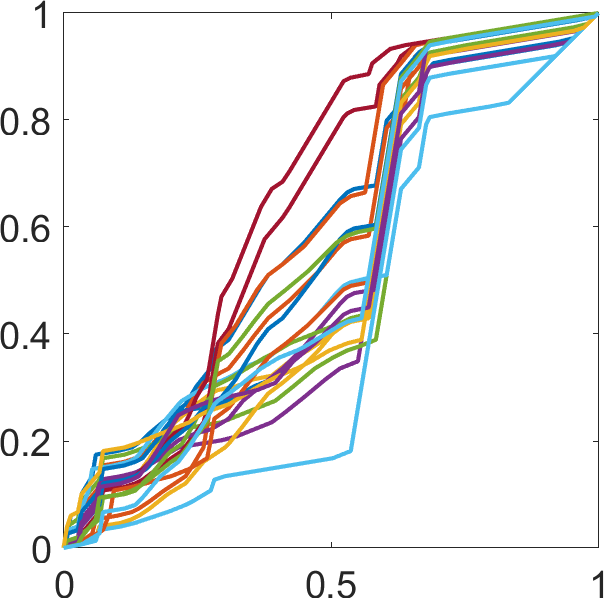}  &  \includegraphics[width = 1 in]{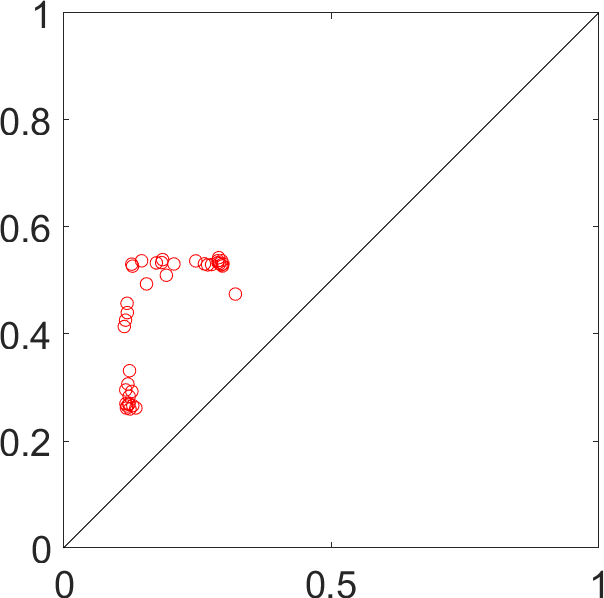} \\
        (h) & (i) & (j) \\
    \end{tabular}
    \end{center}
    \caption{\emph{Same topology with scale and sampling variabilities}: (a)\&(c) Two examples, from 20, of randomly generated point clouds. (b)\&(d) Corresponding persistence landscapes. The two rows correspond to the two component functions in each landsacpe: (e) Persistence landscapes $\{\Lambda_i\}_{i=1}^{20}$ of 20 point clouds. (f) Aligned persistence landscapes $\{\Lambda_i(\gamma_i)\}_{i=1}^{20}$. (g) Mean landscape after (blue) and without (red) alignment. (h) Noisy persistence diagrams $\{(b_{ij},d_{ij})\}_{i=1}^{20}$ from 20 point clouds. (i) Estimated reparameterizations $\{\gamma_i\}_{i=1}^{20}$. (j) Denoised persistence diagrams $\{(\gamma_i^{-1}(b_{ij}),\gamma_i^{-1}(d_{ij}))\}_{i=1}^{20}$.}\label{meanEst:2circles}
\end{figure}

\begin{figure}[!t]
\begin{center}
\begin{tabular}{cccc}
      \includegraphics[width = .6 in]{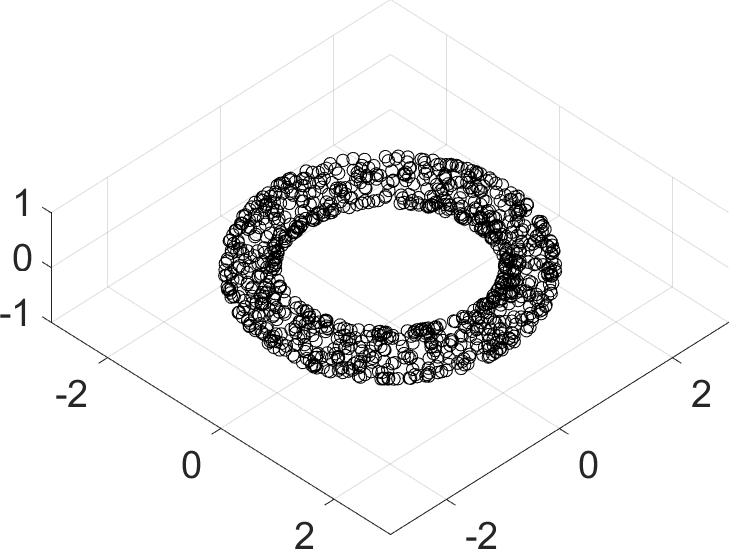} & \includegraphics[width = .7 in]{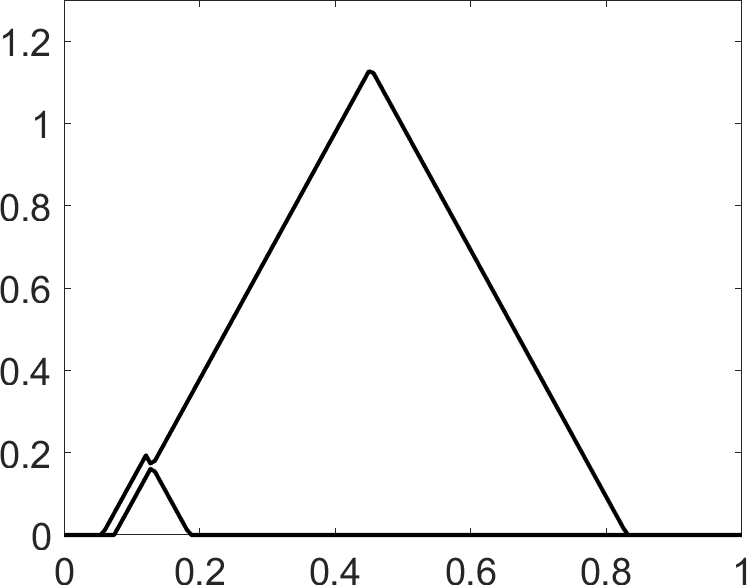}  & \includegraphics[width = .6 in]{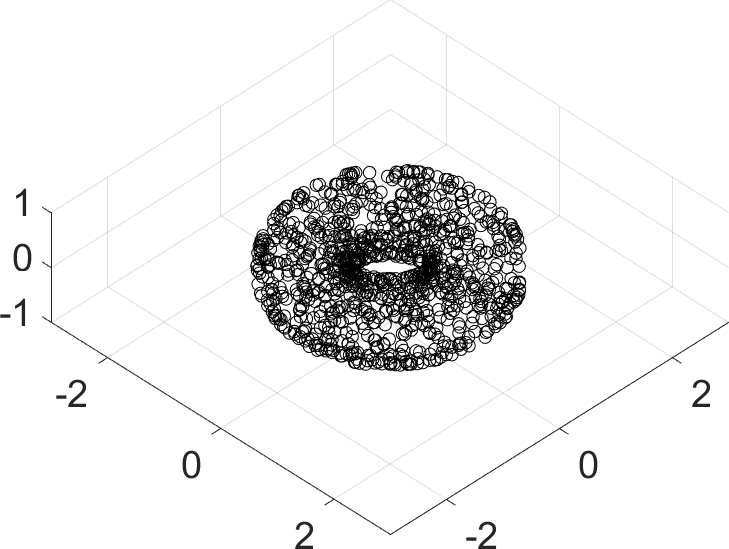} & \includegraphics[width = .7 in]{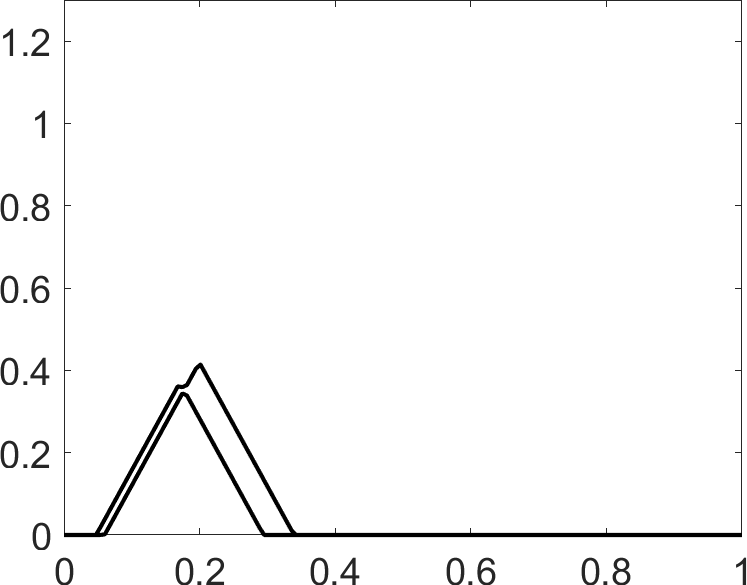}\\
     (a) & (b) & (c) & (d) \\
      \end{tabular}
      
      \begin{tabular}{ccc}
    \includegraphics[width = 1 in]{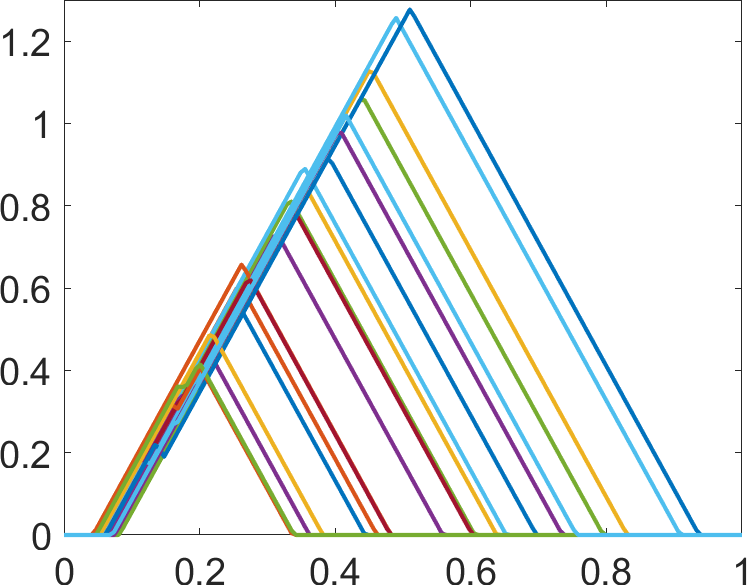} & \includegraphics[width = 1 in]{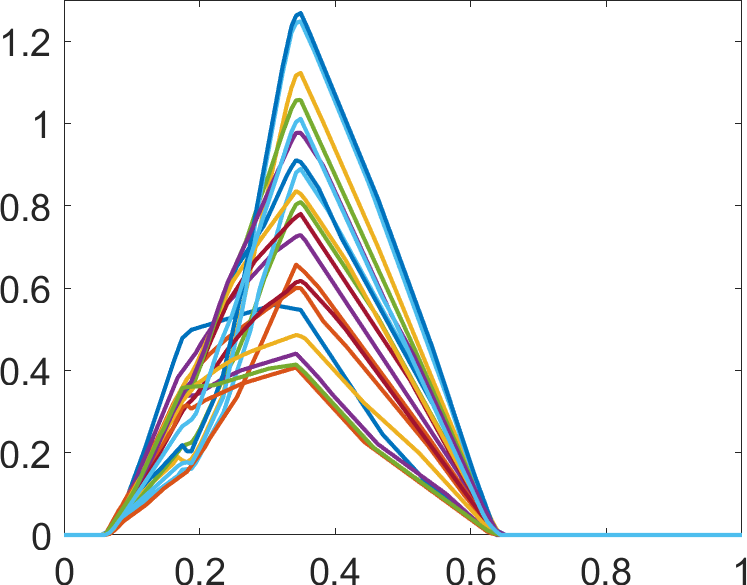}  &  \includegraphics[width = 1 in]{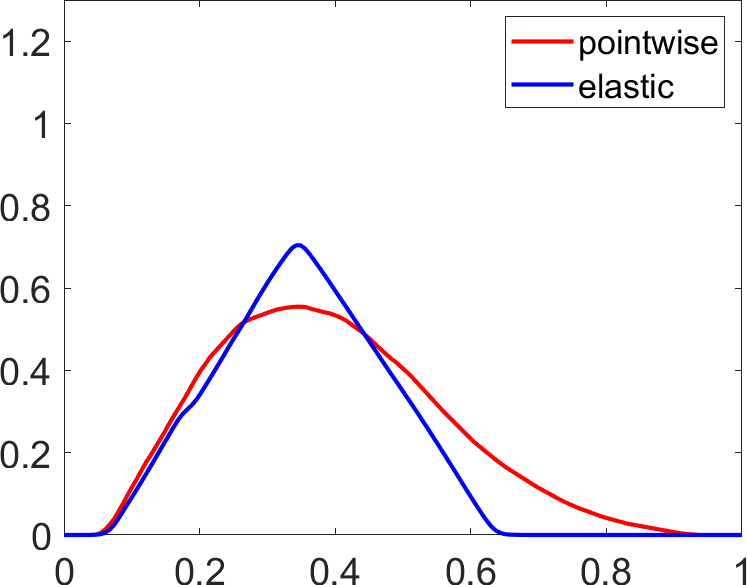} \\
        \includegraphics[width = 1 in]{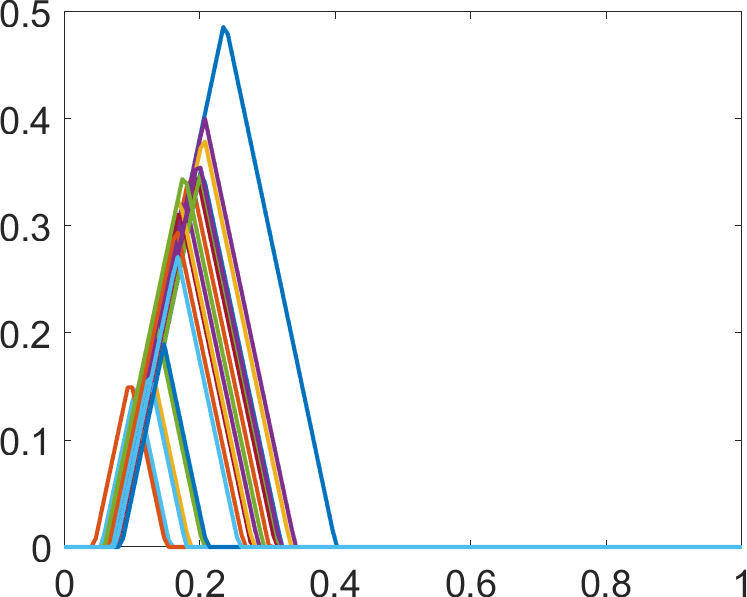} & \includegraphics[width = 1 in]{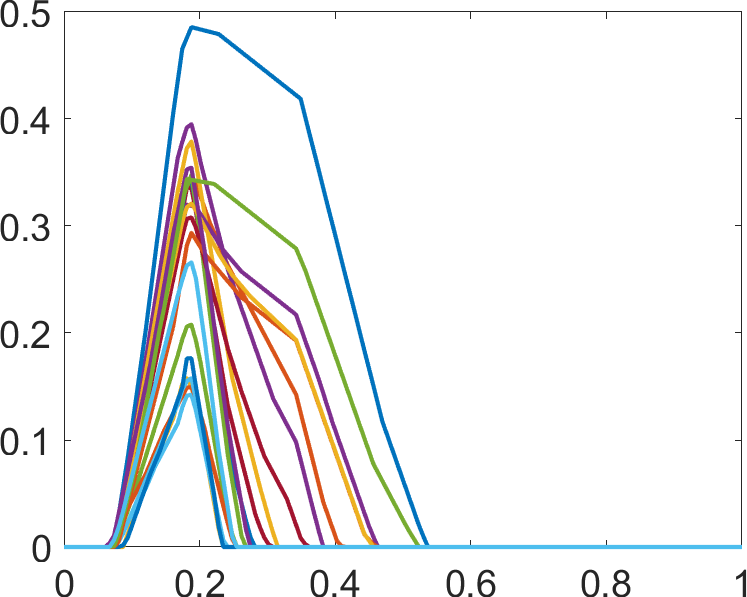}  &  \includegraphics[width = 1 in]{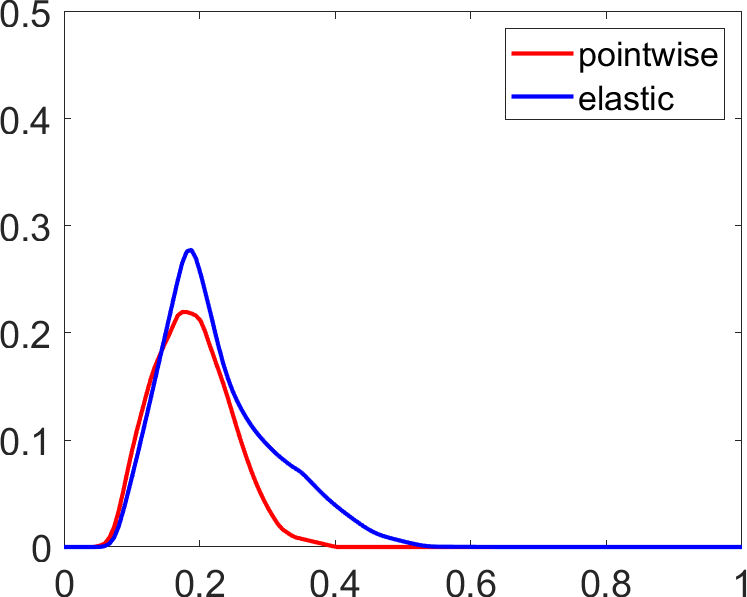} \\
         (e) & (f) & (g) \\
        \includegraphics[width = 1 in]{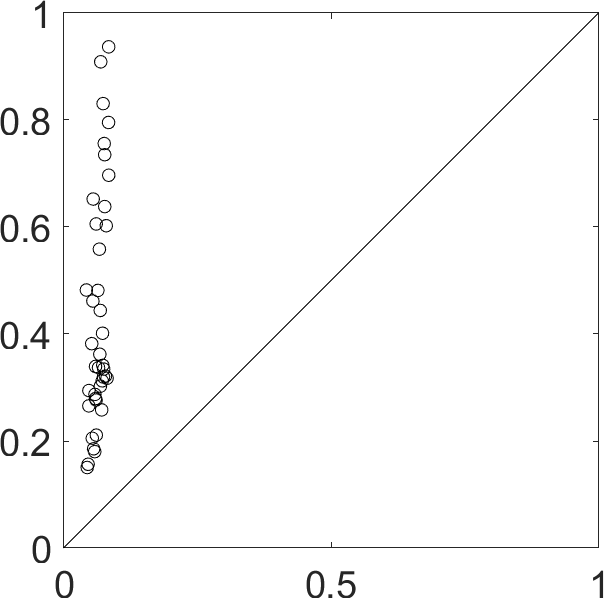} & \includegraphics[width = 1 in]{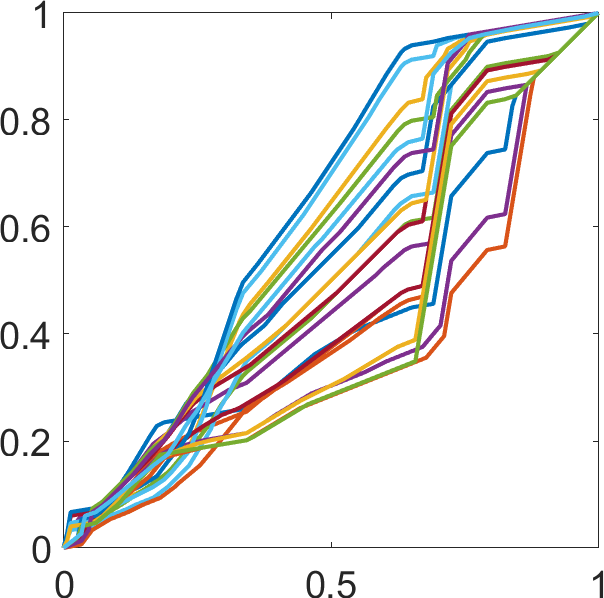}  &  \includegraphics[width = 1 in]{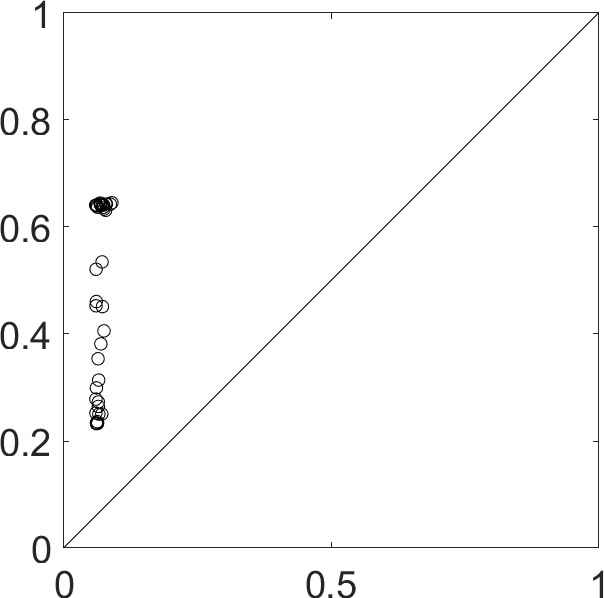} \\
        (h) & (i) & (j) \\
    \end{tabular}
    \end{center}
    \caption{\emph{Same topology with scale and sampling variabilities:} (a)\&(c) Two examples, from 20, of randomly generated point clouds. (b)\&(d) Corresponding persistence landscapes. With two rows corresponding to two component landscapes: (e) Persistence landscapes $\{\Lambda_i\}_{i=1}^{20}$ of 20 point clouds. (f) Aligned persistence landscapes $\{\Lambda_i(\gamma_i)\}_{i=1}^{20}$. (g) Mean landscape after (blue) and without (red) alignment. (h) (Rescaled) Noisy persistence diagram $\{(b_{ij},d_{ij})\}_{i=1}^{20}$ from 20 point clouds. (i) Estimated reparameterizations $\{\gamma_i\}_{i=1}^{20}$. (j) Denoised/transformed persistence diagram $\{(\gamma_i^{-1}(b_{ij}),\gamma_i^{-1}(d_{ij}))\}_{i=1}^{20}$.}\label{meanEst:torus}
\end{figure}

In Figure \ref{meanEst:2circles}, we consider mean estimation based on degree $p=1$, $K=2$-dimensional persistence landscapes computed from 20 point clouds that consist of uniformly sampled points along two circles with different radii. The point clouds in this example are generated in the same way as the data in the red group in Simulated Example 3 in the main article. Panels (a) and (c) show two examples of randomly generated point clouds with their corresponding landscapes in panels (b) and (d). Panels (e)-(g) show all 20 $k=1$ (top) and $k=2$ (bottom) landscape component functions, their alignment, and a comparison of the mean before (red) and after (blue) alignment, respectively. The proposed alignment procedure results in a mean landscape that better preserves the major features along both landscape components. On the other hand, the unaligned mean landscape destroys the prominent two peak structure in the first component.  
In panel (j), the points in the denoised persistence diagrams, corresponding to the two loops in the point clouds, form two separate clusters making the presence of these features in the data much clearer; the noisy diagrams shown in panel (h) do not provide such a distinction. 

In Figure \ref{meanEst:torus}, we consider mean estimation based on degree $p=1$, $K=2$-dimensional persistence landscapes computed from 20 point clouds that consist of 1000 points sampled uniformly on a ringed torus. The major radius of each torus is sampled from a $|N(2,.3^2)|$, while the minor radius is a proportion $\text{Beta}(10,10)$ of the major radius. Two example point clouds are shown in panels (a) and (c). We preprocess the persistence diagrams used to compute the persistence landscapes by disregarding all points except for the two with longest persistence, as these points correspond to the two loops formed by the torii that underlie the point clouds. The landscapes corresponding to point clouds in (a) and (c) are shown in (b) and (d), respectively. Clearly, the estimated landscapes can vary widely depending on the relationship between the major and minor radii. Panels (e)-(g) show the $k=1$ (top) and $k=2$ (bottom) landscape component functions, their alignment, and a comparison of the mean before (red) and after (blue) alignment. The first landscape function is automatically weighted higher during alignment due to the relatively large magnitude of the peak as compared to the second landscape function. In panel (j), the points in the denoised persistence diagrams, using the reparameterizations shown in (i), concentrate to make the presence of the features more clear; the two detected features describe the two loops. In comparison, the noisy diagrams shown in panel (h) do not provide a clear distinction.

\section{Correlation Between Age and Topological Structure of Brain Artery Trees Based on Unscaled Persistence Diagrams}

\begin{figure}[!t]
\begin{center}
\begin{tabular}{c}
\includegraphics[width = 1.5 in]{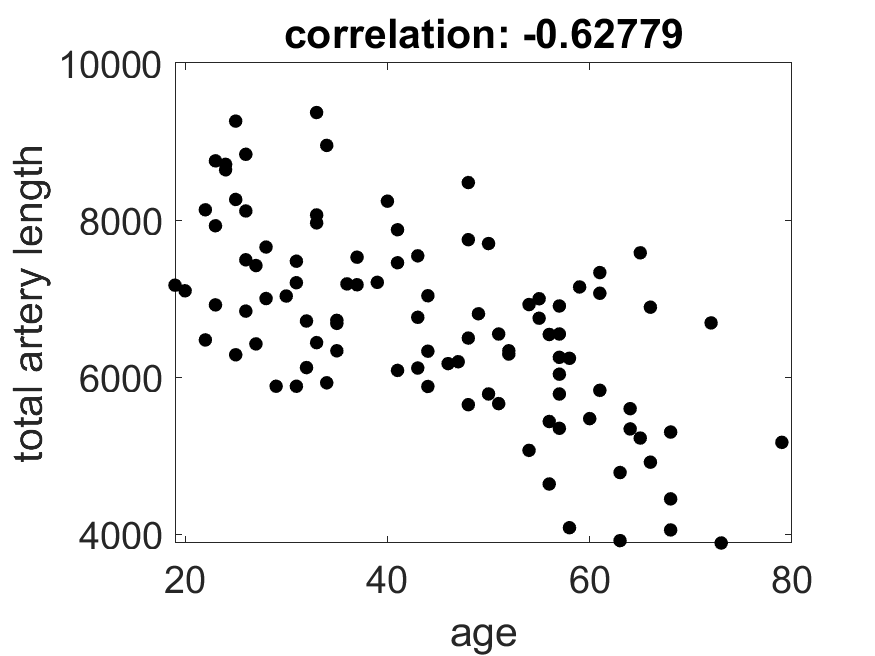}
\end{tabular}
\begin{tabular}{cc}
        \includegraphics[width = 1.5 in]{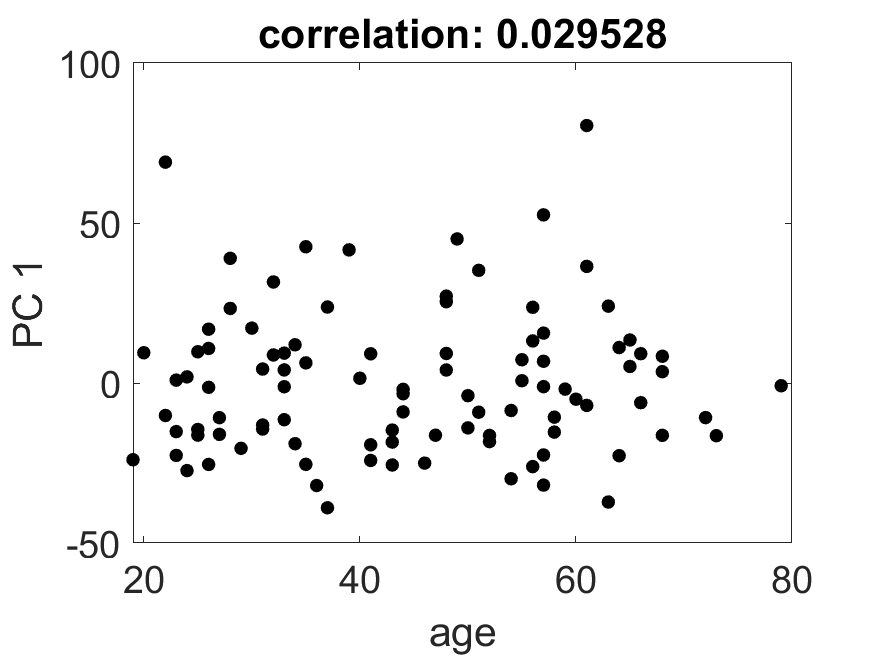} & \includegraphics[width = 1.5 in]{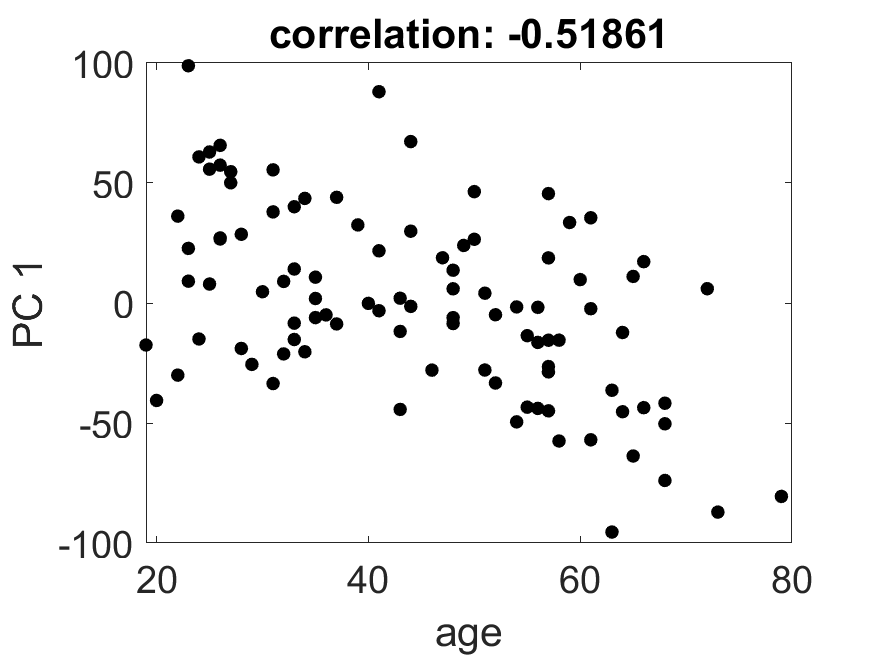}\\
        \includegraphics[width = 1.5 in]{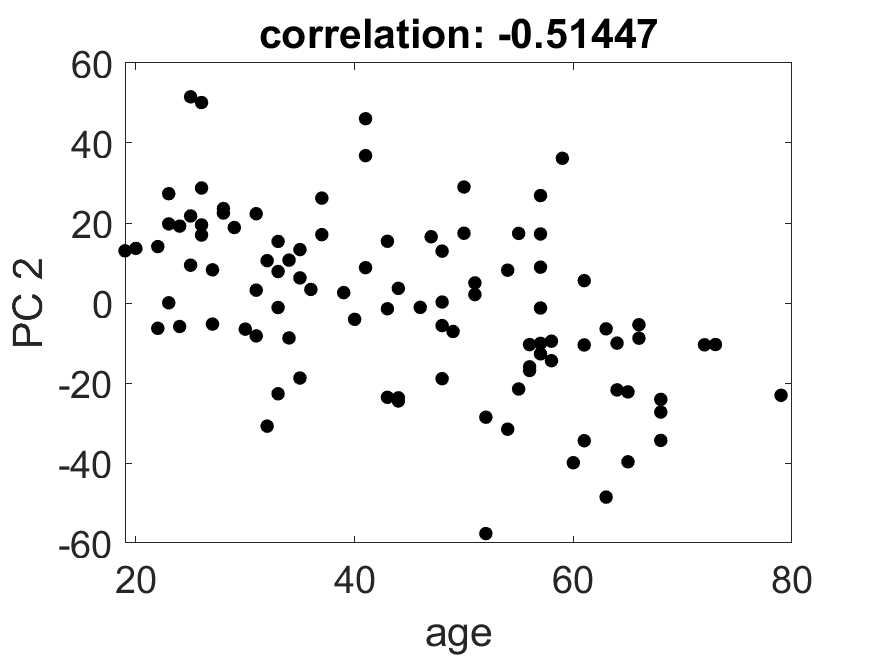}  & \includegraphics[width = 1.5 in]{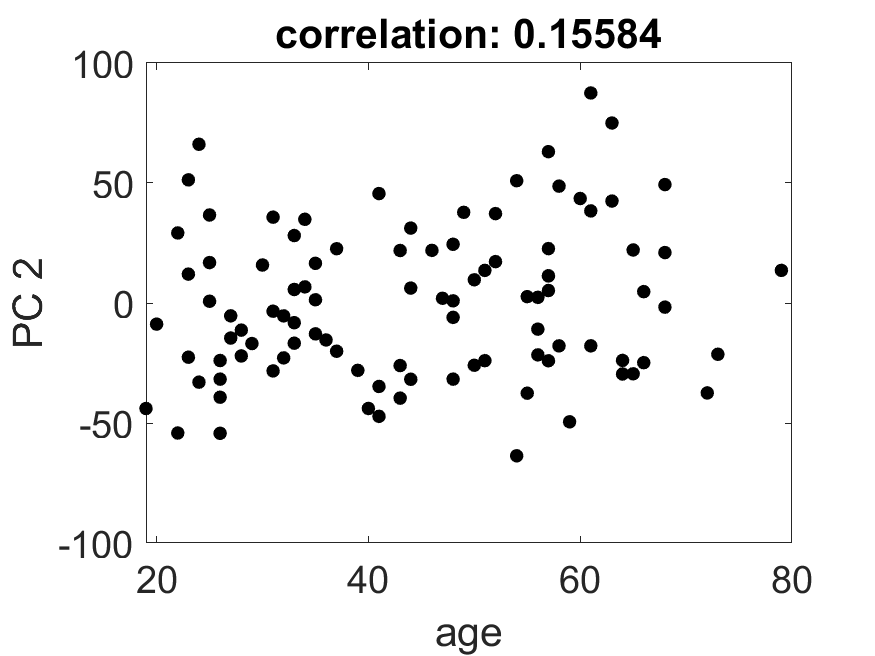} \\
       (b) & (c)\\
           \end{tabular}
    \caption{
    \emph{(a) Correlation and scatterplot of total artery length versus age. Correlations and scatterplots of PC 1 (top) and PC 2 (bottom) estimated using (b) aligned and (c) unaligned landscapes, computed from original persistence diagrams, versus age.}}
    \label{brainH0Pcsunscaled}
    \end{center}
\end{figure}

Finally, we repeat the analysis conducted in Section 4.2 in the main article, but using $p=1$, $K=100$-dimensional persistence landscapes derived from persistence diagrams that were not rescaled using total artery length. First, in Figure \ref{brainH0Pcsunscaled}(a) we show a scatterplot of total artery length versus age; it is clear that they are significantly correlated with total artery length generally decreasing with age. In Figure \ref{brainH0Pcsunscaled}(b), we show the scatterplots of the first (top) and second (bottom) PC scores, derived from aligned persistence landscapes. Figure \ref{brainH0Pcsunscaled}(c) shows the same, but based on unaligned landscapes. It appears that the first PC in panel (c) captures scale differences rather than topological ones (there is a high correlation with age, which is also highly correlated with total artery length). On the other hand, the first PC in panel (b) does not capture such scale differences as the correlation with age is low.

\end{document}